\documentclass{aa}

\usepackage[varg]{txfonts}
\usepackage{subfigure}
\usepackage{dcolumn}     
\usepackage{indentfirst}
\usepackage{hyperref}
\hypersetup{colorlinks=true,citecolor=blue}
\usepackage{breakurl}

\newcommand{\ttt}{\texttt}

\newcommand{\mtt}{\mathtt}
\newcommand{\ti}{$\times$}
\newcommand{\WS}{WISE\ti{}SDSS}
\newcommand{\mi}{$\mu$m}

\bibpunct{(}{)}{;}{a}{}{,}

\begin{document} 

   \title{Towards automatic classification of all WISE sources}

\author{A. Kurcz\inst{1,2}, M.~Bilicki\inst{3,2,4}, A.~Solarz\inst{5,2}, M.~Krupa\inst{1,2}, A.~Pollo\inst{1,5,2} \and  K. Ma{\l}ek\inst{5,2}}

   \institute{Astronomical Observatory of the Jagiellonian University,
ul.Orla 171, 30-244 Cracow, Poland\\
              \email{kurcz.agnieszka@gmail.com}
                           \and
                            Janusz Gil Institute of Astronomy, University of Zielona G\'ora, ul. Szafrana 2, 65-516 Zielona G\'ora, Poland\
                                          \and
Leiden Observatory, Leiden University, Niels Bohrweg 2, NL-2333 CA Leiden, the Netherlands
\and
              Astrophysics, Cosmology and Gravity Centre, Department of Astronomy, University of Cape Town, Rondebosch, South Africa
                            \and
              National Centre for Nuclear Research, ul.Ho\.za 69, 00-681 Warszawa, Poland
              }

   \date{Received ; accepted }

  \abstract
  {The Wide-field Infrared Survey Explorer (WISE) has detected hundreds of millions of sources over the entire sky. Classifying them reliably is, however, a challenging task owing to degeneracies in WISE multicolour space and low levels of detection in its two longest-wavelength bandpasses. Simple colour cuts are often not sufficient;  for satisfactory levels of completeness and purity, more sophisticated classification methods are needed.}
{Here we aim to obtain comprehensive and reliable star, galaxy, and quasar catalogues based on automatic source classification in full-sky WISE data. This means that the final classification will employ only parameters available from WISE itself, in particular those which are reliably measured for the majority of sources.} 
  {For the automatic classification we applied a supervised machine learning algorithm, support vector machines (SVM). It requires a training sample with relevant classes already identified, and we chose to use the SDSS spectroscopic dataset (DR10) for that purpose. We tested the performance of two kernels used by the classifier, and determined the minimum number of sources in the training set required to achieve stable classification, as well as the minimum dimension of the parameter space. We also tested SVM classification accuracy as a function of extinction and apparent magnitude. Thus, the calibrated classifier was finally applied to all-sky WISE data, flux-limited to 16 mag (Vega) in the 3.4 $\mu$m channel.}
  {By calibrating on the test data drawn from SDSS, we first established that a polynomial kernel is preferred over a radial one for this particular dataset. Next, using three classification parameters ($W1$ magnitude, $W1-W2$ colour, and a differential aperture magnitude) we obtained very good classification efficiency in all the tests. At the bright end, the completeness for stars and galaxies reaches $\sim95\%$, deteriorating to $\sim80\%$ at $W1=16$ mag, while for quasars it stays at a level of $\sim95\%$ independently of magnitude. Similar numbers are obtained for purity. Application of the classifier to full-sky WISE data and appropriate a posteriori cleaning allowed us to obtain catalogues of star and galaxy candidates that appear reliable. However, the sources flagged by the classifier as `quasars' are in fact dominated by dusty galaxies; they also exhibit contamination from sources located mainly at low ecliptic latitudes, consistent with solar system objects.}
    {}


\authorrunning{Kurcz et al.}
\titlerunning{Towards automatic classification of all WISE sources}
\maketitle

\section{Introduction}
The Wide-field Infrared Survey Explorer (WISE, \citealt{WISE}) is a space-borne telescope that has scanned the entire sky in four infrared (IR) bands (3.4 -- 23 $\mu$m) and has delivered one of the largest catalogues of astronomical objects to date. It has detected almost 750 million sources, which are compiled in the publicly released AllWISE Source Catalogue \citep{AllWISE}. WISE provides at present the most comprehensive census of the entire sky in the IR, and offers large advancement in comparison to earlier all-sky IR surveys, such as IRAS \citep{IRAS}, 2MASS \citep{2MASS}, or AKARI \citep{AKARI}.

Such a vast amount of data, which gives access to all-sky information in unprecedented volumes, has found multiple astronomical applications starting from our closest neighbourhood (Near-Earth Objects, nearby stars, and brown dwarfs), through the galaxies in the local volume, and up to the largest possible distances of high-redshift quasars \citep{WISE}. All these types of sources are present in the WISE database (within its sensitivity limits), but reliably extracting them in large numbers is challenging. Briefly, the WISE data release does not provide separate catalogues of different objects. What is more, at present there is  no separate point- and extended-source catalogues extracted from this survey, although efforts towards the latter are underway \citep{Cluver14}.

There are several reasons for the lack of comprehensive object identification in WISE. Firstly, this survey is mostly a near-IR selected one. Practically all the sources listed in the WISE catalogue have $S/N>2$ detections in the $W1$ band (3.4 \mi), and 83\% of them have $W2$ (4.6 \mi) measured with this accuracy; the detection rates in $W3$ (12 \mi) and $W4$ (23 \mi) are much lower\footnote{\url{http://wise2.ipac.caltech.edu/docs/release/allwise/expsup/sec2\_1.html\#stats}}. The light emitted at 3.4 and 4.6 \mi{} comes mainly from the photospheres of evolved stars. This means that in the low-redshift universe, where a large part of the extragalactic WISE sources are located, the $W1-W2$ colour of galaxies will be very similar to that of Galactic stars and cannot in general provide a comprehensive criterion for distinguishing  one from the other. This is readily seen in WISE colour-colour diagrams such as those provided by \cite{Jarrett11}. These diagrams also show that even adding the W3 band, which traces dust such as silicates and PAHs, is not sufficient to unambiguously distinguish  stars from elliptical galaxies. Last but not least, such colour-colour plots assume negligible photometric errors, while in reality significant scatter will occur, as is the case for the $W3$ and $W4$ bands, for which most of the WISE sources have only upper limits or are not even detected. 

Another way to separate  galaxies from stars in WISE could be  identifying extended sources in the sample. Similarly to  the case of the 2MASS Extended Source Catalogue (XSC, \citealt{2MASSXSC}), such sources in WISE are expected to be mainly extragalactic, especially at sufficiently high Galactic latitudes. However, despite much better sensitivity of WISE with respect to 2MASS (e.g.\ 0.054~mJy in $W1$ vs. 2.7~mJy in $K_s$) and the lack of atmospheric nuisances in the former, lower angular resolution of WISE and higher background levels mean that the eventual all-sky WISE XSC is expected to contain a similar number of sources to the 2MASS catalogue  ($\sim30$ per square degree; \citealt{Jarrett16}). This will be a very small percentage of all WISE galaxies: \cite{Cluver14} showed that only 3--4\% of WISE sources matched with the Galaxy And Mass Assembly (GAMA, \citealt{Driver11}) survey are resolved in $W1$, while  the WISE\ti{}GAMA sample itself is already much shallower than    expected from the full WISE galaxy catalogue (\citealt{WISC}; \citealt{Jarrett16}). One possible avenue leading to a WISE-based all-sky catalogue of galaxies of a similar depth to those in GAMA is to cross-match WISE sources with SuperCOSMOS data \citep{SCOS}, as was first discussed by \cite{2MPZ}. Such a catalogue has been recently compiled  \citep{WISC,Krakowski16}, but it includes only a part of the WISE galaxies owing to limitations of the SuperCOSMOS scans of photographic plates (both in depth and in the colour space). 

Until now, most of the studies dealing with WISE source classification were based on cross-matching this catalogue with other samples and using multiband  magnitudes and colours as discriminants. \cite{Stern12}, who paired up WISE with COSMOS, have proposed using $W1-W2 \geq 0.8$ mag (Vega) to identify WISE active galactic nuclei (AGNs). \cite{Assef13} have extended this work to a larger and deeper NOAO Deep Wide-Field Survey Boötes field and showed that this criterion is no longer optimal at fainter magnitudes. A more comprehensive effort has been undertaken by \cite{Yan13}, where WISE sources were cross-matched with SDSS to derive colour cuts for object selection. The \cite{Stern12} AGN identification has been confirmed and WISE colours (especially $W1-W2$ vs. $W2-W3$) have been shown to be sufficient to separate  star-forming galaxies from AGNs and stars from some galaxies,   although this was not the case for early-type, low-redshift  galaxies, which occupy practically the same region in the $W1$--$W2$--$W3$  colour-colour space as stars. More recently, \cite{Ferraro14} have defined their own colour cuts to identify galaxies and quasars from the WISE database;  however, this left position-dependent contamination visible in all-sky maps. \cite{Nikutta14} have explored WISE colours of Galactic and other nearby sources, while \cite{Mateos12} have used the XMM-Newton survey to define a WISE colour-based selection of luminous AGN.  The latter criterion has recently been applied to the all-sky WISE data by \cite{Secrest15} to select a sample of 1.4 million AGN candidates. We note, however, that their criterion of $\mtt{w1,2,3snr}\geq 5$ (signal-to-noise ratios in $W1$, $W2$ and $W3$) eliminates about 95\% of AllWISE sources from the parent catalogue, mostly owing to a very low level of WISE detection in the $W3$ channel. Last but not least, \cite{Jarrett16} have identified various source types in the G12 equatorial field by calibrating WISE magnitude and colour cuts on GAMA and SDSS spectroscopic data.

Some other WISE source classification studies where additional colours from external surveys were used include \cite{EdMa12} and \cite{Wu12} for QSOs/AGNs, \cite{TuWa13} for asymptotic giant branch stars, and \cite{KoSz15} for general star-galaxy separation in a WISE -- 2MASS PSC cross-match. Finally, \cite{Anderson14} compiled a catalogue of Galactic H II regions from WISE, based on their mid-IR morphology.

In the present paper, we go beyond simple colour and magnitude cuts and explore a more sophisticated classification of WISE sources. Our approach is based on automatised procedures of machine learning, and we use a specific algorithm -- the support vector machines (SVM) -- which has proven its aptitude for similar tasks within AKARI \citep{Solarz12} and VIPERS \citep{Malek13} surveys. A similar idea was explored in the independent study by \cite{KoSz15} where multiband photometry of those WISE sources that had cross-matches with the 2MASS Point Source Catalogue (PSC) was used for SVM-based source classification. Our analysis is  more general, as we do not limit the final source selection to a cross-match with an external catalogue. In addition, for the classification we use only the two shortest WISE bands in order to retain the highest all-sky completeness possible. By training the classifier on WISE data cross-matched with the tenth spectroscopic release of the Sloan Digital Sky Survey (SDSS, \citealt{SDSS.DR10}), we make the first step towards building reliable and comprehensive WISE catalogues of stars, galaxies, and AGNs/QSOs. At present, this classification is limited by the spectroscopic data that we used for training, but the methodology can be  extended with forthcoming data from different star, galaxy, and quasar catalogues. Solar system bodies should probably be included, as we find hints of them contaminating especially our final quasar candidate dataset.

The paper is organised as follows. Section \ref{Sec: data} describes the data used in our analysis: the photometric sample extracted from WISE (Sec. \ref{Sec: WISE}) and the spectroscopic sample from SDSS (Sec. \ref{Sec: SDSS}) used by the classification algorithm. In Section \ref{Sec: SVM}, we present the support vector machines classifier and how to apply it to imbalanced datasets (Sec. \ref{Sec: imbalanced}) such as ours. Various tests of the SVM method on WISE data are shown in Section \ref{Sec: tests}. Section \ref{Sec: All-sky} presents the application of the SVM classifier to a full-sky sample drawn from WISE data. We summarise our work in Section \ref{Sec: summary}.

All the WISE magnitudes in this paper will be given in the Vega system. For transformation to AB see \citet{Jarrett11}.

\section{Data selection}   
\label{Sec: data}

\subsection{WISE}
\label{Sec: WISE}

The WISE is a Medium Class Explorer mission funded by NASA and  launched in December 2009. With the use of a 40 cm space-based telescope, WISE has mapped the whole sky (with a total 47x47 arcmin field of view) in four infrared bands $W1$ -- $W4$, centred respectively at 3.4, 4.6, 12, and 23 \mi\footnote{ Recalibration of the  $W4$ effective wavelength   from 22 \mi{} was carried out by \cite{BJC14}.}, with an angular resolution of $6.1"$, $6.4"$, $6.5"$, and $12.0"$, respectively. Its $5\sigma$ point source sensitivities exceeded 0.054, 0.071, 0.73, and 5 mJy in the four respective bands;    in the $W3$ channel, for instance,  this is more than a hundred times better than that of IRAS at similar wavelengths. The publicly available AllWISE catalogue contains positional, photometric, quality, and reliability information and motion fit parameters for over 747 million sources \citep{AllWISE}.

The  goal of the present study is to obtain a comprehensive source classification for as many WISE objects as possible, and so for object selection we  decided to rely uniquely on the parameters provided in the WISE database as there is no other all-sky survey of  comparable depth currently available. The basic dataset we employ is the `AllWISE' data release\footnote{Available from the NASA/IPAC Infrared Science Archive at \url{http://irsa.ipac.caltech.edu/}.} \citep{AllWISE}, which was made publicly available in 2013 and combines data from the WISE cryogenic and NEOWISE \citep{NEOWISE} post-cryogenic survey stages. This dataset offers enhanced photometry and astrometry in comparison to the earlier WISE `All-Sky' release and includes estimates of source apparent motions.

We wanted to have the greatest sky coverage and depth  possible, and we thus chose to be flexible in the preliminary source selection for our catalogue. Regarding the photometry,  we use only the two shortest WISE bands, $W1$ and $W2$, and we employ additional quality parameters to ensure reliability of the sources. Our catalogue includes the sources that match the following criteria in the WISE database: $\mtt{w1snr}\geq 5$;  $\mtt{w2snr}\geq 2$; $\mtt{w?sat}\leq 0.1$ (no more than 10\% of saturated pixels in the respective bands, where \ttt{?} stands for 1 or 2); \ttt{cc\_flags[?] $\neq$ `DPHO'} (no severe artefacts). These  criteria ensure that the sources are detected in the two bands with reliable photometry (most of the objects have $S/N$ much higher than the limits used for the preselection; see below). There are over 606 million such sources in WISE; however, a large number of these are concentrated in the Galactic Plane, where the WISE data suffer from severe blending and saturation due to the enhanced source density. Our ability to classify data at low Galactic latitudes is thus very much compromised, and practically impossible within the Galactic Plane and Bulge.

An important caveat here is that the AllWISE catalogue is not complete at the very bright end ($W1< 8$~mag and $W2<7$~mag) owing to the saturation of such sources and also in two strips at ecliptic longitudes of $45^\circ <  \lambda < 55^\circ$ and $231^\circ < \lambda < 239^\circ$. Both these issues are related to instrumental limitations\footnote{For details see \url{http://wise2.ipac.caltech.edu/docs/release/allwise/expsup/sec2\_2.html\#cat_phot} and  \url{http://wise2.ipac.caltech.edu/docs/release/allwise/expsup/sec2\_2.html\#w1sat}.}, while the survey strategy causes additional patterns related to Moon avoidance manoeuvres\footnote{\url{http://wise2.ipac.caltech.edu/docs/release/allwise/expsup/sec1\_2.html\#survey}}. This will be reflected in the  all-sky maps that we are producing.

\begin{figure}
\centering
\includegraphics[width=0.49\textwidth]{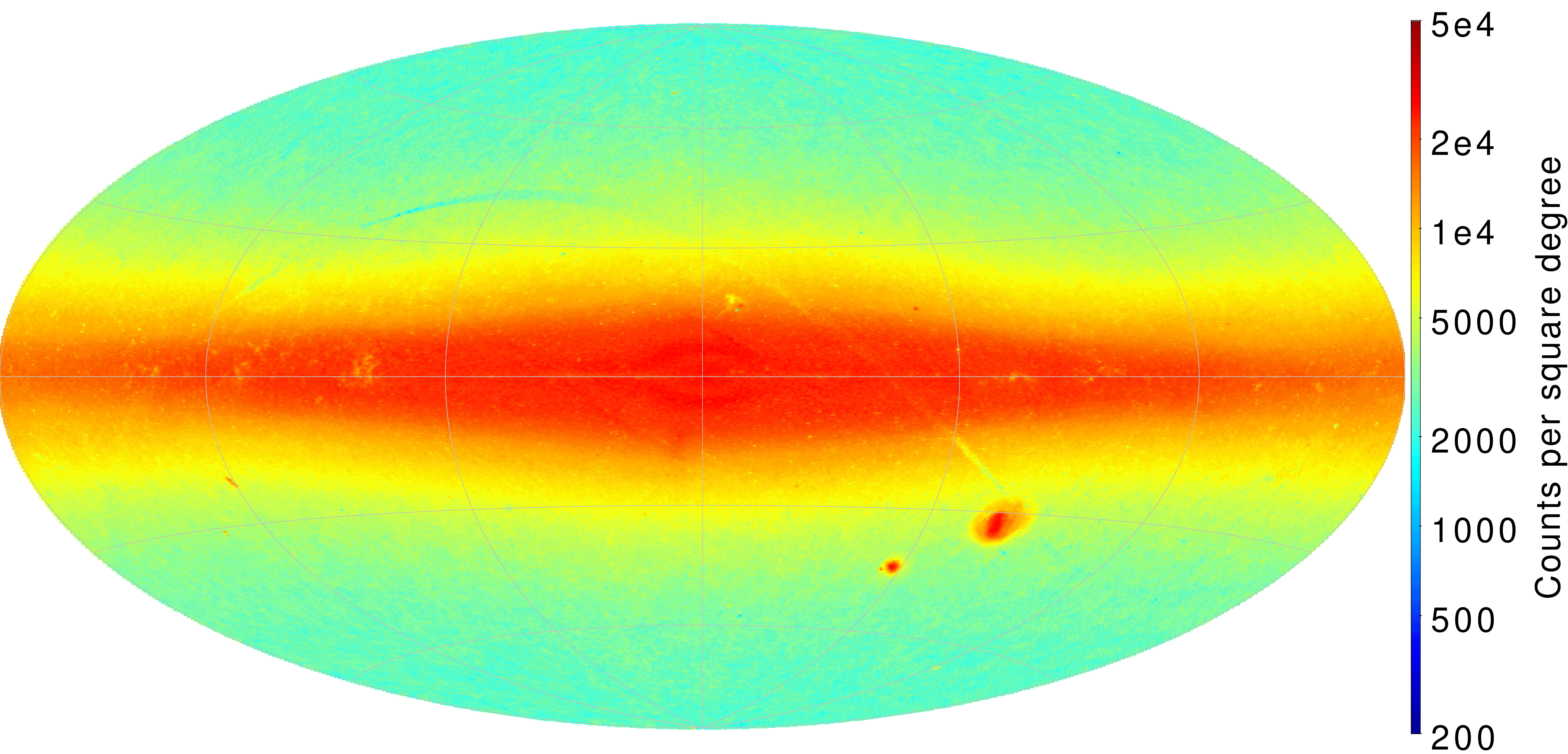} 
\caption{\label{Fig: AllWISE Aitoff}Aitoff projection in Galactic coordinates of 314 million sources in the AllWISE catalogue, flux-limited to $W1<16$ mag.}
\end{figure}

Here we are not  able to comprehensively classify all the WISE sources preselected as discussed above owing to the limitations brought about by the training set from the spectroscopic SDSS DR10 that we use. Namely, that dataset cross-matched with WISE practically does not provide galaxies fainter than Vega $W1<16$ mag\footnote{The situation has not improved in the final SDSS-III Data Release 12 \citep{SDSS.DR12}.}. This is one magnitude brighter than the average all-sky photometric completeness of WISE, so the present analysis will need to be extended once deeper training data become available. This should be possible in the coming years thanks to the plethora of spectroscopic surveys currently   underway. The all-sky sample of preselected $W1<16$ mag AllWISE sources  includes 314 million sources and is illustrated in the Aitoff projection in Fig.\ \ref{Fig: AllWISE Aitoff}.  We note  the logarithmic scaling of the counts and one order of magnitude larger source density in the Galactic Plane than at high latitudes. We also note that at this flux limit most of our sources have very reliable photometry, especially in the $W1$ channel. The median signal-to-noise ratios in $W1$ and $W2$ are respectively 31.3 and 16, and more than 99\% of the sources have magnitude errors smaller than $0.08$ mag for $W1$ and $0.28$ mag for $W2$.

In the machine learning procedure of source classification described later in the text we  use the following parameters provided by WISE: 
\begin{enumerate}
\item magnitude $\mtt{w1mpro}$ measured with profile-fitting photometry in the $W1$ band (hereafter $W1$);
\item colour $W1-W2$ defined as the difference in the $\mtt{w1mpro}$ and $\mtt{w2mpro}$ (hereafter $W2$) profile-fitting magnitudes;
\item a concentration parameter defined as the difference of two circular aperture magnitudes in the $W1$ channel, $\mtt{w1mag}\_1 - \mtt{w1mag}\_3$, measured respectively in radii $5.5"$ and $11"$ centred on the source; we note that these apertures were fixed, independent of the actual  size or shape of the sources, and were not corrected for contamination or bad pixels, thus they cannot be  used on their own as reliable measurements of fluxes for resolved sources.
\end{enumerate}
As already mentioned, measurements in the $W1$ band in our flux-limited sample have typically very high signal-to-noise ratios; the other two parameters used for the classification are somewhat noisier. The error in the $W1-W2$ colour is mostly driven by the less accurate $W2$ channel, and respectively 90\% (99\%) of the sources have $\delta(W1-W2)<0.16$ mag ($<0.29$ mag). For the concentration parameter, the same percentiles are $\delta(\mtt{w1mag}\_1 - \mtt{w1mag}\_3)<0.17$ mag ($<0.41$ mag).

We  also tested the usefulness of apparent motions for source classification. These motions, as provided in the AllWISE database ($\mtt{pmra}$ and $\mtt{pmdec}$), are composed of source proper motions and those due to the parallax, and  are expected to be different for various source types. We note, however, different caveats related to their measurements by WISE, discussed by \cite{Kirkpatrick14} and in the data description\footnote{\url{http://wise2.ipac.caltech.edu/docs/release/allwise/expsup/sec2\_6.html}}. The accuracy and signal-to-noise of AllWISE proper motions are strongly correlated with the source's flux, which means  they cannot  be reliably used for the whole catalogue (about 2\% of WISE objects have no motion measurements at all). We consider using them as a proof of concept for future datasets and surveys that will bring much more precise and comprehensive motion measurements, such as the MaxWISE proposal \citep{Faherty15}, Gaia \citep{Gaia} or LSST \citep{Ivezic2008}. The fourth parameter used in such a case was
\begin{enumerate}
\item[4.] apparent motion defined as $\mtt{pm} =(\mtt{pmra}^2+\mtt{pmdec}^2)^{1/2}$, where $\mtt{pmra}$ and $\mtt{pmdec}$ are the apparent motion in right ascension and declination, respectively.
\end{enumerate}
All the tests involving the apparent motions were carried out with the imposed condition on their signal-to-noise being larger than 1: $\mtt{pm}>\mtt{sigpm}$, where $\mtt{sigpm} = (\mtt{sigpmra}^2+\mtt{sigpmdec}^2)^{1/2}$ is the motion accuracy as provided in the database. This condition introduces a selection effect as it removes mostly faint, point-like sources. This constraint is  avoided in the final classification procedure when proper motions are not  used.

One can further increase the number of parameters used for machine learning classification; however,  it should   first be noted that every new parameter considerably extends computation time. In addition, many of the WISE database parameters are available or are sufficiently reliable only for a subset of all the sources. For instance, the two longer WISE bands, $W3$ and $W4$, which are often also used for source classification (\citealt{Ferraro14}; \citealt{KoSz15}), have much worse sensitivity than $W1$ and $W2$. Most of the WISE sources are not detected at the longer wavelengths or have only upper limits of $S/N<2$. In view of  classifying a considerable number  of  WISE objects (selected as discussed above), we have thus decided to limit ourselves to the basic information from the $W1$ and $W2$ bands. A possible extension employing more WISE parameters would first require determining  which of them are optimal, for instance thorough a principal component analysis \citep{Soumagnac15}. In addition, similarly to other applications of SVM in astronomy, our study does not take  the observational errors explicitly into account. For the present sample this should be a good approximation, as the noise level of the parameters we use for classification is relatively low (or even very low for the $W1$ magnitude). In Section \ref{Sec: summary} we  discuss how this issue can be dealt with in future work by using what is known as fuzzy logic.

\subsection{Training sample: WISE $\times$ SDSS DR10}
\label{Sec: SDSS}
Machine learning methods for source classification, like  the one we employ here, rely on the availability of a training sample that has relevant classes already identified. Ideally, this dataset should be as typical of the whole sample as possible. At present, however, such samples drawn from WISE are not available, and the only solution is to cross-match the WISE data with an external dataset that has relevant source types listed. Such an auxiliary dataset is provided by the Sloan Digital Sky Survey (SDSS, \citealt{SDSS}), which in its third phase (SDSS III, \citealt{SDSS.III}) comprises several dedicated star, galaxy, and quasar surveys; however, these three classes are  available only for the spectroscopic part of the SDSS (the photometric classes are  `stars', i.e.\ point-like, and `galaxies', i.e.\ resolved). For this reason we  chose to use only those SDSS sources that have spectra \citep{Bolton12}. Here we use the spectroscopic sample in the SDSS Data Release 10 (DR10, \citealt{SDSS.DR10}), which includes almost 3.4 million sources,   26\% of which are classified by the SDSS pipeline as stars, 59\% are galaxies, and the remaining 15\% are quasars/AGNs (class `\ttt{QSO}' in SDSS). We have cross-matched this sample with the WISE catalogue selected as described above, using a $1"$ matching radius, which gave us 2.1 million sources (18\% stars, 72\% galaxies, and 10\% QSOs). However, not all of them had SDSS spectra of sufficient quality, so in order to maintain the reliability of the training sample, we  filtered the sources according to redshift (velocity) quality, keeping only those with $\mtt{zWarning}=0$. Additional visual inspection of redshift and error distributions of \WS\ sources led us to eliminate the following outliers: $\mtt{zErr}>0.001$ for stars, $\mtt{zErr}>0.001$ or $\mtt{zErr/z}>0.1$ for galaxies, and $\mtt{zErr}>0.01$ for QSOs. This filtering left us with about 390,000 stars, 1.5 million galaxies, and 190,000 quasars in our training sample (i.e.\ present both in SDSS and in WISE), which reduces further to 120,000 stars, 620,000 galaxies, and 55,000 QSOs if the condition on the apparent motions to have $S/N>1$ is applied (Sect. \ref{Sec: WISE}).

\begin{figure}
\centering
\includegraphics[width=0.5\textwidth]{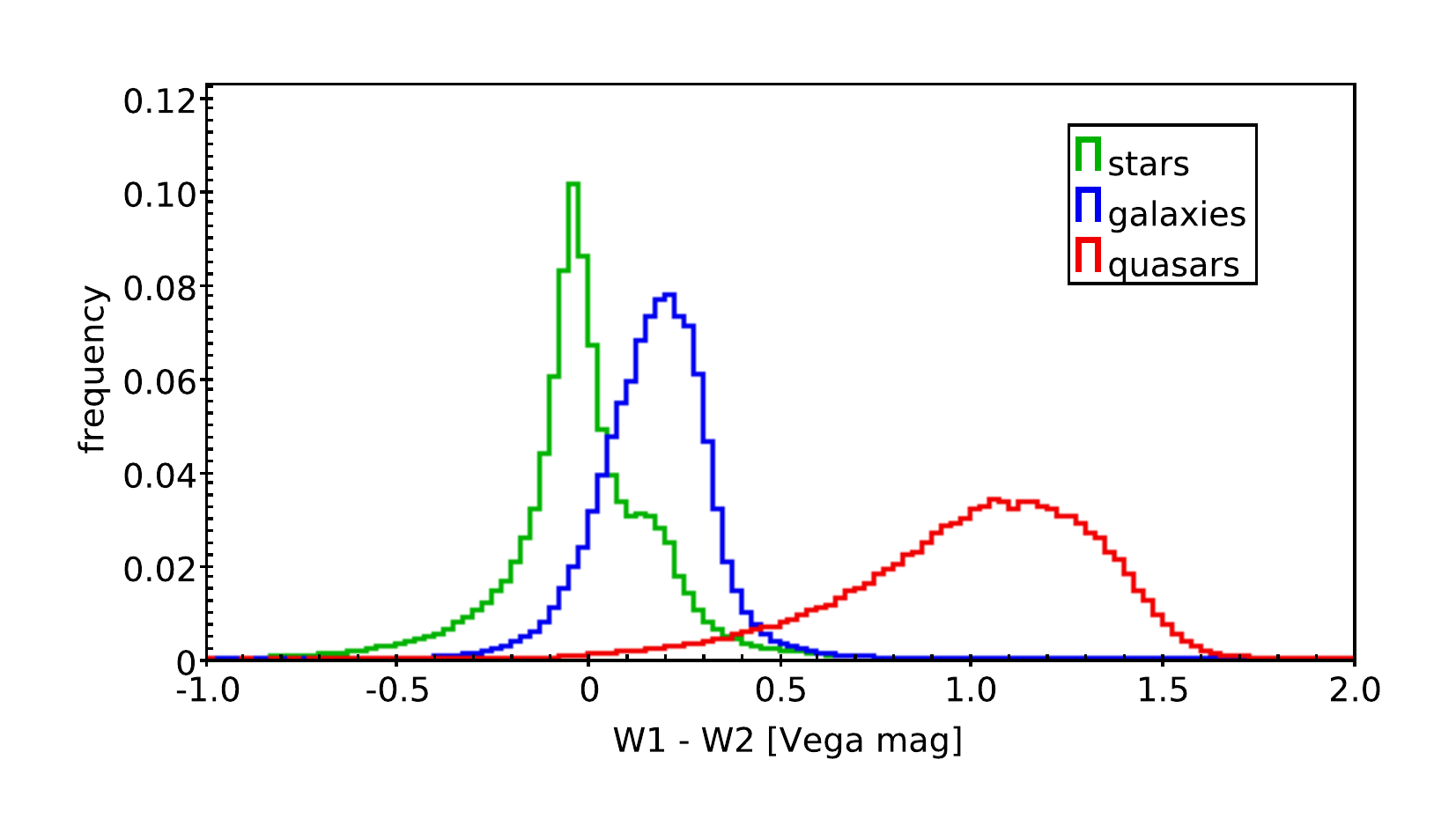}
\caption{\label{Fig: WISExSDSS W1-W2}Distributions of the observed $W1-W2$ colour for stars, galaxies, and quasars in the cross-matched \WS\ DR10 sample.}
\end{figure}

Figure\ \ref{Fig: WISExSDSS W1-W2} presents the $W1-W2$ colour histogram of our training sources (as observed, i.e.\ without extinction- or $k$-corrections applied). It clearly shows that while a simple selection in this colour may allow  a large fraction of WISE quasars to be identified (although the $W1-W2>0.8$ mag cut proposed by \citealt{Stern12} will miss some of them), it is not sufficient to reliably separate stars from galaxies. For instance, a constant $W1-W2$ colour cut will not produce samples that are both complete and pure at the same time. Maps presented in \cite{Ferraro14} also indicate  that applying fixed colour cuts to WISE data may leave position-dependent contamination. In the WISE$\times$SuperCOSMOS dataset  (\citealt{WISC}) this issue was partly alleviated by varying the star-galaxy colour separation as a function of distance from the Galactic Centre. Here we move beyond this simple methodology.

\begin{figure}
\centering 
\includegraphics[width=0.5\textwidth]{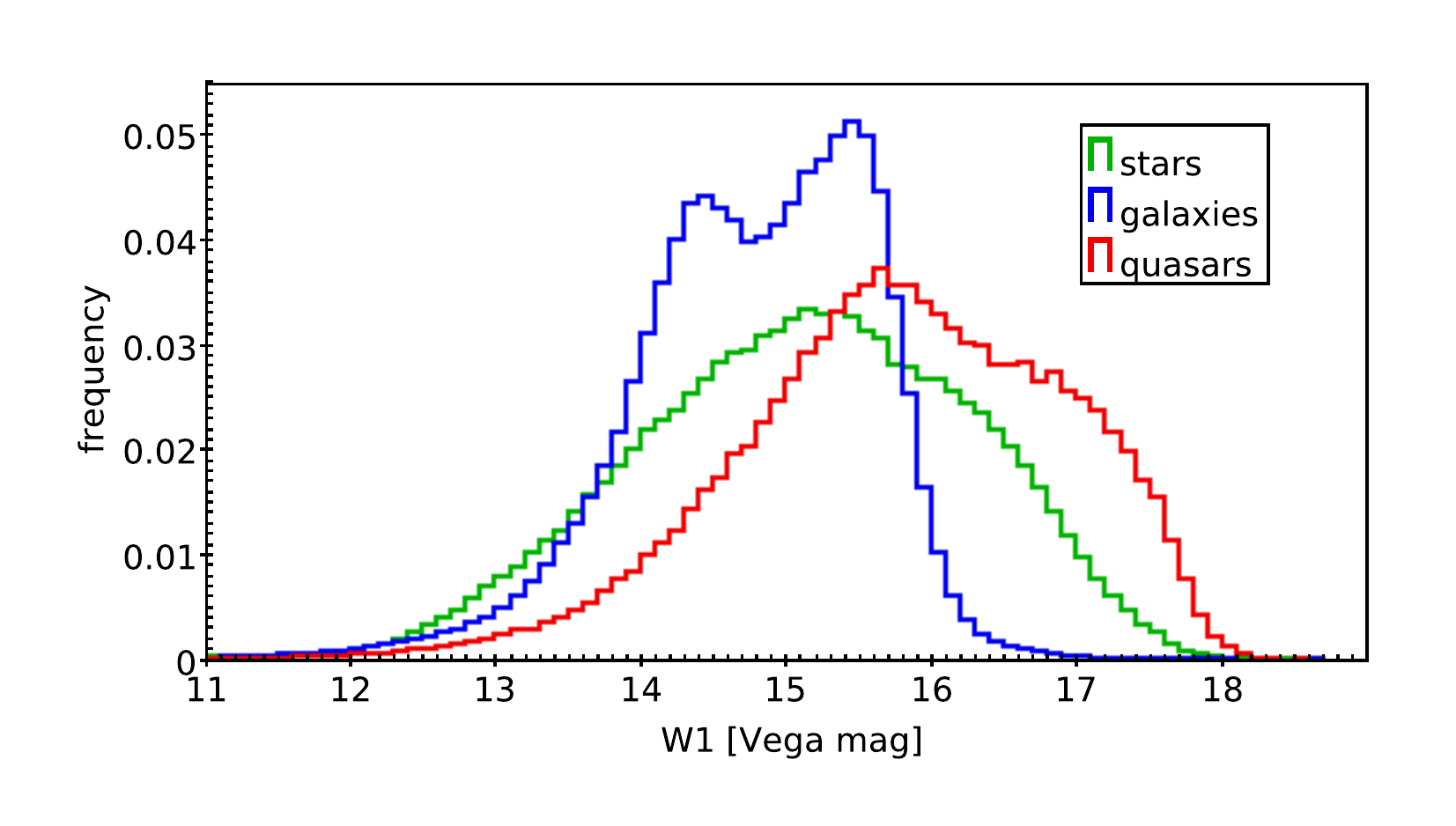}
\caption{\label{Fig: WISExSDSS W1 counts}Normalised apparent $W1$ magnitude counts for galaxies, quasars, and stars in the WISE\ti{}SDSS DR10 sample.}
\end{figure}

\begin{figure*}
\begin{minipage}{0.5\textwidth}
\centering
\includegraphics[width=0.8\textwidth]{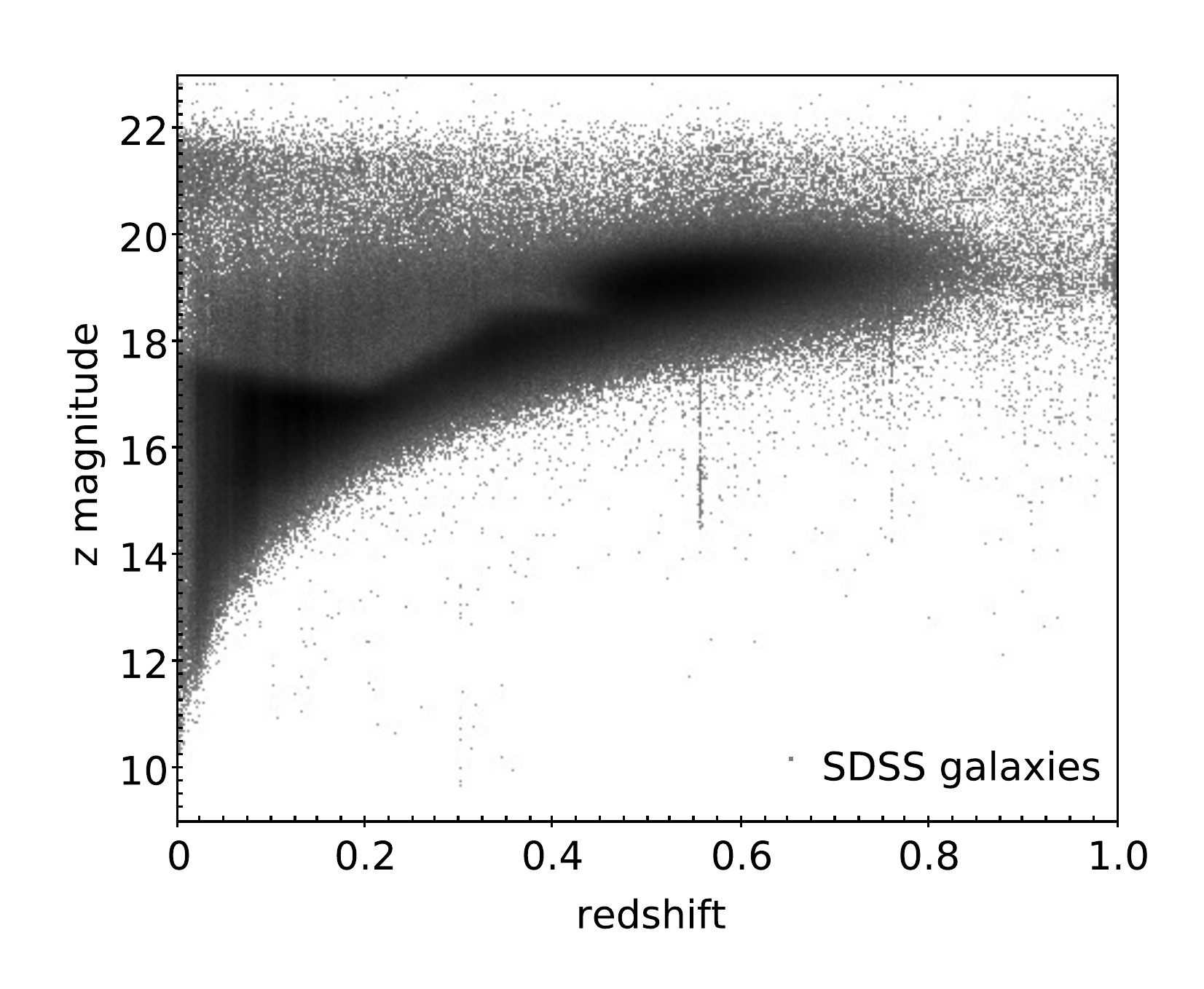}
\end{minipage}
\begin{minipage}{0.5\textwidth}
\centering
\includegraphics[width=0.8\textwidth]{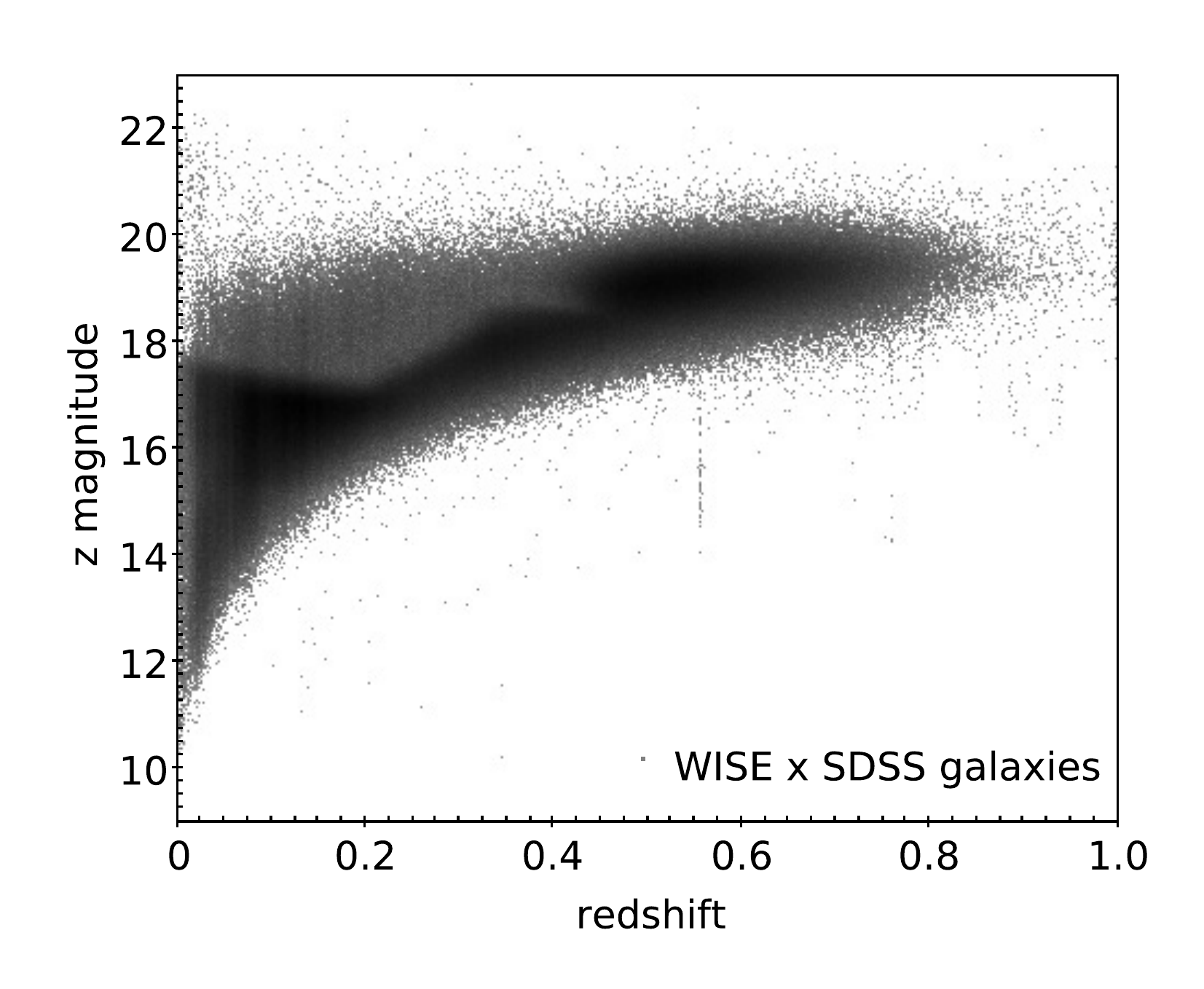}
\end{minipage}
\caption{Redshift -- apparent magnitude diagrams for the SDSS $z$ band: SDSS-only galaxies (left panel) and galaxies from the \WS\ cross-match (right panel).}
\label{Fig: redshift vs z mag}
\end{figure*}

\begin{figure}
\centering 
\includegraphics[width=0.41\textwidth]{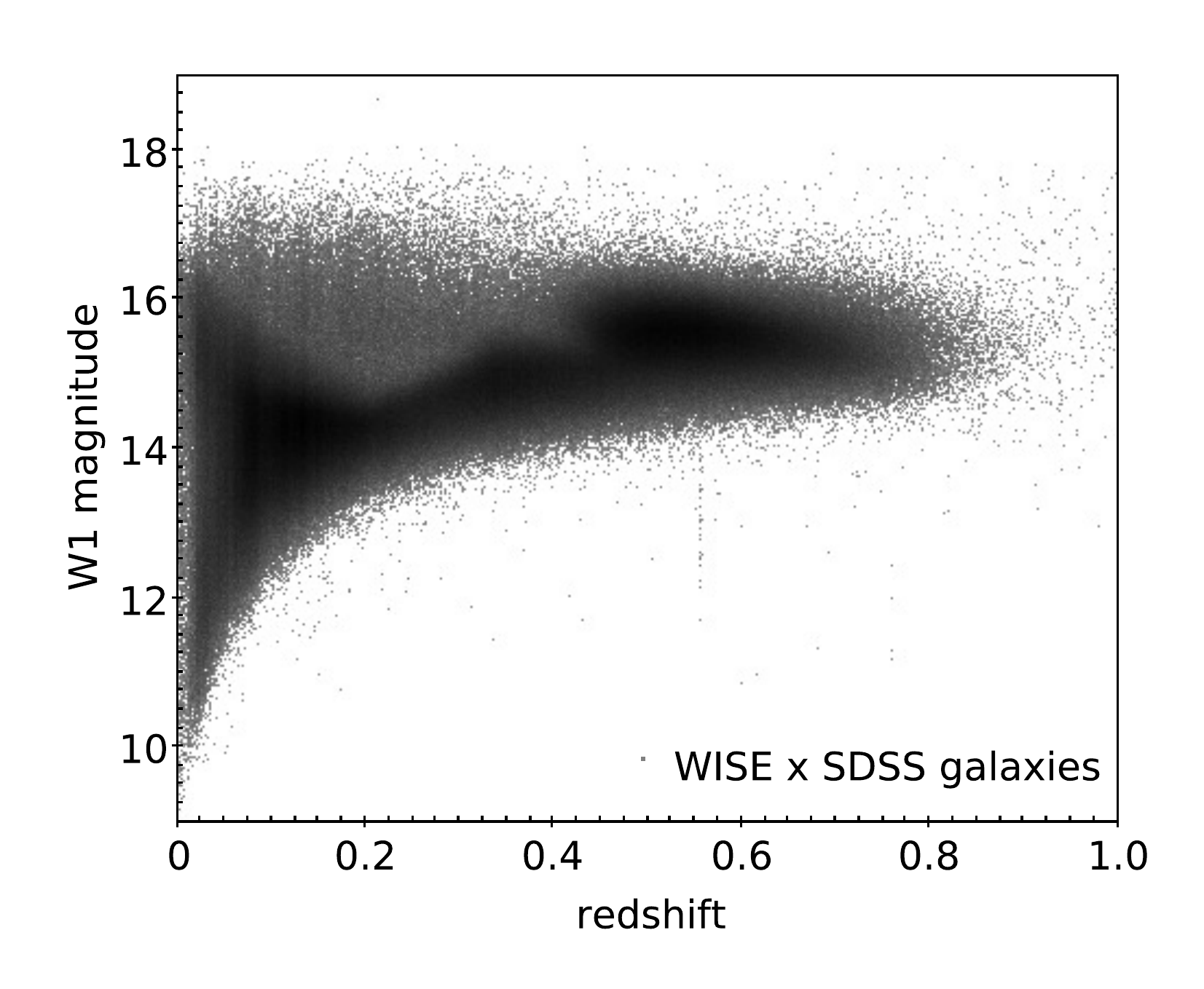}
\caption{Redshift -- apparent magnitude  diagram for the WISE $W1$ band for the galaxies from the \WS\ cross-match.}
\label{Fig: redshift vs W1 mag}
\end{figure}

\begin{figure}
\centering 
\includegraphics[width=0.43\textwidth]{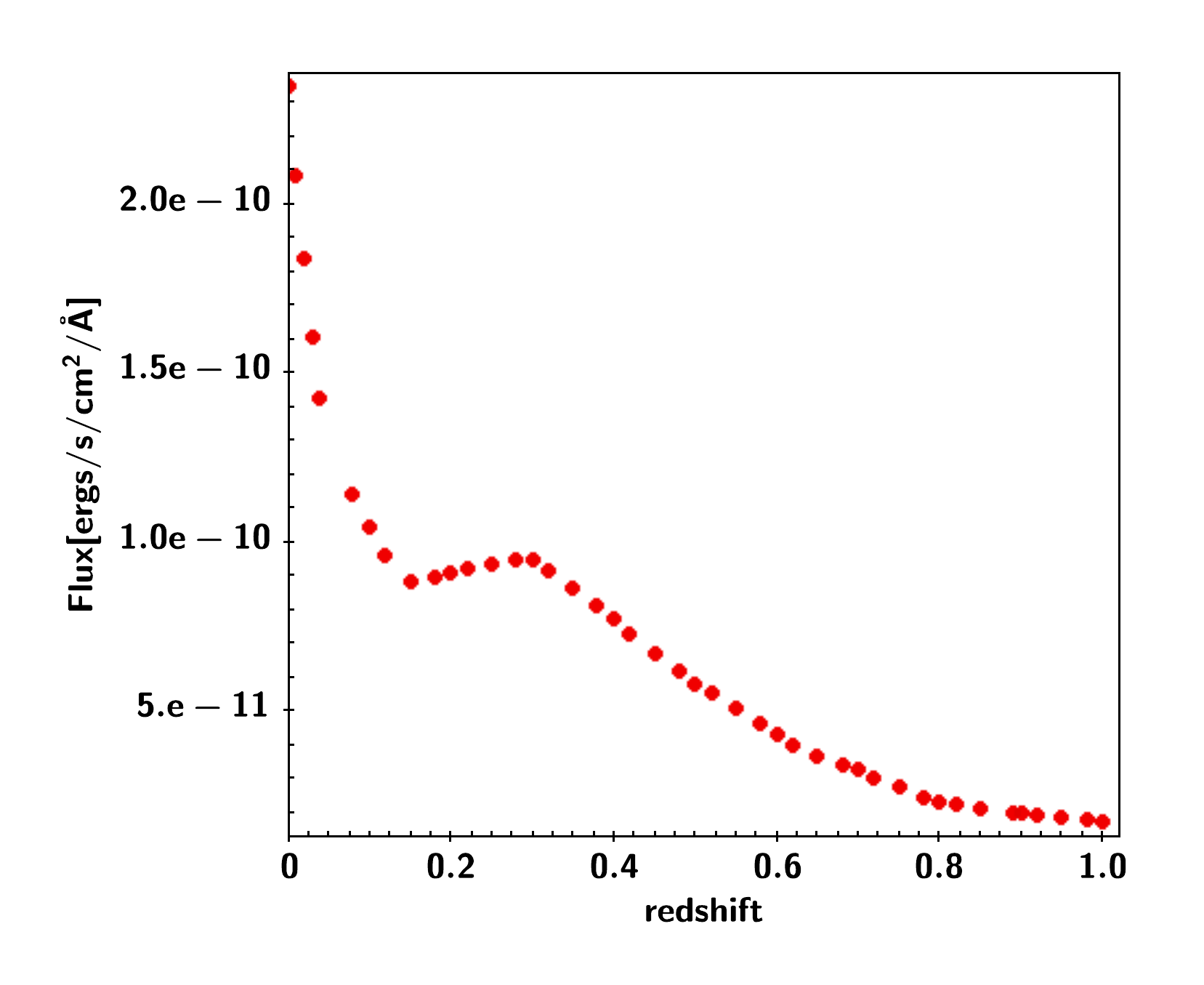}
\caption{Convolution of the $W1$ filter with a template spectrum of a typical spiral galaxy as a function of redshift.}
\label{Fig: W1 convolution}
\end{figure}

As already mentioned in Sec.\ \ref{Sec: WISE}, using the SDSS as the training sample imposes restrictions on the depth up to which we can classify WISE sources in the present application. As shown in Fig.\ \ref{Fig: WISExSDSS W1 counts}, presenting normalised $W1$ counts for the three source types in the \WS\ cross-match, there are hardly any galaxies  fainter than $W1=16$ mag, so we are not  able to create reliable training samples beyond this magnitude, although both stars and galaxies are present in the \WS\ sample at considerably fainter fluxes. The histogram for galaxies also displays  two clear peaks. This shape for the galaxy counts can be attributed to the combination of two effects:  one  results from the heterogeneity of the SDSS dataset itself (due to preselections of the Main, CMASS, and BOSS samples) and the other  is related to the selection of the \WS\ catalogue, which is also  based on the detections in the $W1$ filter. First, as demonstrated in Figs.\ \ref{Fig: redshift vs z mag} and \ref{Fig: redshift vs W1 mag}, the SDSS-based galaxy selection does not sample all the regions of the redshift-magnitude space equally. In particular, as illustrated in the left panel of Fig.\ \ref{Fig: redshift vs z mag} which shows the redshift--magnitude diagram for the $z$ band (the longest wavelength measured by SDSS), in the pure SDSS data we observe a clear depletion  of faint red galaxies at $z \le 0.3$ in comparison to higher redshifts. This effect persists after adding the $W1$-based selection of the \WS\ sample, as shown in the right panel of Fig.\ \ref{Fig: redshift vs z mag}. The enhancement of this effect can be attributed to the properties of the $W1$ filter, which at low redshifts probes the part of the galaxy spectrum where strong PAH features are very prominent. Second, in Fig.\ \ref{Fig: W1 convolution} the convolution of the $W1$ filter with a spectrum of a typical spiral galaxy at various redshifts (taken from the \citealt{brown} library) demonstrates that observing in $W1$ we can expect a selection function which does not change monotonically with redshift. In particular,  it has a minimum at $z \sim 0.15$, and then rises again until $z \sim 0.28$. The combination of these effects results in a relatively complex sampling of galaxies in the redshift -- $W1$ magnitude space in the \WS\ data, as demonstrated in Fig.\ \ref{Fig: redshift vs W1 mag}.

Finally, we cannot hope for reliable classification also at the very bright end. The WISE $W1<8$ mag or $W2<7$ mag sources  are saturated; in addition,    the \WS\ sample  does not include galaxies or quasars brighter than $W1\sim9.5$ mag. These brightest objects are thus removed from our final samples, which  has a minor influence on the results because the $W1<9.5$ mag WISE sources are concentrated mostly in the Galactic Bulge (i.e. they are stars and blends thereof).

We are aware of all these biases, and we would like to note that introducing them is a trade-off if we want to use the training sample providing star, galaxy, and quasar spectral classifications.  At the depths we are  interested in, these spectral classifications are currently available only from the SDSS. As a final caveat, the training sample applied in this study does not include solar system bodies; however, they   are present in the WISE database. Our final classification  thus ignores this contamination, which  affects mostly the sources flagged as quasars by our classifier.

\section{Classification method: support vector machines}
\label{Sec: SVM}

In general, classification is a process that uses pattern recognition. A classifier is a function that maps a feature vector of a given object's characteristics into a discriminant vector containing likelihoods that the objects  belong to the different considered classes. Classification schemes rely on choosing a feature space where different classes occupy different volumes with minimal overlapping. This approach has been used to develop machine learning algorithms -- statistical methods which constitute a branch of artificial intelligence and are based on creating and exploiting systems which learn from data.

In this work for the task of identifying object types we adopt the support vector machines (SVM) algorithm. This supervised method based on kernel algorithms (\citealt{shawetaylor04}) was designed to extract structures from data, and thanks to its excellent ability to deal with multidimensional samples combined with its high accuracy, it has been extensively applied to many diverse astronomical problems. To name a few, SVMs have been used to solve problems like classifying different structures in the interstellar medium \citep{beaumont11}, pinpointing active galactic nucleus (AGN) candidates \citep{cavuoti14}, or distinguishing different subclasses of specific spectral type stars \citep{bu14}. Last but not least, and of particular relevance here, SVM has been proven efficient in classifying different objects, such as stars, quasars, and galaxies (e.g. \citealt{saglia12,Solarz12,Malek13,KoSz15}).

In what follows, we draw the general outline of the nature of the algorithm; for an in-depth discussion we refer the reader to \citet{vapnik99,crist00,hsu03}.

Each training object can be described by a number of quantities,  $N$, which determine its discriminating properties.
The SVM regards the values of the quantities as a position of a given object in an $N$-dimensional parameter space; in other words, the algorithm maps the feature vector from the input space $X$ to a feature space $H$ using a non-linear function $\phi : X \rightarrow H$. In the feature space $H$, the discriminant function, which will determine the boundary, takes the form of
\begin{equation}
f(x)=\sum^{n}_{i=1} \alpha_i k(x,x')+b.
\label{Eq: discrim function}
\end{equation}
Here $k(x,x')$ represents the kernel function, which returns the inner product of the mapped vectors; $\alpha_i$ is a linear coefficient; and $b$ is a perpendicular distance called bias, which translates the discriminant function into a given
direction.

With a substantial amount of feature vectors representing different classes of objects, the algorithm searches for boundaries segregating those classes with the biggest possible distance from each data point (a margin). The objects lying closest to the boundary are called {\it support vectors}. In other words, the SVM algorithm searches for a decision boundary $B$ that will maximise a fitness function $F$ 
\begin{equation}
F=M-C\sum\limits_{i}\xi_{i}(B,M),
\end{equation}
where $M$ denotes the margin of the boundary. The number of training examples violating this criterion is given by $\xi_{i}(B,M)$. If a position of a point $i$ is found within a distance higher than $M$ from $B$, then $\xi_{i}=0$. In the opposite case, $\xi_{i}$ will be equal to the distance that point $i$ should be shifted  so that the condition is satisfied.
To set a trade-off between the large margins $M$ and misclassifications $\xi_{i}$, an adjustable cost parameter is used (more details in \citealt{beaumont11}).

Using kernel functions allows a shift into a higher dimensional parameter space, where  the data are usually  much more simply separated than in lower dimensional space. There are many possible kernel functions that can be used (such as polynomials, exponential radial basis functions, or multilayer perceptrons; \citealt{crist00}). The choice of the proper kernel   suitable for a given problem is crucial; the usual procedure is to try several, beginning from the simplest cases (to avoid overfitting and to save on parameter tuning time) and then to move towards more complex ones in order to gain accuracy. For this particular dataset we tested two kernel functions to obtain the most reliable classification outcome: the Gaussian radial basis function (GRB) and the polynomial function.
The GRB is given by
\begin{equation}
 k({\bf x},{\bf x}')=\exp(-\gamma ||{\bf x}-{\bf x}'||^2),
\label{Eq: GRB function}
\end{equation}
where ${\bf x}$ and ${\bf x}$' represent feature vectors in the input space, $||\cdot||$ denotes the Euclidean distance, and $\gamma$ is the adjustable kernel width parameter, which is responsible for the curvature of the decision surface. 
The polynomial kernel is defined as
\begin{equation}
 k({\bf x},{\bf x}')=(\gamma({\bf x}\cdot{\bf x'})+c_0)^{d},
\label{eqpoly}
\end{equation}
where ${\bf x}$ and ${\bf x}$' represent feature vectors in the input space, ${\bf x}\cdot{\bf x'}$ is their inner product, $d$ stands for the degree of the polynomial function, and $c_0$ is a constant coefficient. 

Therefore,  for the GRB kernel there are two adjustable parameters, $\gamma$ and $C$, which will determine the separation boundary and complete the training of the SVM classifier. 
In the case of the polynomial kernel, the number of adjustable parameters increases to four: in addition to $\gamma$ and $C$, the degree $d$ and the coefficient $c_0$ have to be known.
Then, after the most efficient kernel function is chosen, a classification of the new data points depends on their  position relative to the boundary: SVM will assign a type to unknown objects based on which side of the separation hyperplane they fall.

Furthermore, instead of assigning discrete class labels, it is possible to determine  the class probability for a given object. In the case of binary SVM  this can be done by implementing  Platt's a posteriori probabilities (\citealt{platt2000}; \citealt{linlinweng}): once the decision values $f$ of the SVM classifiers are computed, a sigmoid function
\begin{equation}
 P(i|i \mbox{ or }j,{\bf x}) =\left(1+e^{Af+B}\right)^{-1}
\end{equation}
 is fitted (where $i$ and $j$ represent two classes). Then, $A$ and $B$ are estimated by minimising the negative log-likelihood function. 
In order to extend the probabilities of classes to a three-class problem, all class probabilities from output of binary classifiers are combined \citep{Wu03}. The probabilities calculated this way are  used in the final classification to eliminate sources that have low probabilities of belonging to any class (meaning that each class has $p<0.5$, and in some cases the three probabilities are $p\sim0.33$).

Support vector machines are currently available in a variety of software packages; the most widely used  is \texttt{libsvm} \citep{libsvmcite}\footnote{\url{http://www.csie.ntu.edu.tw//~cjlin/libsvm/}}, which provides robust implementation of SVM for both classification and regression. 
In this work we use the \texttt{R} \citep{rcite}\footnote{\url{http://www.R-project.org}} implementation of SVM included in the \texttt{e1071} package \citep{e1071cite}, which provides an interface to \texttt{libsvm}.

\subsection{SVM on imbalanced datasets: oversampling}
\label{Sec: imbalanced}

\begin{table*}

\caption{\label{Table: Oversampling}
        Numbers of objects before and after oversampling in the bins for which oversampling was applied. }
\centering
\begin{footnotesize}
                \begin{tabular}{|c|c|c|c|c|c|c|c|c|c|c|}
                \hline
        Magnitude limit & \multicolumn{8}{|c|}{W1$<$14}& \multicolumn{2}{|c|}{W1$<$15}\\
                \hline
                        Extinction $I_{100}$& \multicolumn{2}{|c|}{$\langle $0;1)}&\multicolumn{2}{|c|}{$\langle $1;2)}&\multicolumn{2}{|c|}{$\langle $2;3)}&\multicolumn{2}{|c|}{$\langle $3;10)}&\multicolumn{2}{|c|}{$\langle $3;10)}\\
                \hline
        Oversampling    &before& after&before& after&before& after&before&after&before&after\\
        \hline
         Number of galaxies&36801&36801&54417&54417&22916&22916&10732&10732&33720&33720\\
        \hline
        Number of stars&6113&6113&10612&10612&4757&4757&6409&6409&13822&13822\\
        \hline
         Number of QSOs&2161&29598&2556&43998&1141&18398&525&8798&1498&27198\\
                \hline
        $\sigma_{W}$ [mag] & \multicolumn{2}{|c|}{0.025} & \multicolumn{2}{|c|}{0.025} & \multicolumn{2}{|c|}{0.026} & \multicolumn{2}{|c|}{0.026}& \multicolumn{2}{|c|}{0.028}\\      
        \hline
        $\sigma_{pm}$ [mas/yr] & \multicolumn{2}{|c|}{84} & \multicolumn{2}{|c|}{98} & \multicolumn{2}{|c|}{99} & \multicolumn{2}{|c|}{113}& \multicolumn{2}{|c|}{165}\\
        \hline
        
\end{tabular}
\end{footnotesize}
 \tablefoot{For a sample with $W1<14$ we applied oversampling in all the extinction bins, while for $W1<15$ only for extinction in the range $I_{100}\in\langle 3;10) $.  Oversampling was implemented for quasars only.  Parameters $\sigma_{W}$ and $\sigma_{pm}$ denote specific $\sigma$ values of Gaussian distributions, respectively for magnitudes and proper motions.}       
\end{table*}

Our training dataset is characterised by low numbers of bright objects in the QSO sample.
It was then necessary to address the problem of the accuracy decreasing as a result of  this imbalance. 
This feature is common in many classification schemes, especially  those which aim for the maximisation of  accuracy, like SVMs \citep{Akabani04}. If not accounted for properly, this can result in making a simple decision that is the basis of   the maintenance of the highest success rate: assigning the most common class to the test objects.
There are two ways of addressing this problem: it  can be solved eiher through rebalancing the dataset or by altering the algorithm itself.
The first solution works at the level of manipulating the data, where the under-represented population(s) can be oversampled, or the dominant class can be undersampled. In the latter case, when reducing the number of  objects contained within the majority class,  distributional assumptions on the data must be made: some crucial information may be lost or additional noise may be introduced.
On the other hand, changing the algorithm mainly relies on  cost-sensitive learning where a higher penalty is assigned to the misclassifications, resulting in a shift of the classifiers  towards the minority class, which improves the detection accuracy.

Since SVM decision making relies solely on the support vectors, it  works well against any noise in the data and any light imbalance. 
Therefore, if the distribution of the training sets is very skewed, the number of  support vectors in the majority class overweighs the ones from the minority class. In the case of the WISE data we  decided to perform oversampling of the under-represented class of QSO, i.e. additional artificial objects were created. The number of missing objects ($\rm{X_{missing}}$) needed to be added to
QSO training samples was calculated using the  equation \citep{Malek13}  
\begin{equation}
\label{Eq: oversampling}
\rm{ \lceil X_{missing}\rceil_{10}=NG\times0.8-X, }
\end{equation}
where  $\rm{ \lceil\rceil_{10} }$ stands for rounding the value up to the nearest ten, X corresponds to the number of original 
QSOs in the sample, and NG is the number of galaxies. This strategy provides   fully balanced training samples, which are essential for building an effective classifier.

\begin{figure}
\centering 
\includegraphics[width=0.4\textwidth]{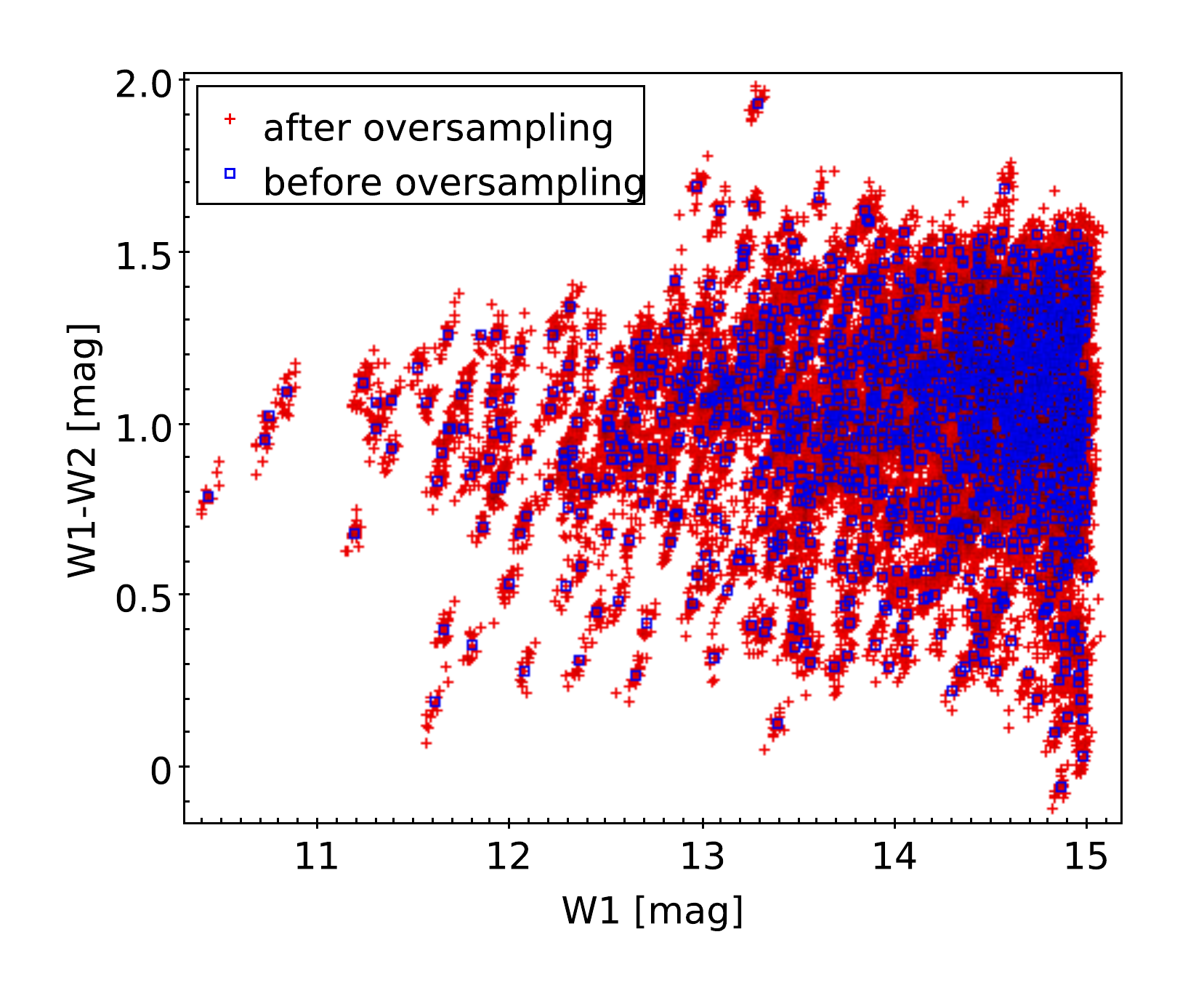}
\caption{Representative colour-magnitude diagram before and after oversampling for quasars.}
\label{Fig: oversampling}
\end{figure}

We created mock samples of missing objects by slight changes in the real parameters. In the first step a real QSO was randomly chosen, and then its parameters were reassigned by shifting the real ones by an amount drawn from a Gaussian distribution with specific standard deviations $\sigma$. Different values of $\sigma$ were used, depending on the type of parameter: $\sigma_{pm}$ for proper motions and $\sigma_W$ for magnitudes. They were calculated as the median values of the proper motion and magnitude uncertainties, respectively.  These values are given in Table \ref{Table: Oversampling}, which lists the cases in which the oversampling was applied, providing numbers of objects before and after the oversampling (see  Sect.\ \ref{Sec: tests} for  details on the magnitude and extinction divisions). Figure\ \ref{Fig: oversampling} presents an example of a colour-magnitude diagram for quasars before and after oversampling. The distribution of objects after the oversampling closely mimics the real one, as designed.

\section{Calibrating the SVM classifier for the WISE data}
\label{Sec: tests}

In this section we present various tests of the SVM algorithm performed on the \WS\ training data to verify and optimise the performance of the classifier.
In particular, we tested the algorithm's efficiency as a function of the following: 
i) choice of the kernel,
ii) number of sources in the training samples,
iii) number of parameters used for the classification,
iv) Galactic extinction,
v) limiting magnitude of the sample, and
vi) use of source apparent motions. 
This information  allowed us to prepare the SVM for the  application to the all-sky WISE dataset (Section \ref{Sec: All-sky}).

\begin{table*}
\caption{\label{polyvsrgb}
Comparison of SVM performances for two kernels, polynomial and radial, for the self-check and cross-test.}
\centering
\begin{footnotesize}
\begin{tabular}{|c|c|c|c|c|c|c|}
\hline
& \multicolumn{ 6}{c|}{SELF-CHECK} \\ \hline
kernel & \multicolumn{ 3}{c|}{polynomial} & \multicolumn{ 3}{c|}{radial} \\ \hline
 & \multicolumn{1}{l|}{Completeness} & \multicolumn{1}{l|}{Purity} & \multicolumn{1}{l|}{Contamination} & \multicolumn{1}{l|}{Completeness} & \multicolumn{1}{l|}{Purity} & \multicolumn{1}{l|}{Contamination} \\ \hline
Galaxy & 82.9 & 79.4 & 20.6 & 86.3 & 79.9 & 20.1 \\ \cline{ 1- 4}\cline{ 5- 7}
Stars & 81.0 & 84.1 & 15.9 & 81.0 & 87.1 & 12.9 \\ \cline{ 1- 4}\cline{ 5- 7}
QSO & 96.9 & 97.8 & 2.2 & 96.9 & 97.9 & 2.1 \\ \hline
& \multicolumn{ 6}{c|}{CROSS-TEST} \\ \hline
kernel & \multicolumn{ 3}{c|}{polynomial} & \multicolumn{ 3}{c|}{radial} \\ \hline
 & \multicolumn{1}{l|}{Completeness} & \multicolumn{1}{l|}{Purity} & \multicolumn{1}{l|}{Contamination} & \multicolumn{1}{l|}{Completeness} & \multicolumn{1}{l|}{Purity} & \multicolumn{1}{l|}{Contamination} \\ \hline
Galaxy & 81.0 & 77.9 & 22.1 & 84.0 & 75.7 & 24.3 \\ \cline{ 1- 4}\cline{ 5- 7}
Stars & 78.0 & 83.9 & 16.1  & 77.0 & 86.5 & 13.5 \\ \cline{ 1- 4}\cline{ 5- 7}
QSO & 97.0 & 94.2 & 5.8 & 96.0 & 96.0 & 4.0 \\ \hline
\end{tabular}

\end{footnotesize}
\end{table*} 

In each of the tests, we used a ten-fold cross-validation technique: we divided the training set into ten subsets of equal size,  selected nine of the  subsets to train the classification model, and then tested it on the  remaining subset; this was repeated ten times, leaving out a different subset each time. We then counted the training objects whose nature was correctly identified by SVM: TS (true star), TG (true galaxy), TQ (true QSO), and those   misclassified by the algorithm: FG (false galaxy), FS (false star), and FQ (false QSO). 

Then we define {\it completeness c, contamination f}, and {\it purity p} for the three classes of objects (e.g.\ \citealt{Soumagnac15}). For galaxies we have
\begin{equation}
{c_G=\rm \frac{TG}{TG+FGS+FGQ}}\;,
\label{Eq: completeness}
\end{equation}
\begin{equation}
{f_G=\rm \frac{FSG+FQG}{TG+FSG+FQG}}\;,
\label{Eq: contamination}
\end{equation}
\begin{equation}
p_G=1-f_G=\rm \frac{TG}{TG+FSG+FQG}\;,
\end{equation}
where TG, FGS, and FGQ stand for galaxies classified respectively as galaxies, stars, and quasars, and FSG (FQG) define stars (quasars) misclassified as galaxies. Analogous definitions are used for {\it completeness, contamination}, and {\it purity } of stellar ($c_S$, $f_S$, $p_S$) and quasar ($c_Q$, $f_Q$, $p_Q$) samples.

In each case we measure completeness, purity, and contamination for two variants: a self-check and a cross-test. For the self-check we classified the same objects  used in the given training sample. In the cross-test we classified objects from the subsamples that were not in the current training set.

The tests described below were performed for different combinations of magnitude limits and Galactic extinction levels. Three flux limits were adopted: $W1<14$ mag, $W1<15$ mag, and $W1<16$ mag. In each  data were also binned according to the measured Galactic dust emission. Here we chose  the 100-micron intensity ($I_{100}$) sky map made from a combination of COBE/DIRBE and IRAS 100 \mi{} measurements \citep{schlegel}. We preferred the $I_{100}$ parameter over the commonly applied $E(B-V)$ because the former was directly measured from data, while the latter was derived. We adopted four extinction bins: $I_{100}<1$, $1 \leq I_{100} < 2$, $2 \leq I_{100} < 3$, and $3 \leq I_{100} < 10$ [MJy/sr]. Above $I_{100} = 10$ MJy/sr, which constitutes about 1\% of the \WS\ catalogue, there are practically no galaxies or quasars in the training set. In general, the $I_{100} \geq 10$ MJy/sr areas cover about 17\% of the full sky, practically only in the Galactic Plane and regions of high dust obscuration where our classification is not expected to be reliable. 

In some cases, the above splitting of the full sample  left us with very small numbers of quasars in the relevant training sets, and the oversampling methodology had to be applied (see Sect.\ \ref{Sec: imbalanced}).


\subsection{Kernel performance comparison}
\label{Sec: kernel performance}
The first test  served to determine the optimal kernel for our application. We compared the performance of the two kernel functions described in Sec.\ \ref{Sec: SVM}, polynomial and radial (see Table~\ref{polyvsrgb}), 
and analysed the so-called univariate histograms of projections from the self-check and cross-test of the known data. 
A univariate histogram of projections is a graphical representation of the training data for a given binary classification (in the case of a three-class classifier we have three two-class classifiers) and the decision boundary SVM provides given the data.
To obtain the best efficiency of the classification, it is  standard practice to divide the training set into two subsets: one is used for actual training and the other is a validation subset used  as a verification of the accuracy of the created hyperplane against other known objects, even if not used for training. 
In this test, the training set contains 99\% of the total number of sources with known classification, while the validation test is composed of the remaining 1\% of known objects.

For non-linear SVM kernels, projected values $f(x)$ (Eq.\ \ref{Eq: discrim function}) are  obtained though  the kernel representation in a dual space. This means that there are three support vector machines (in the case of \WS\ data), each of which has its own decision function. Projection of an object $x_{{\bf k}}$ from a training set onto the normal direction of a nonlinear SVM boundary can be written as  
\begin{equation}
f(x_{k})=\Sigma_{i\in s,\nu} \alpha_{i} y_{i}K(x_{i},x_{k})+b,
\end{equation}
 where $x_{i}$ denotes a support vector, and classification of each example is determined by the sign of this function. The soft margin of a classifier can be written as 
\begin{equation}
 y_{f(x_{k})}=sign(f(x_{k})) (1-exp^{|f(x_{k})|}),
\end{equation}
 and the boundaries of the soft margin are then~$y_{f(x_{k})} \in [-1;1]$.
Then, $y_{f(x_{k})}$ describes two aspects of each example. The first  comes from the sign: it encodes a `hard' decision whether the example $x_{k}$ belongs to a given class or not. The second  comes from its absolute value: it represents how strong the decision is. This means that the farther a given example $x_{k}$ falls from the decision boundary, the more certain the decision is.
\begin{figure*}
\centering
 \begin{subfigure}{ \includegraphics[scale=0.4]{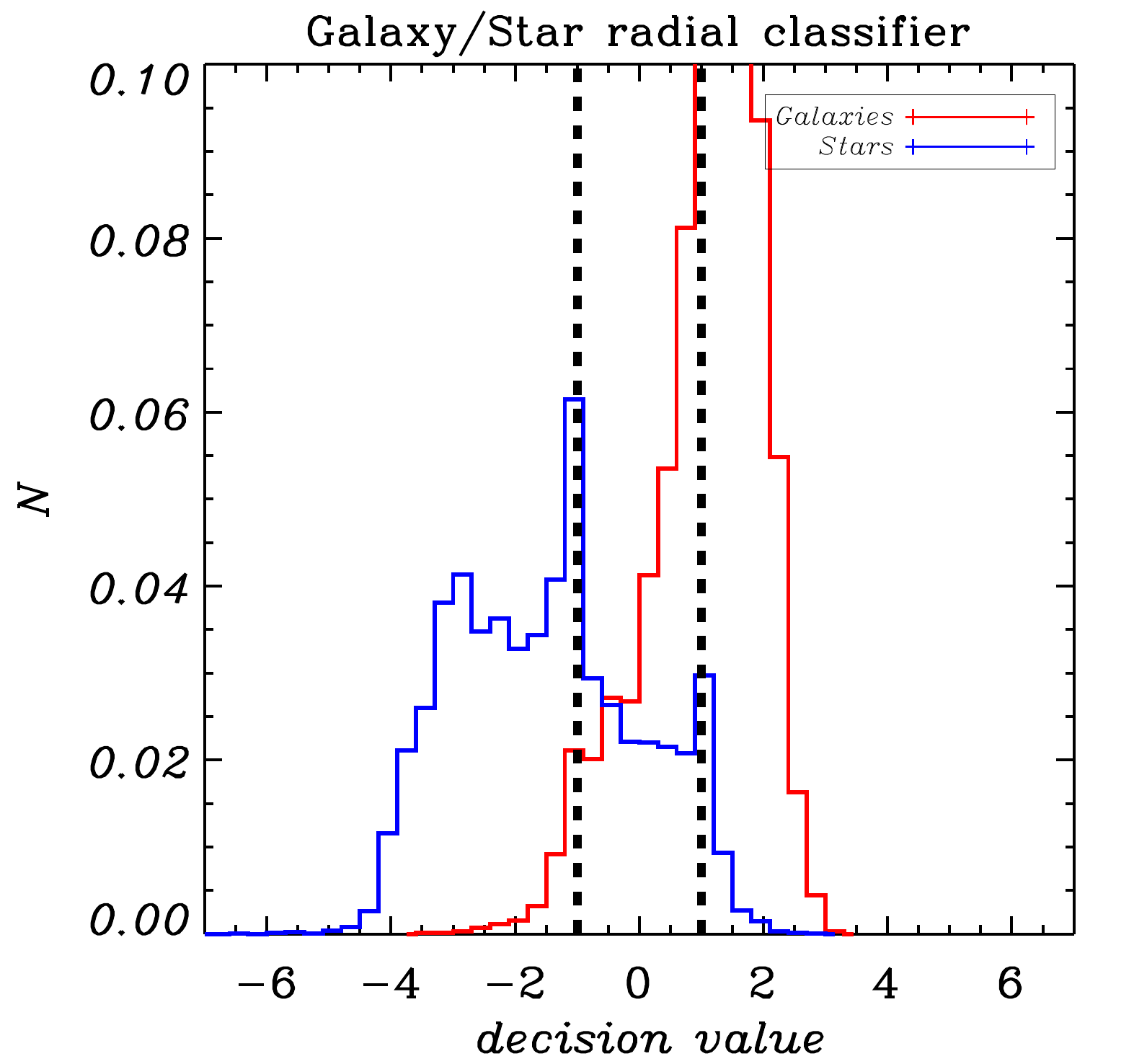} } 
\end{subfigure}
\begin{subfigure}{ \includegraphics[scale=0.4]{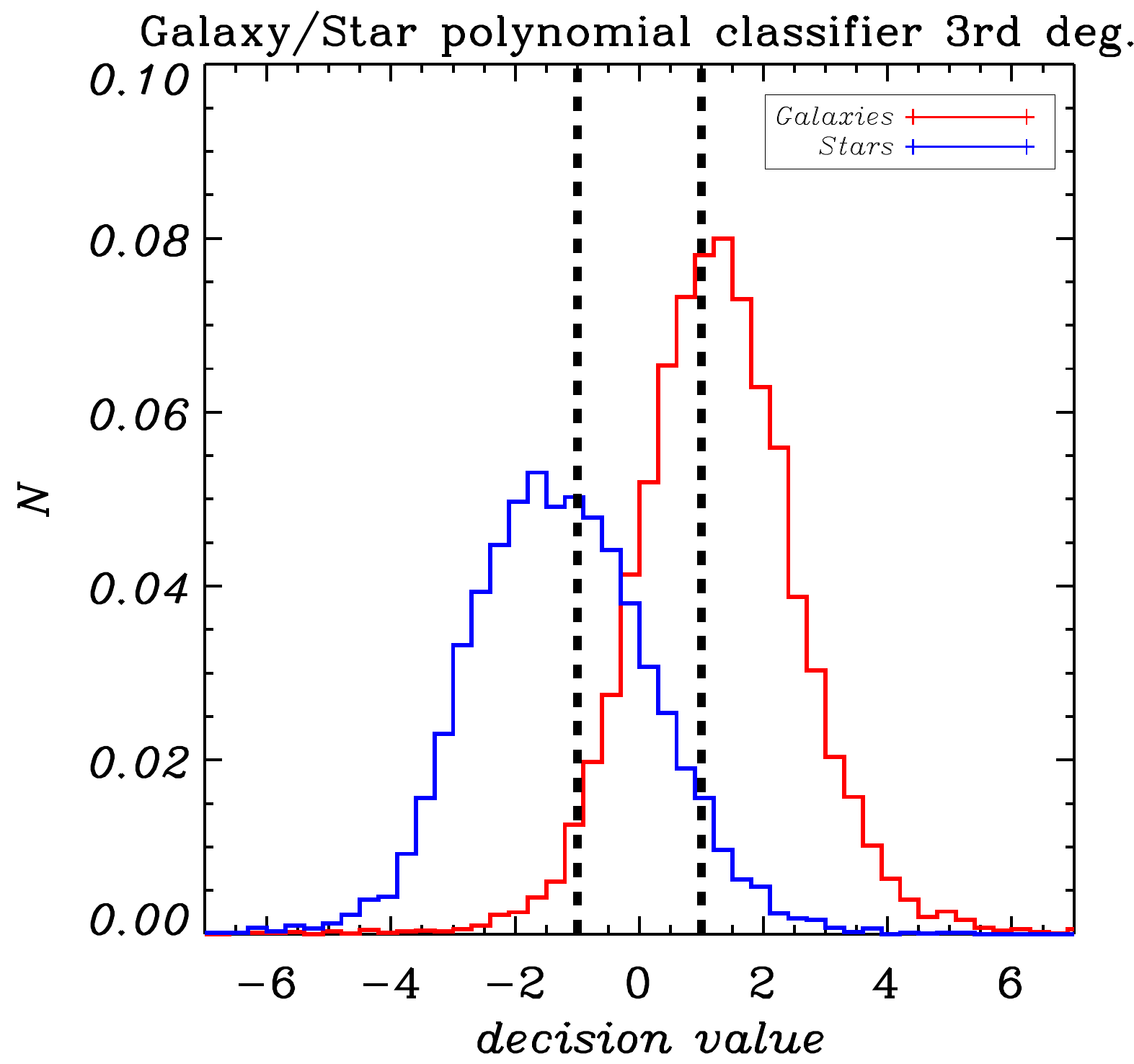} }
\end{subfigure}\\
\begin{subfigure}{  \includegraphics[scale=0.4]{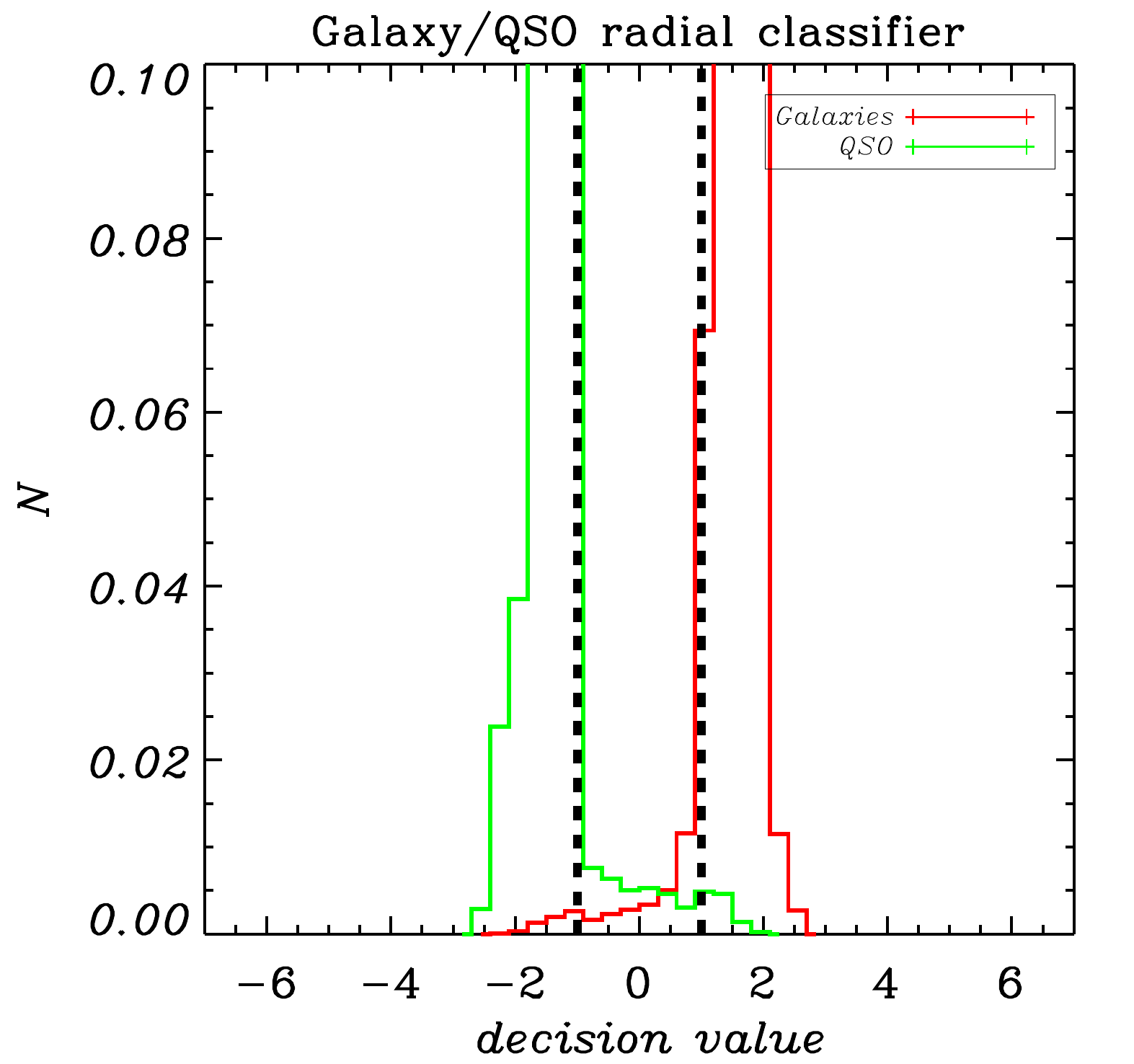}
} 
\end{subfigure}
\begin{subfigure} { \includegraphics[scale=0.4]{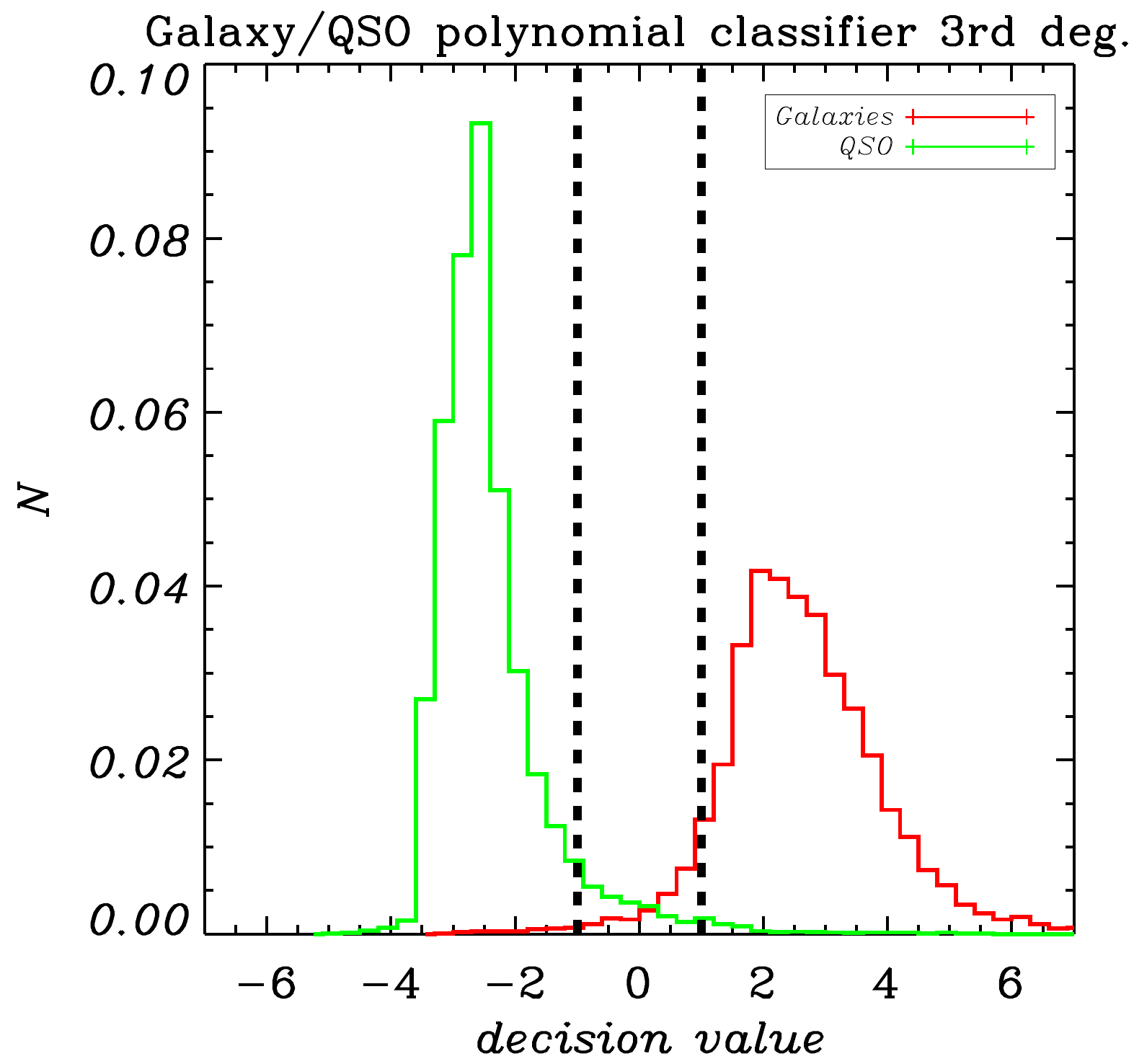}
}
\end{subfigure}\\
\begin{subfigure} { \includegraphics[scale=0.4]{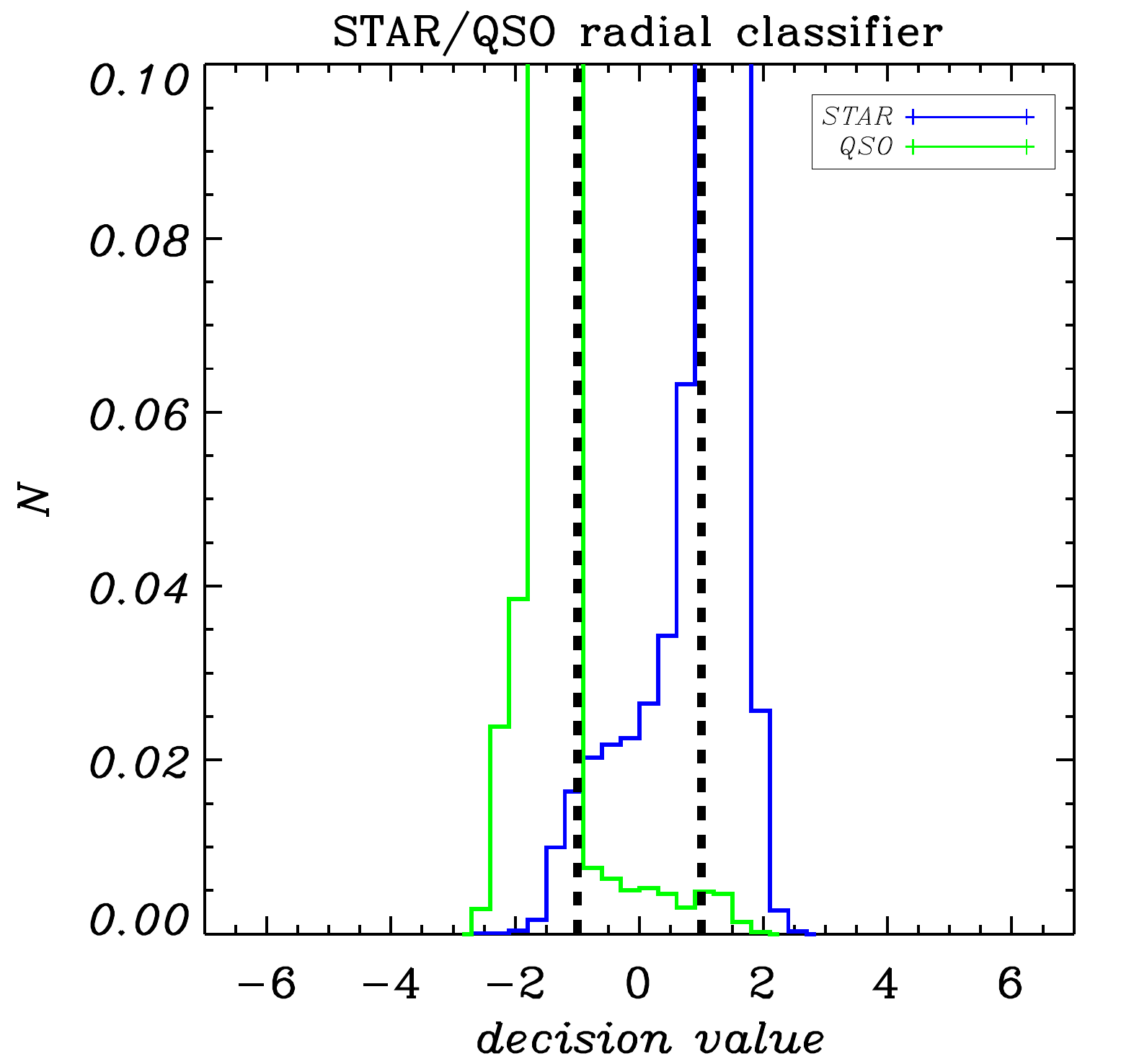}}
\end{subfigure}
\begin{subfigure} { \includegraphics[scale=0.4]{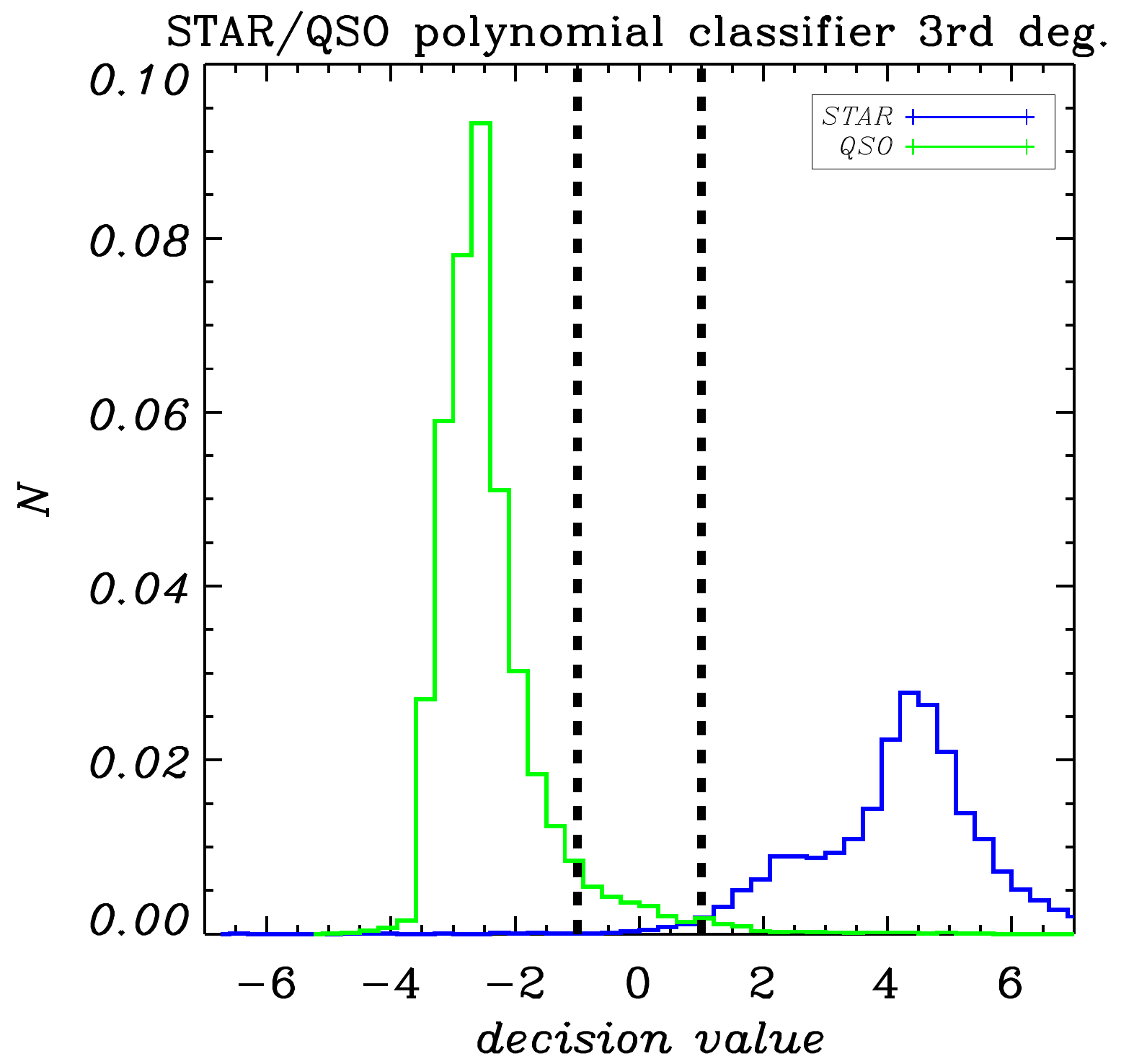}}
\end{subfigure}
 
\caption{Comparison of the histograms of projection values of the training data samples onto the normal direction of the SVM decision boundary for a radial (left column) and polynomial kernel (right column). The top row represents the division between galaxies and stars (red and blue, respectively), the middle row between galaxies and quasars (red and green), the bottom row between stars and quasars. Vertical lines represent the boundaries of the decision hyperplane margins for the two classes (+1 and -1), and    0  marks the position of the hyperplane itself.   \label{Fig: rgbvspoly}
}
\end{figure*} 

\begin{figure*}
\centering
\begin{subfigure}{\includegraphics[scale=0.4]{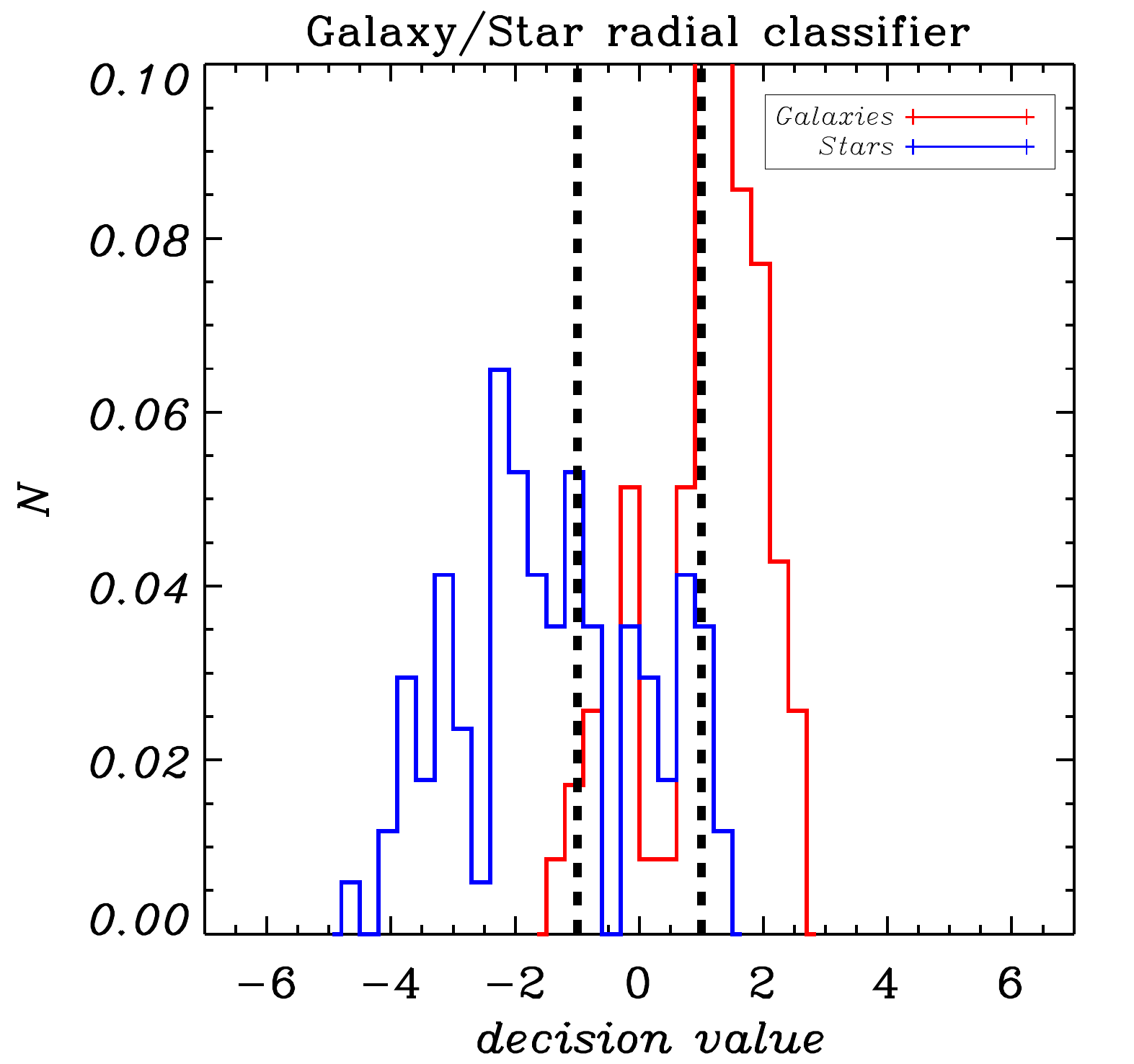}}
\end{subfigure}
\begin{subfigure}{\includegraphics[scale=0.4]{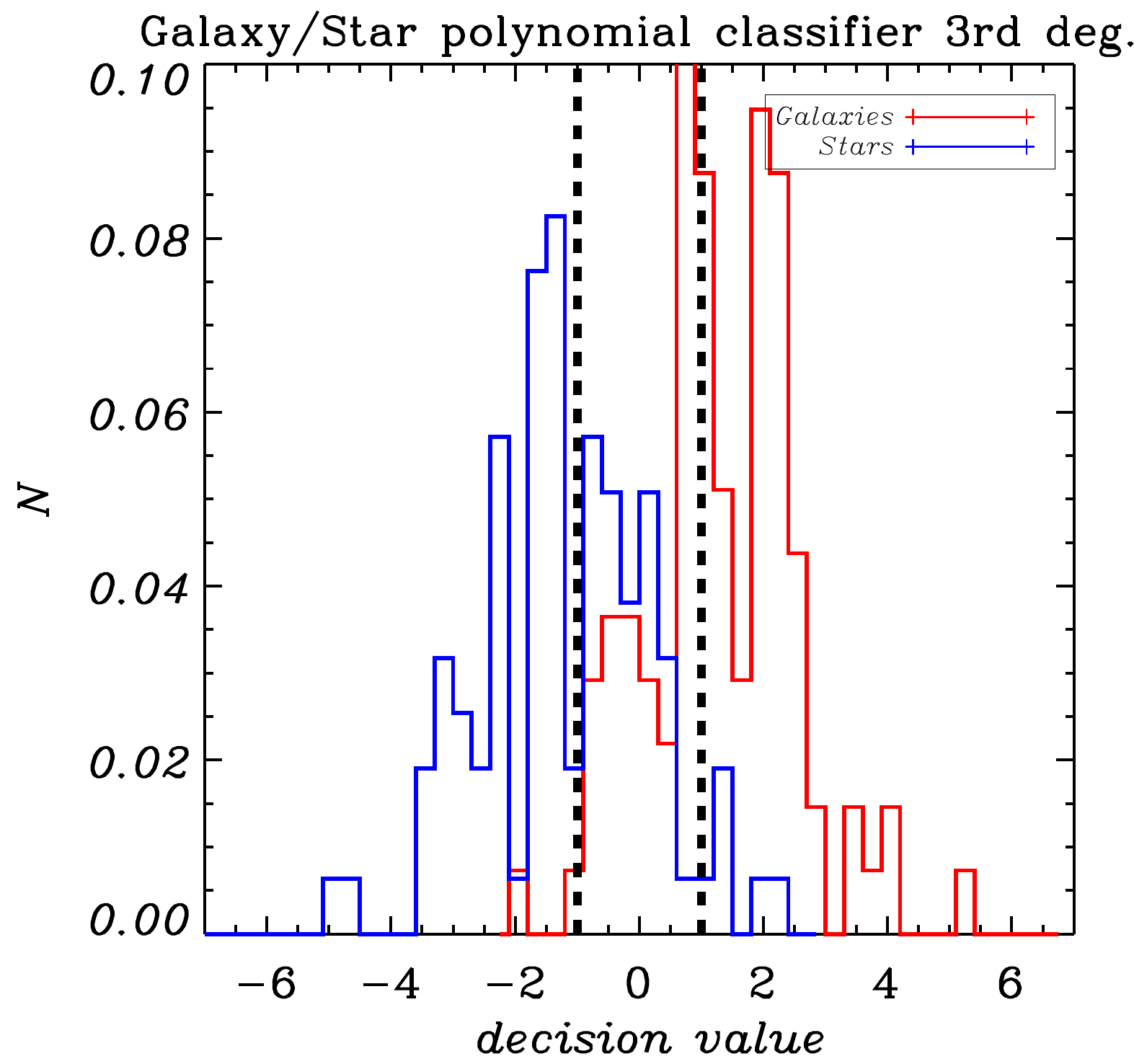}}
\end{subfigure}
\\
\begin{subfigure}{\includegraphics[scale=0.4]{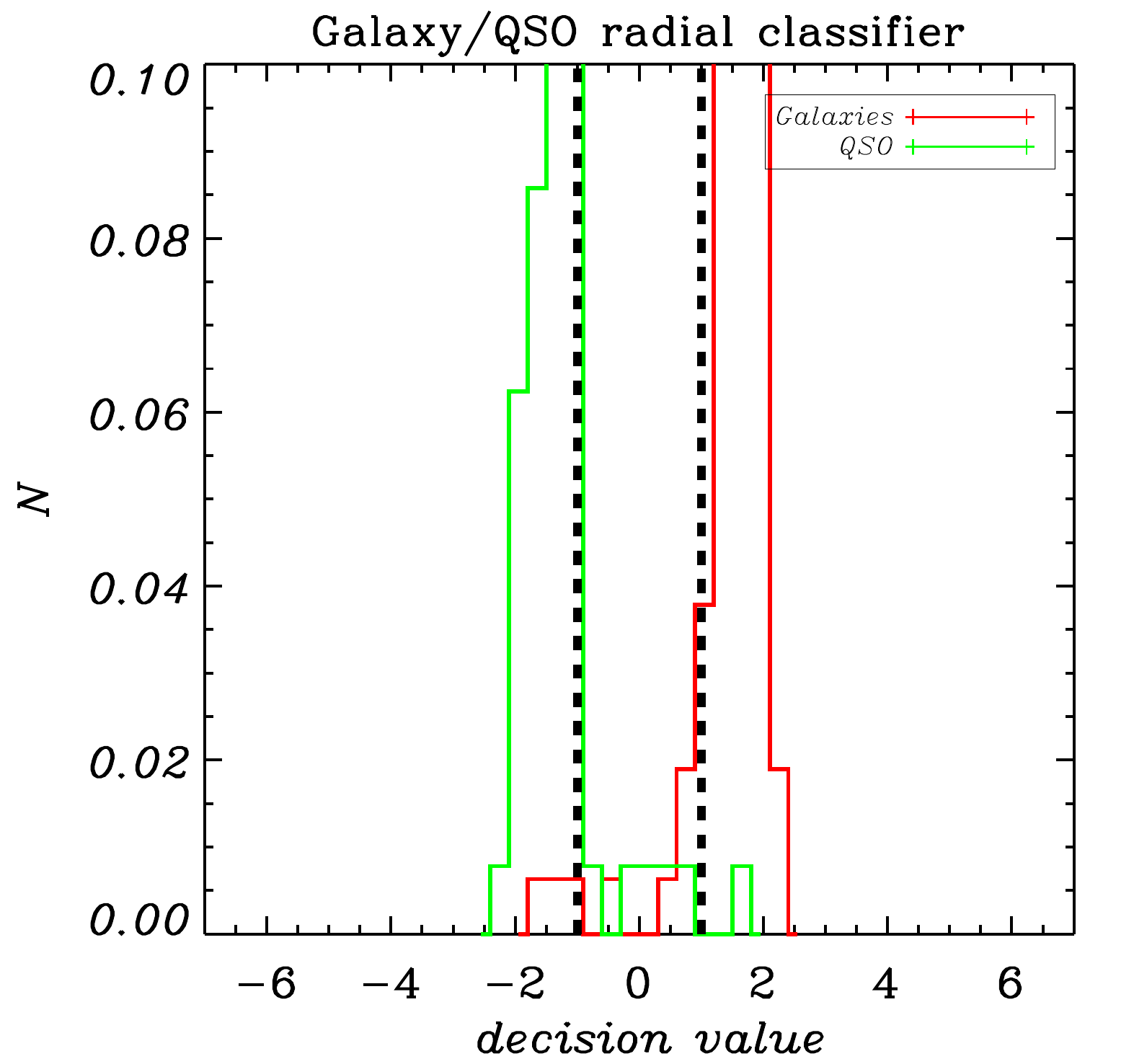}}
\end{subfigure}
\begin{subfigure}{\includegraphics[scale=0.4]{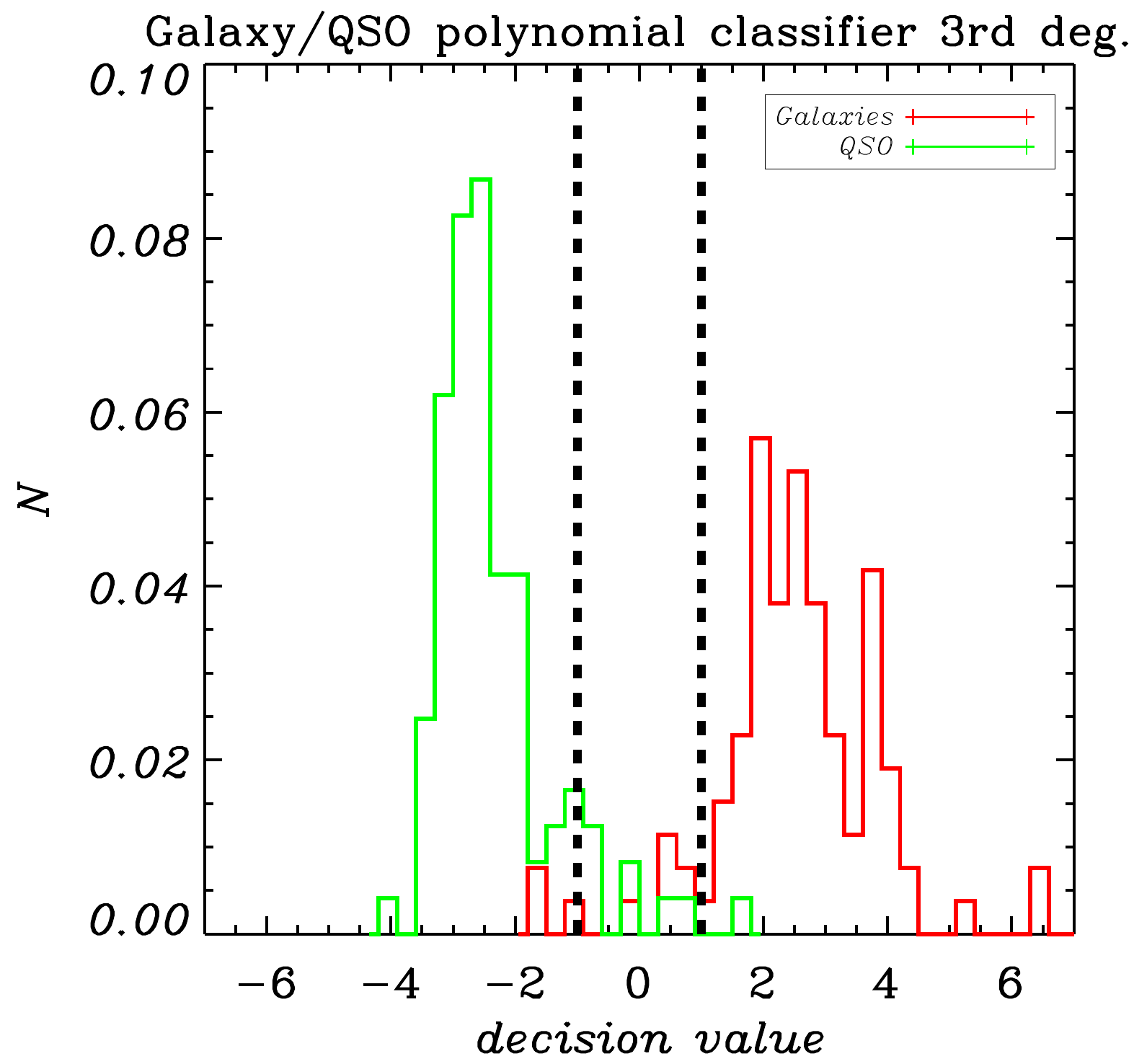}}
\end{subfigure}
\\
\begin{subfigure}{  \includegraphics[scale=0.4]{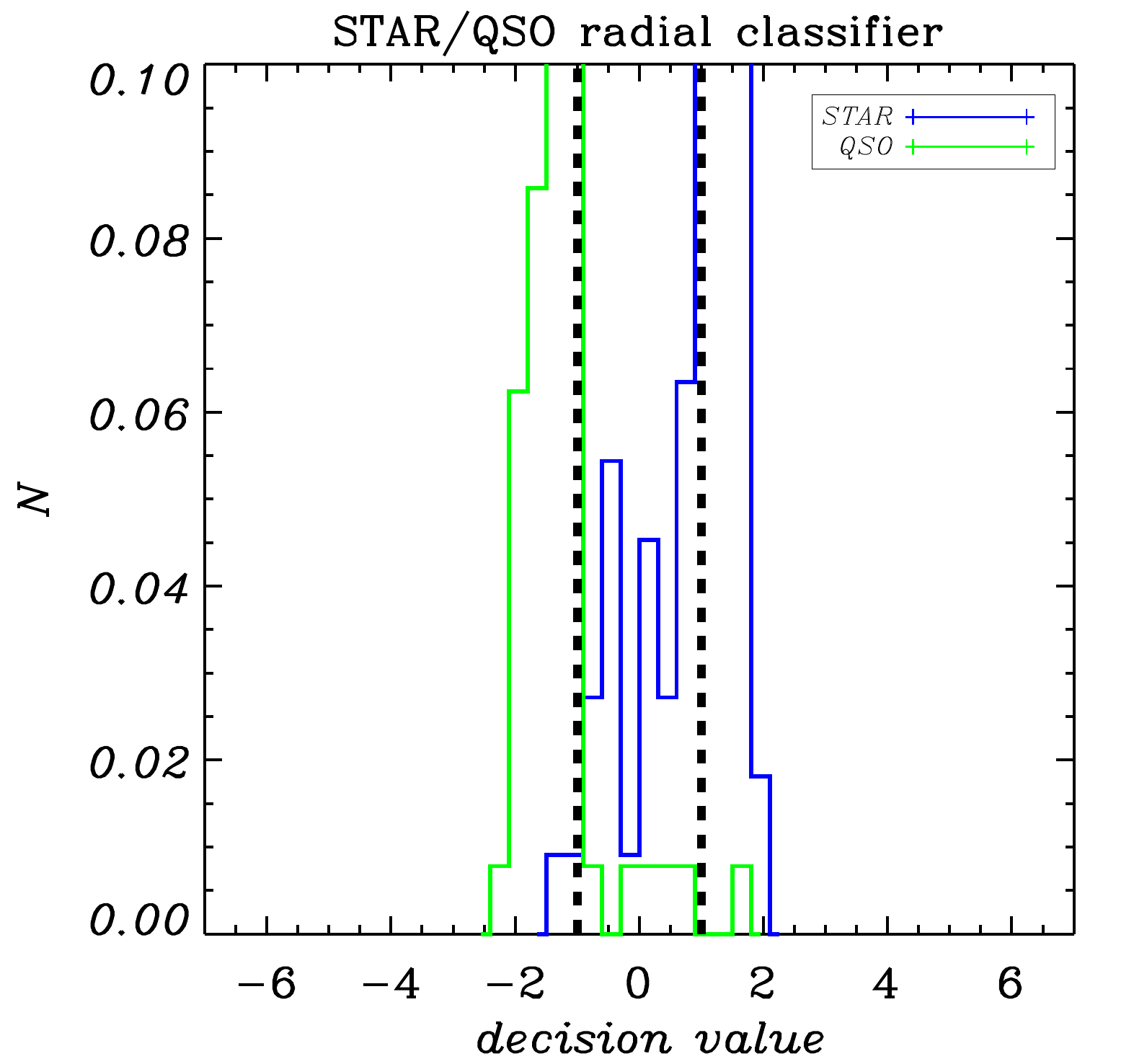}}
\end{subfigure}
\begin{subfigure}{  \includegraphics[scale=0.4]{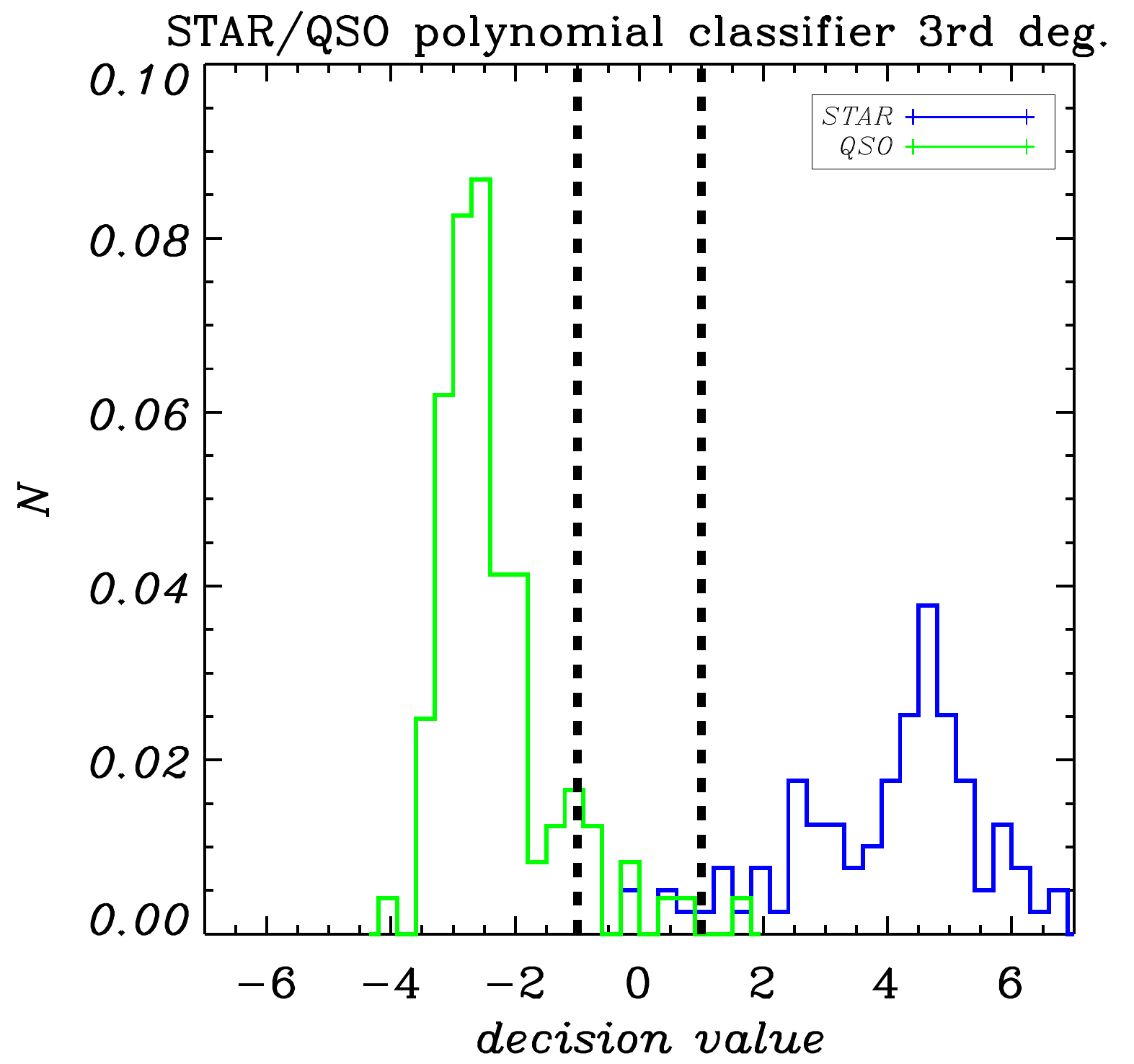}}
\end{subfigure}

\caption{Same as Fig.~\ref{Fig: rgbvspoly}, but for the validation dataset.     \label{Fig: rgbvspolytest}}
\end{figure*}

As can be seen from Table~\ref{polyvsrgb}, the differences between  completeness and purity of the training  samples and  validation datasets are small for the two kernels considered. 
However, when we compare the histograms of projections of each point with respect to the boundary we see clear differences (Figs.\ \ref{Fig: rgbvspoly} and \ref{Fig: rgbvspolytest}).
For the radial kernel we observe an effect of data piling on the margins, which is typical for high-dimensional data \citep{cherkassky07}. Separability of the set on which the classifier was trained does not imply that the validation or test sets will be equally well separated (see Figs.~\ref{Fig: rgbvspoly} and \ref{Fig: rgbvspolytest}, left columns). As the SVM optimisation aims at high separability of the training data, it penalises the data that fall into the soft margin of the separation boundary. However, the aim is  also to have a good separation of the validation set (which in turn should improve the separability of the test sample), which is why the preferred model should allow  data points to fall into the soft margin. Moreover, it is desirable to have as  decision values that are as strong as possible; therefore, the data piling effect should be avoided, which is why the kernel that displays a clearer division of validation -- the polynomial kernel in this case -- is  preferable, as is shown in Fig.~\ref{Fig: rgbvspolytest}.

\subsection{Optimal number of objects in the training samples}

\begin{figure*}
\centering 
\begin{subfigure}{
\includegraphics[scale=0.3]{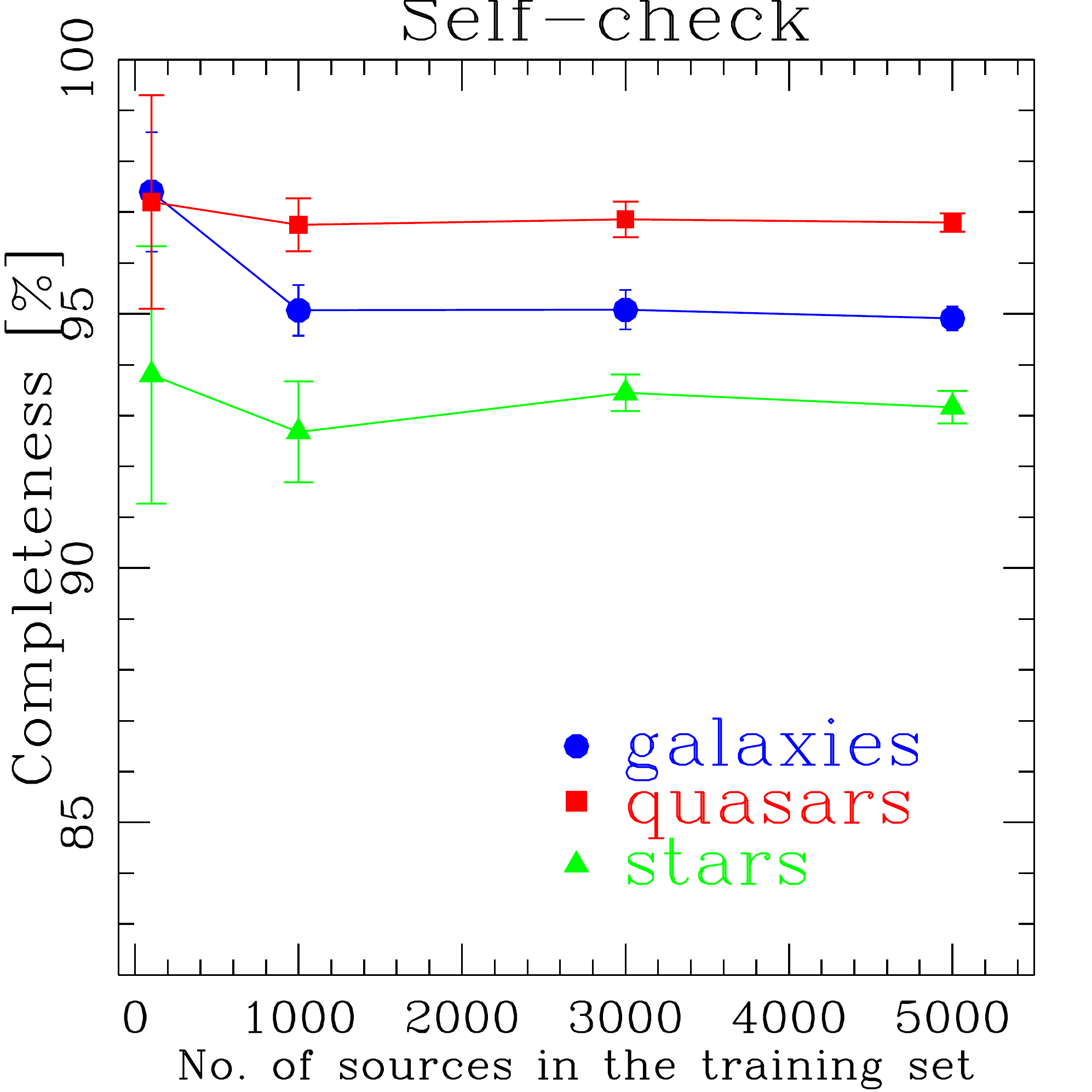}}
\end{subfigure}
\begin{subfigure}{
\includegraphics[scale=0.3]{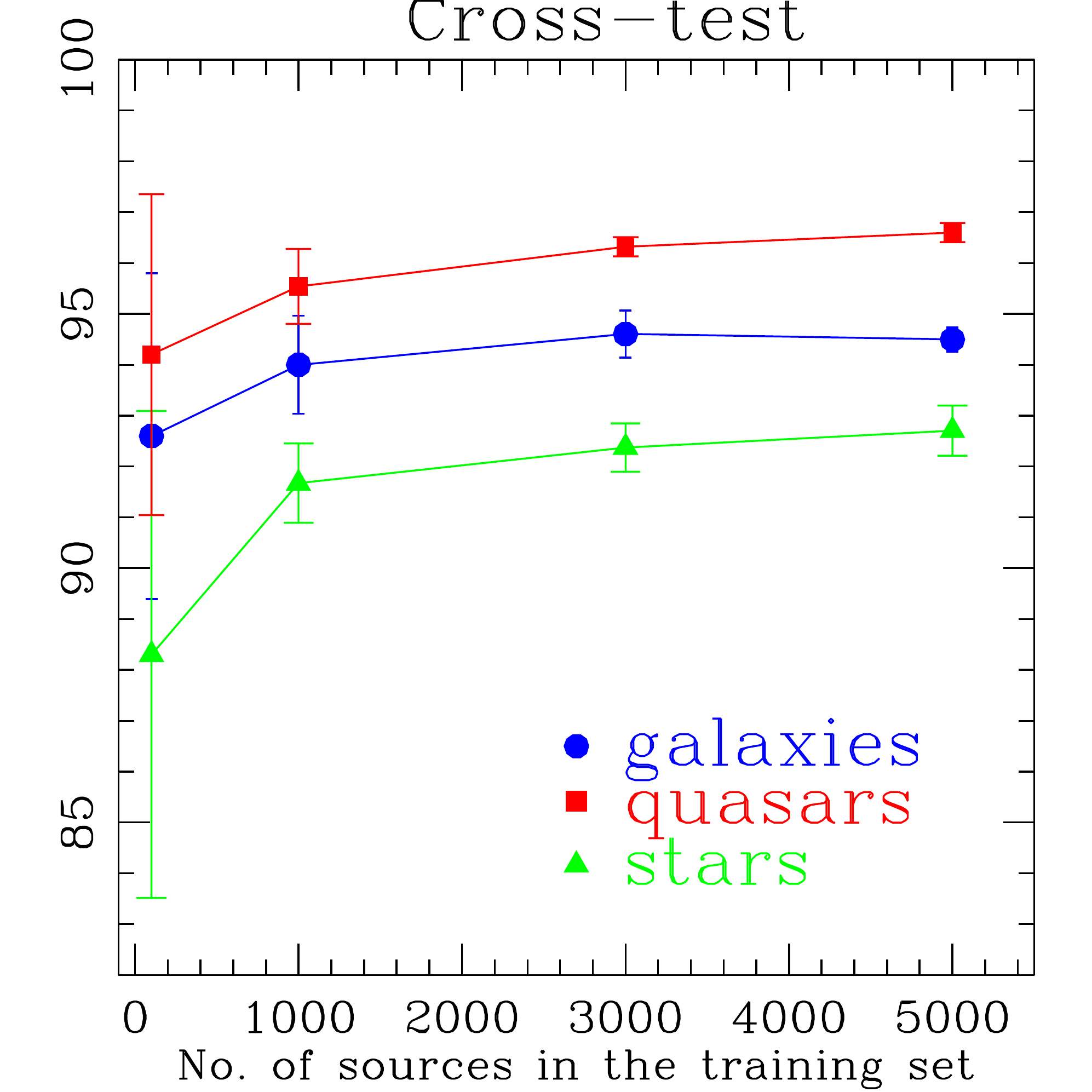}}
\end{subfigure}
\\
\begin{subfigure}{
\includegraphics[scale=0.3]{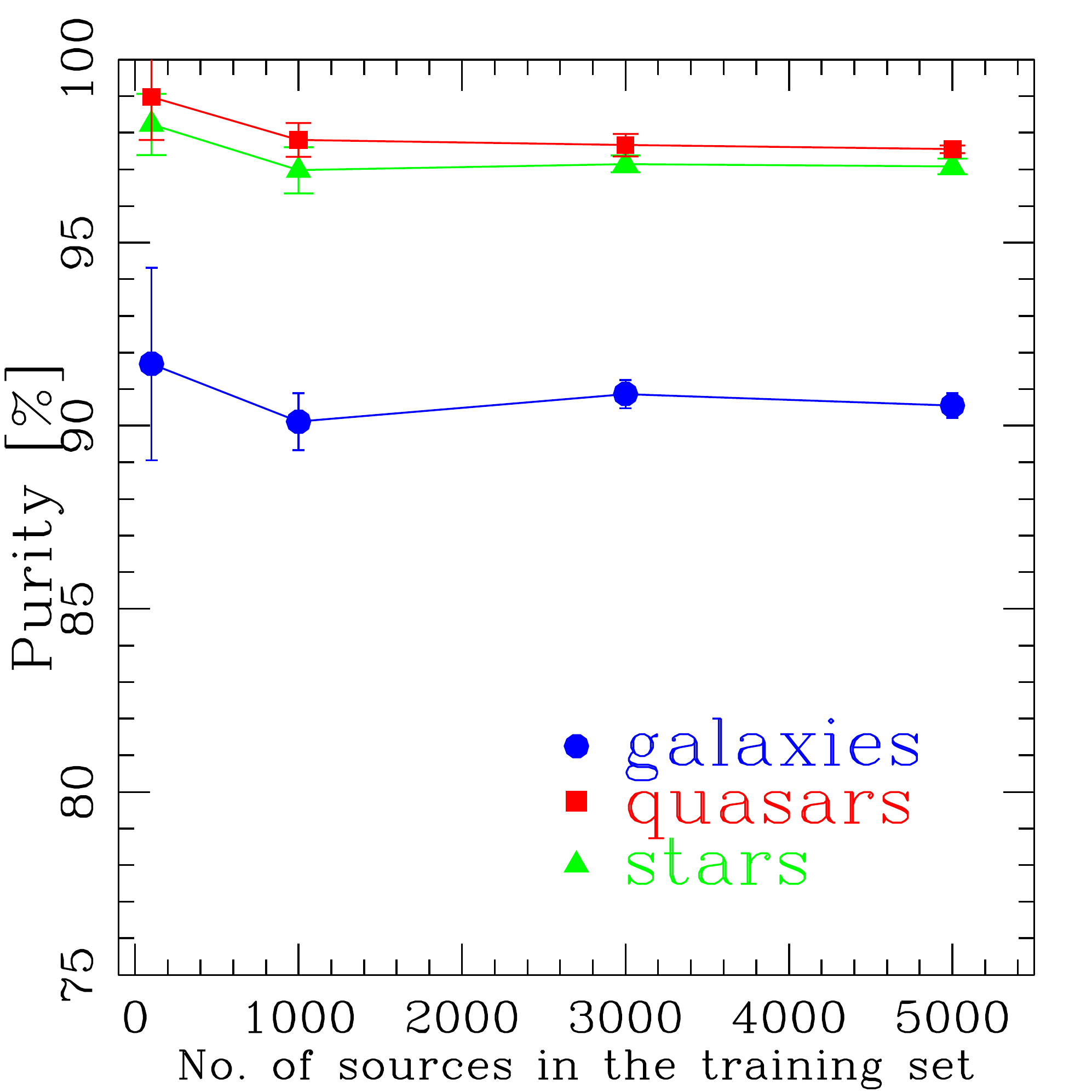}}
\end{subfigure}
\begin{subfigure}{
\includegraphics[scale=0.3]{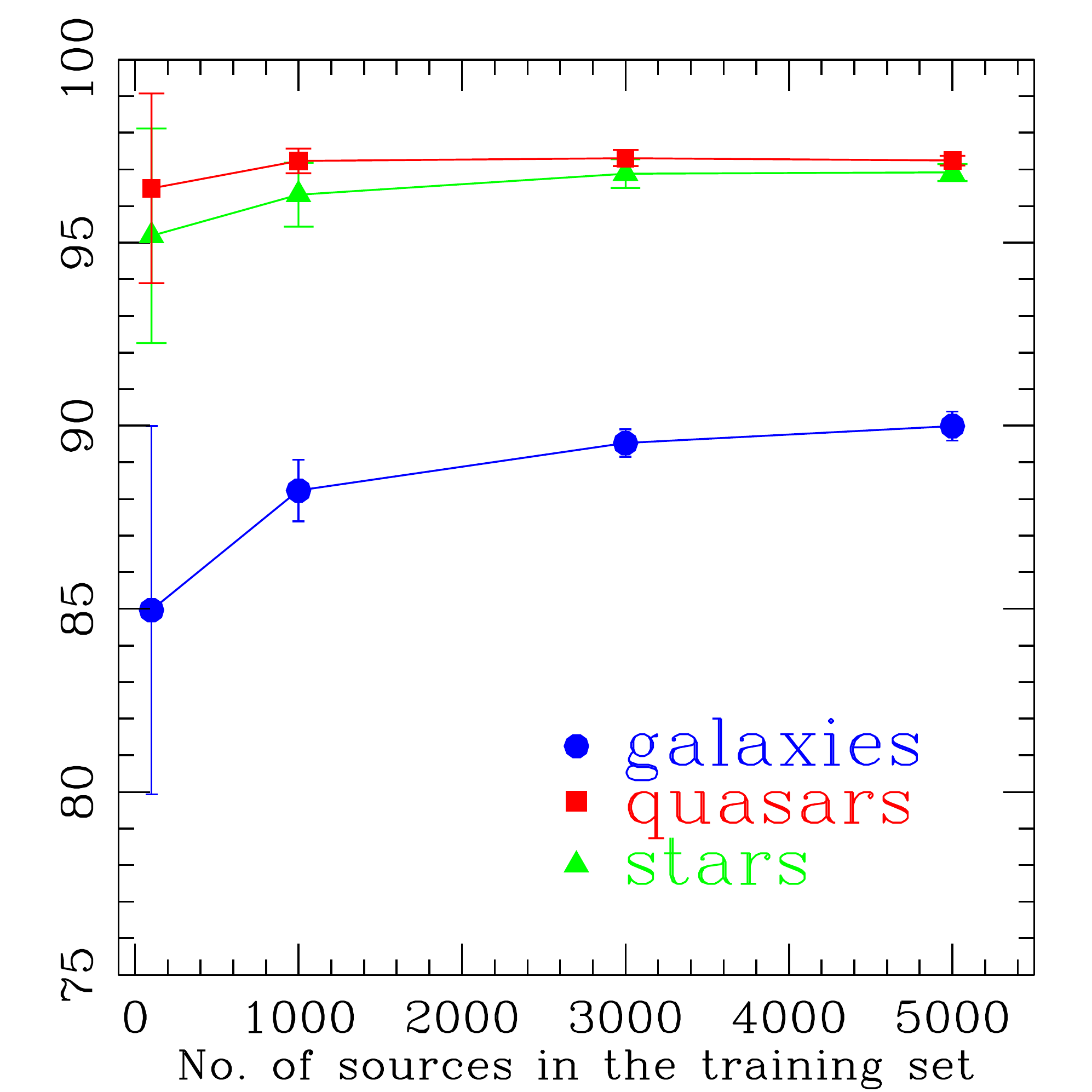}}
\end{subfigure}
\caption{Dependence of the completeness (upper panels) and purity (lower panels) on the number of objects in the training set of a $W1 < 14$ mag flux-limited sample, for the self-check (left panels) and the cross-test (right panels) cases. Relevant contamination levels are 100\%$-$Purity. }
\label{Fig: training set size}
\end{figure*}

As our \WS\ training set is much  larger than in earlier SVM classification applications (e.g. in AKARI by \citealt{Solarz12} or in VIPERS by \citealt{Malek13}), in the first step, after deciding which kernel function to use, we performed a series of tests to check whether we were able to calibrate our SVM method on smaller subsamples without deteriorating the results. We conducted four tests for each of the three flux-limited samples (in this case there was no extra division according to the extinction), where we randomly chose 100, 1000, 3000, and 5000 objects for each class (i.e. 100 galaxies, 100 stars, 100 quasars, etc.), and we used these training sets to compute relevant statistics as defined above. Each test was repeated ten times, and in all the cases the error bars provided represent the standard deviation from the mean of the ten tests.

Fig.\ \ref{Fig: training set size} shows, as an example, the dependence of the completeness and purity on the number of training objects for a bin $W1 < 14$ mag for the self-check and cross-test. The results for other flux limits are similar. As seen in this figure, our results stabilise for subsamples with 3000 randomly chosen  objects from each class. Based on these results, the following tests were applied for these numbers of objects. This allowed us to significantly save on computation time in the tests, as it scales highly non-linearly with the size of the training set.

\begin{figure*}
\centering 
\begin{subfigure}{
\includegraphics[scale=0.3]{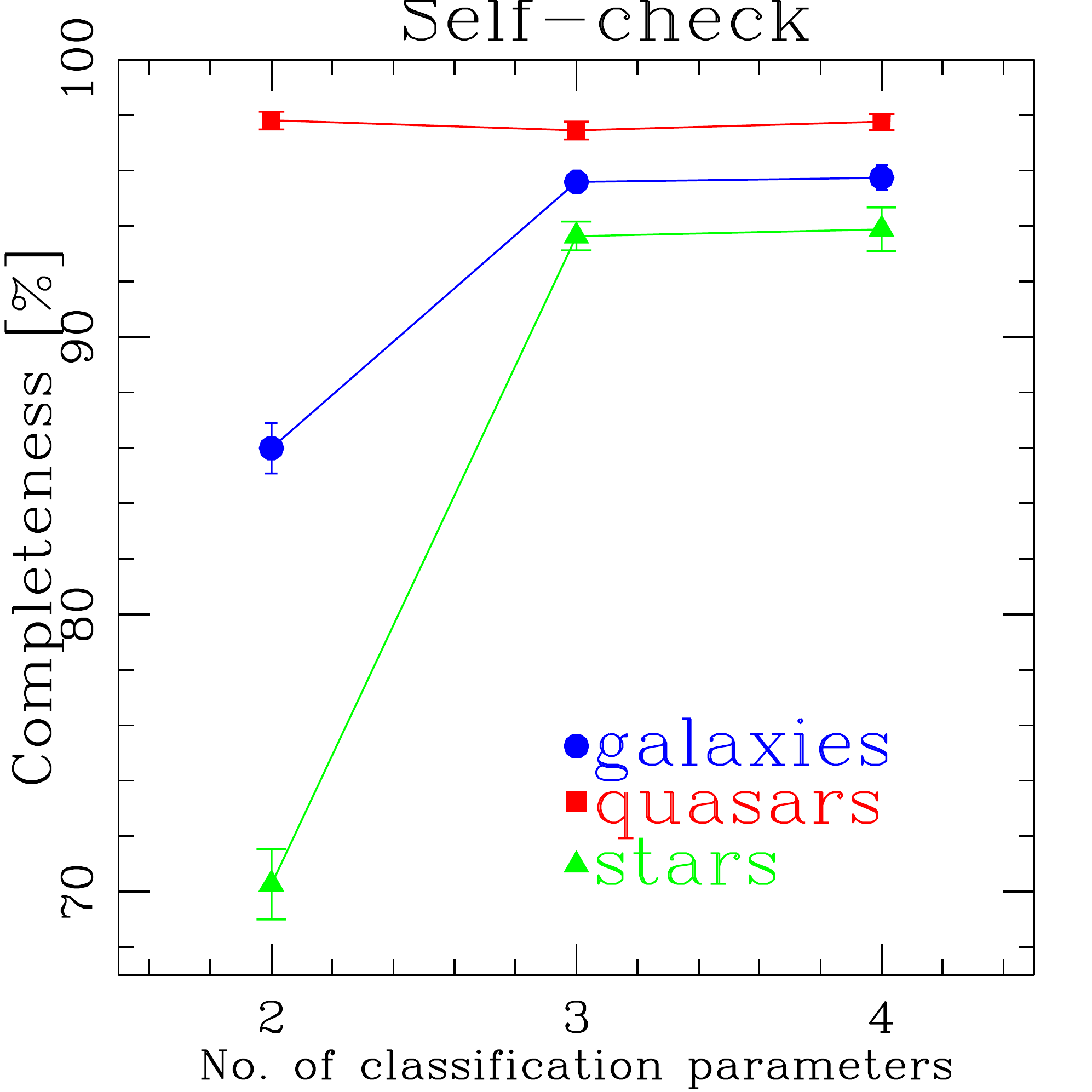}}
\end{subfigure}
\begin{subfigure}{
\includegraphics[scale=0.3]{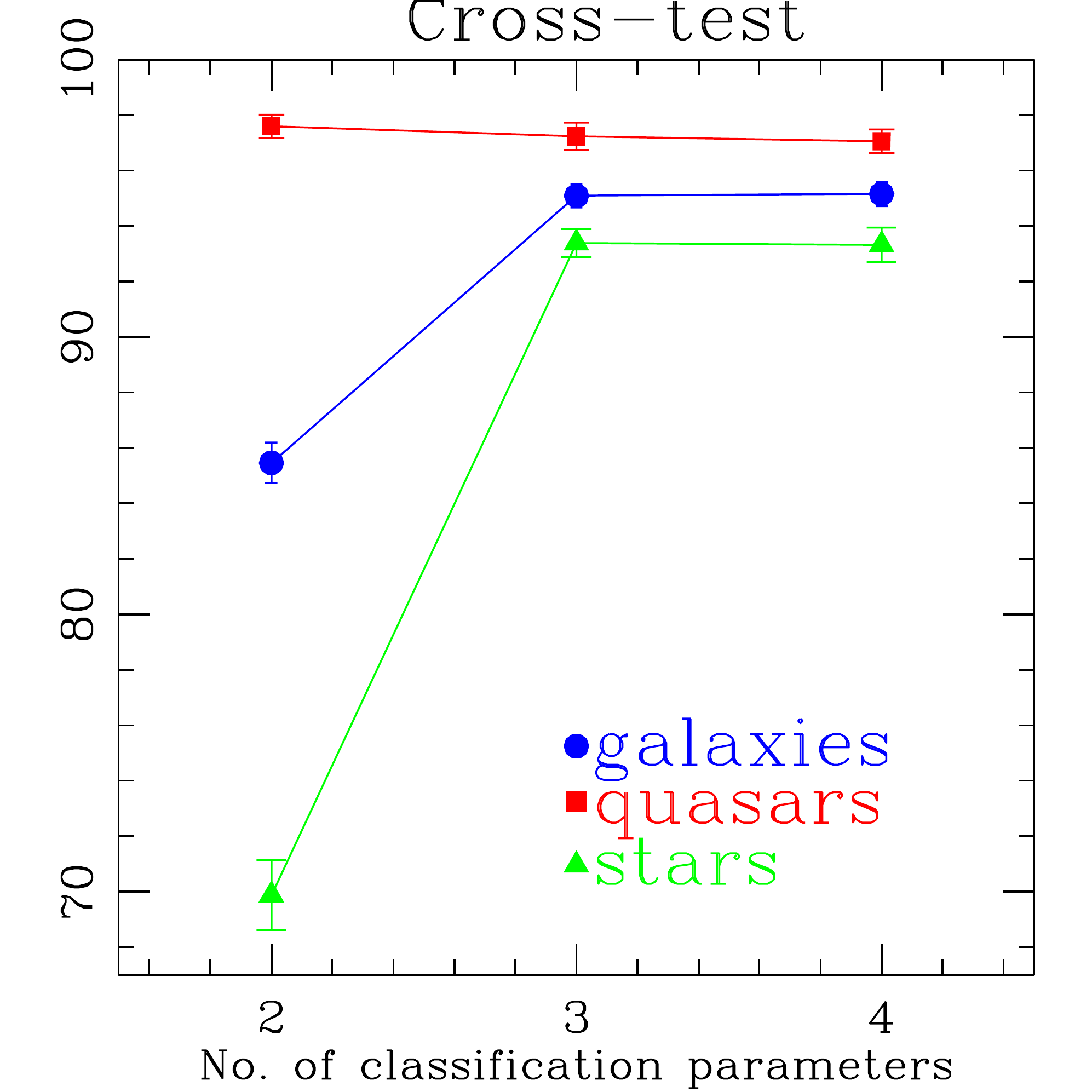}}
\end{subfigure}
\\
\begin{subfigure}{
\includegraphics[scale=0.3]{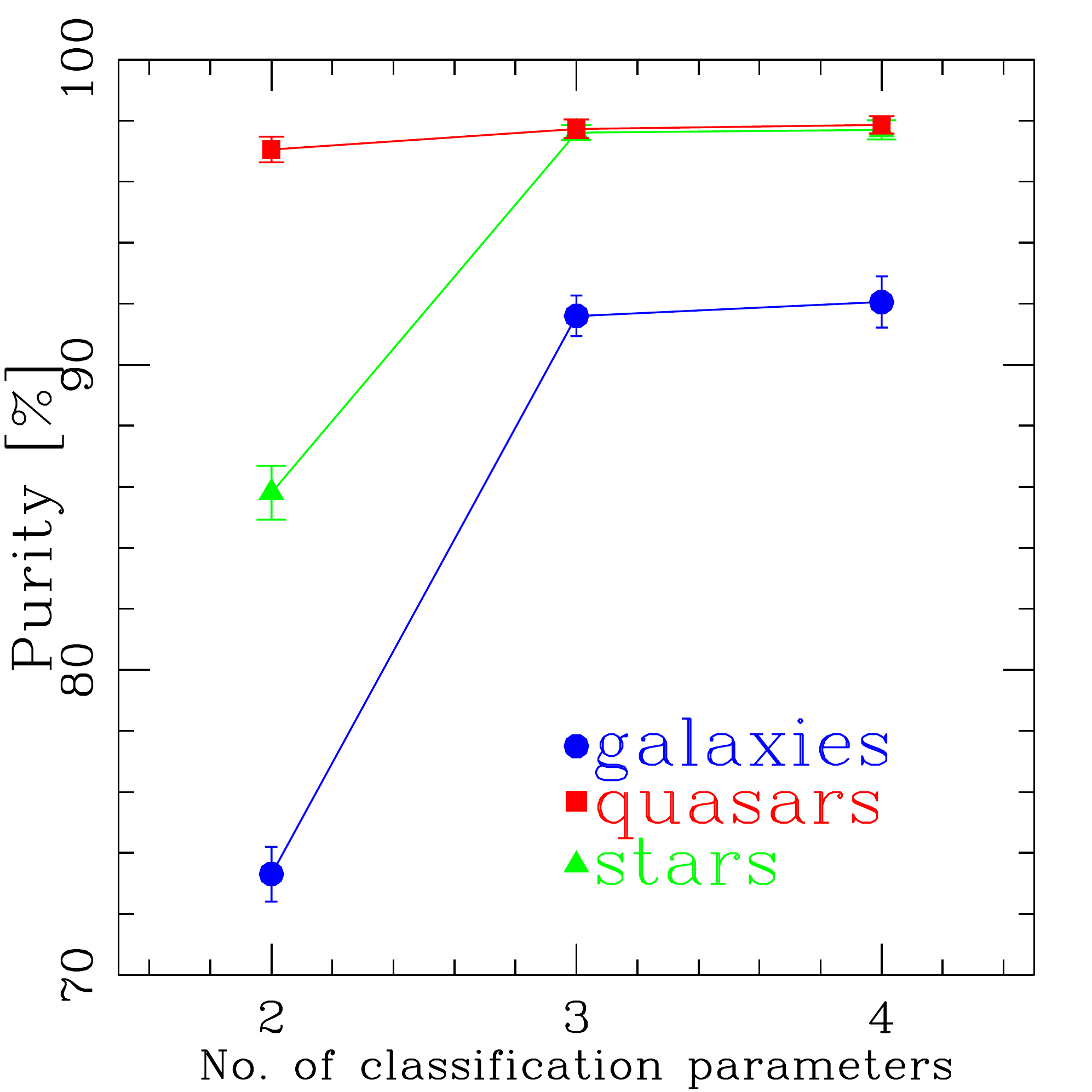}}
\end{subfigure}
\begin{subfigure}{
\includegraphics[scale=0.3]{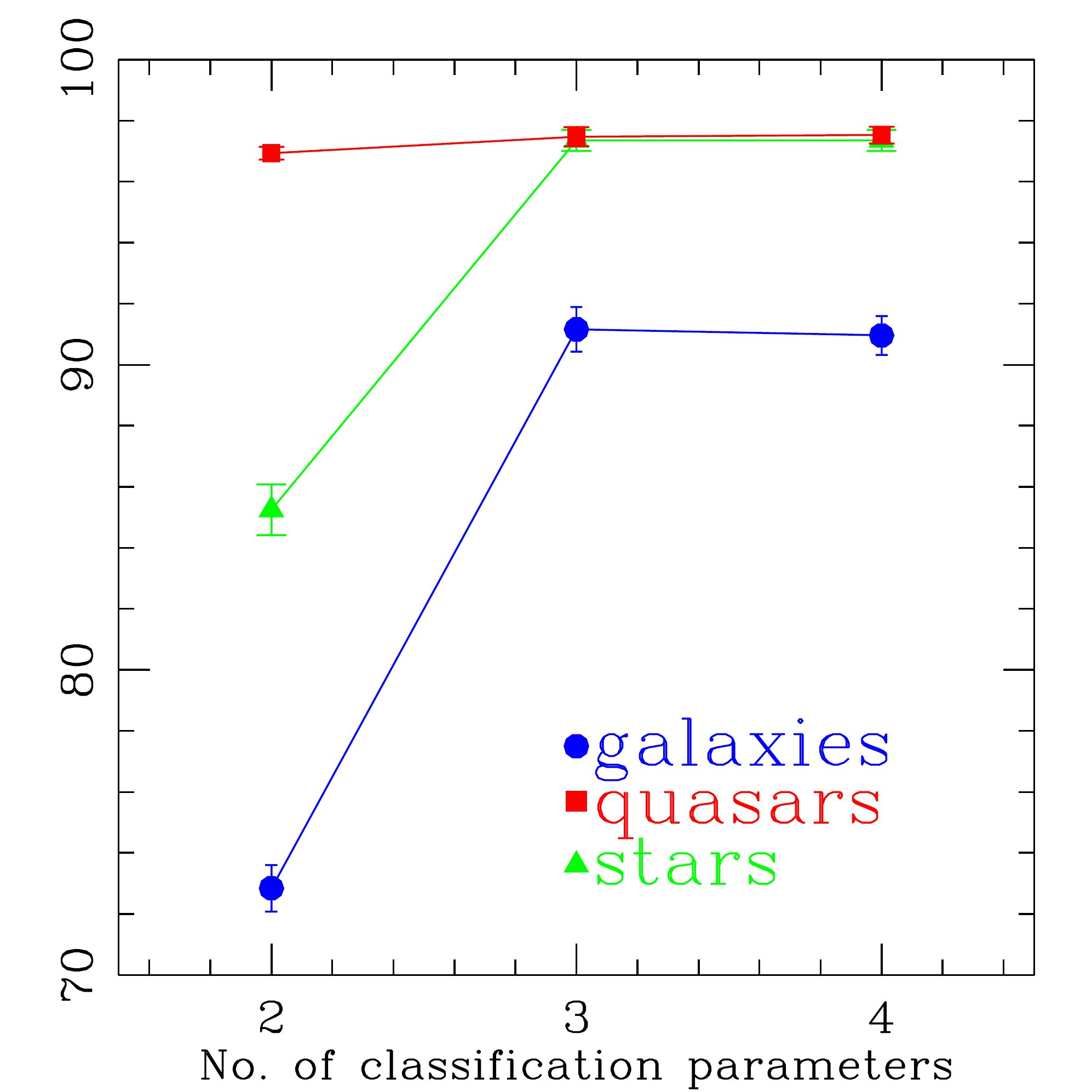}}
\end{subfigure}
\caption{Dependence of the completeness (upper panels) and purity (lower panels) on the number of parameters used for SVM training of a sample with $W1 < 14$ mag and $I_{100}<1$ [MJy/sr], for the self-check (left panels) and the cross-test (right panels) cases. The specific classification parameters for the training were $W1$ (1st), $W1-W2$ (2nd), $\mtt{w1mag\_1} - \mtt{w1mag\_3}$ (3rd), and apparent motions (4th). See text for details. Relevant contamination levels are 100\%$-$Purity.}
\label{Fig: number of training parameters}
\end{figure*}

\subsection{Parameter space for classification}

After establishing the optimal size of the training samples, we performed a series of tests to determine the minimum number of parameters sufficient for optimal classification. Each additional parameter significantly extends computation time, while it does not necessarily   improve overall accuracy, for instance in cases when it is noise-dominated and/or not related to the source type. We thus started by using just two, the $W1$ magnitude and the $W1-W2$ colour, and then extended the parameter space by first adding the differential aperture magnitude $\mtt{w1mag\_1} - \mtt{w1mag\_3}$, and then  the apparent motions $\mtt{pm}$ (cf.\ Sec.\ \ref{Sec: WISE} for parameter descriptions). For a proper comparison, these tests were applied on sources with `detected' proper motions, i.e.  $\mtt{pm}>\mtt{sigpm}$. In addition, they were employed in various combinations of magnitude cuts and extinction bins, as described earlier.

Fig.\ \ref{Fig: number of training parameters} shows how completeness and purity change when the number of implemented parameters increases. This particular example is for $W1 < 14$ mag and $ I_{100} <1 $ [MJy/sr] , but the results were qualitatively the same for each of the magnitude cut -- extinction combinations. While quasars were already very accurately classified  for the two parameters used ($W1$ and $W1-W2$), for stars and galaxies both completeness and purity significantly increased after the differential aperture magnitude was used as the third parameter. On the other hand, the proper motions did not bring any improvement, and sometimes even a slight deterioration in accuracy was observed once they were applied. This is hardly surprising given all the caveats associated with WISE apparent motion measurements \citep{Kirkpatrick14}: they are not accurate enough for our type of analysis, and from now on we will thus focus on tests not using proper motions. We can expect, however, that longer time baselines and/or better photometric accuracy possible with the NEOWISE data \citep{NEOWISE}, once combined into the MaxWISE data product \citep{Faherty15}\footnote{See also \url{http://wise5.ipac.caltech.edu/posters/Eisenhardt.pdf}.}, and with future surveys (such as the LSST) should allow  proper motions to become a useful parameter for the classification of sources, including extragalactic ones. This would also identify  a fourth type of source, one that  we have not considered here as we do not have them in the training samples, although they are certainly  present in the WISE database, namely minor bodies of the solar system. As is shown in Sec.\ \ref{Sec: All-sky}, they most likely contaminate the quasar candidate sample in the final all-sky classification.

\subsection{Dependence of classification accuracy on extinction and limiting magnitude}

\begin{figure}
\centering 
\begin{subfigure}{
\includegraphics[scale=0.3]{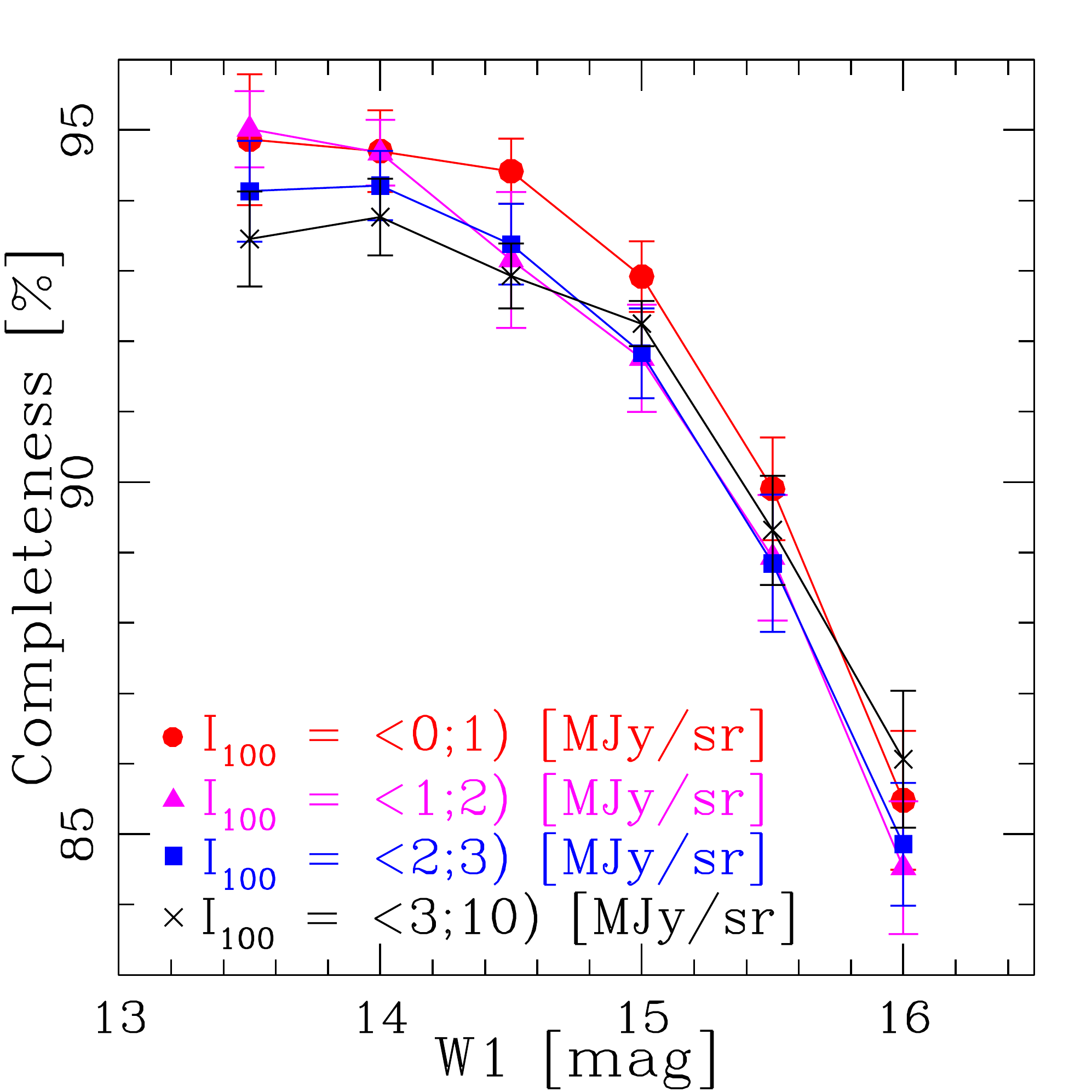}}
\end{subfigure}
\\
\begin{subfigure}{
\includegraphics[scale=0.3]{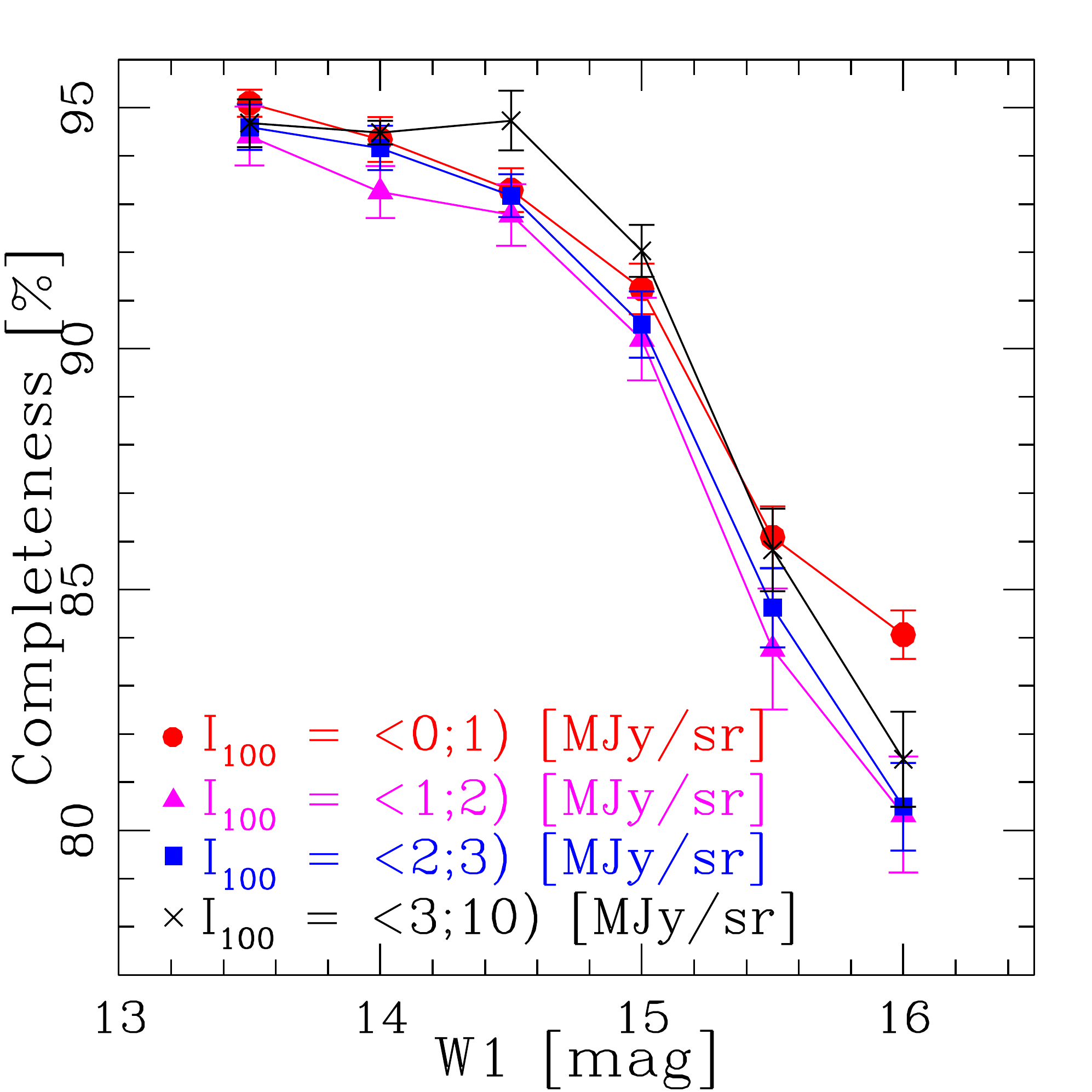}}
\end{subfigure}
\\
\begin{subfigure}{
\includegraphics[scale=0.3]{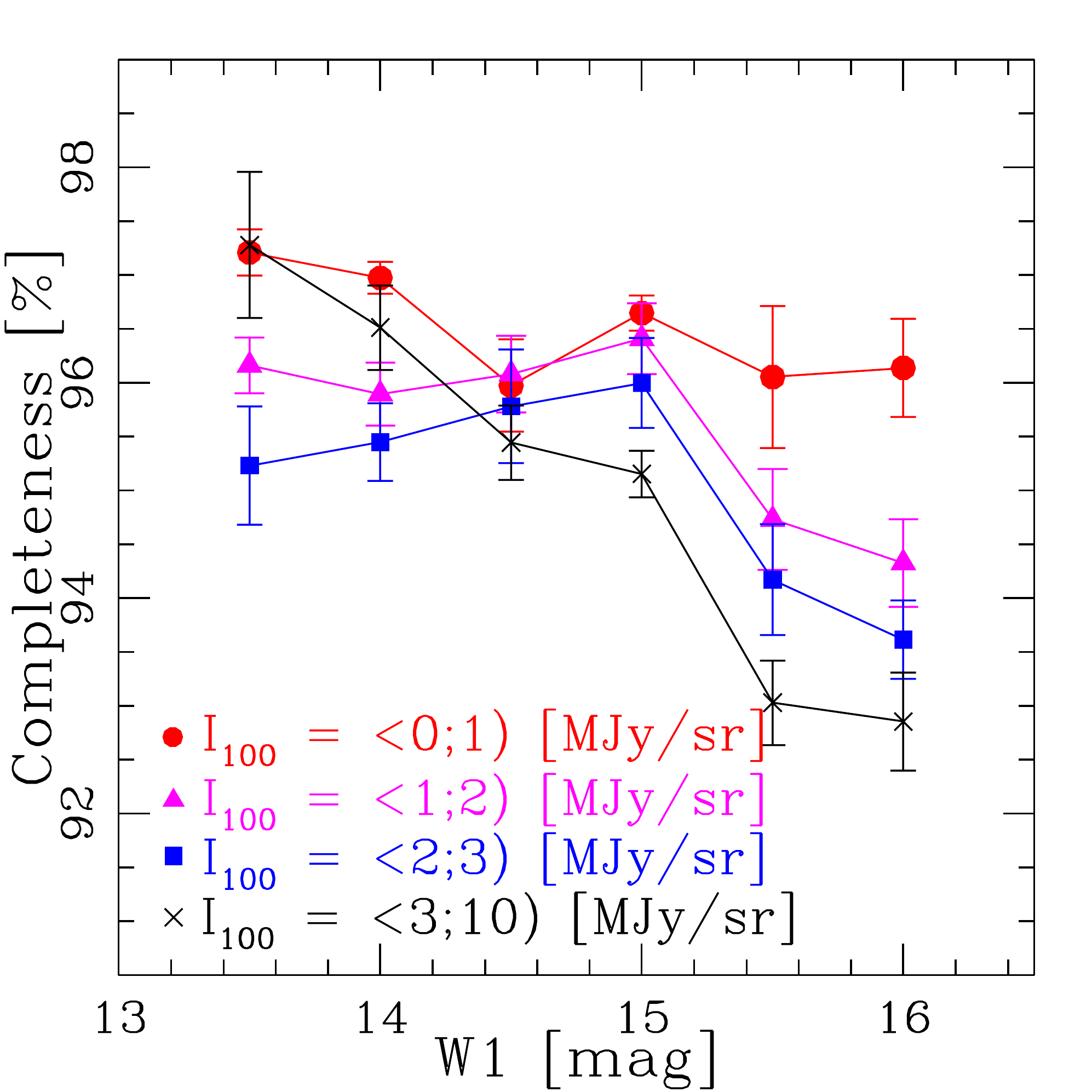}}
\end{subfigure}
\caption{Dependence of the completeness on the magnitude for four $I_{100}$ bins when using three classification parameters ($W1$, $W1-W2$ and $\mtt{w1mag\_1} - \mtt{w1mag\_3}$) for galaxies (upper panel), stars (middle panel) and quasars (lower panel) from the \WS\ sample. These results are for the cross-test. }
\label{Fig: compl ext-mag dependence}
\end{figure}

\begin{figure}
\centering 
\begin{subfigure}{
\includegraphics[scale=0.3]{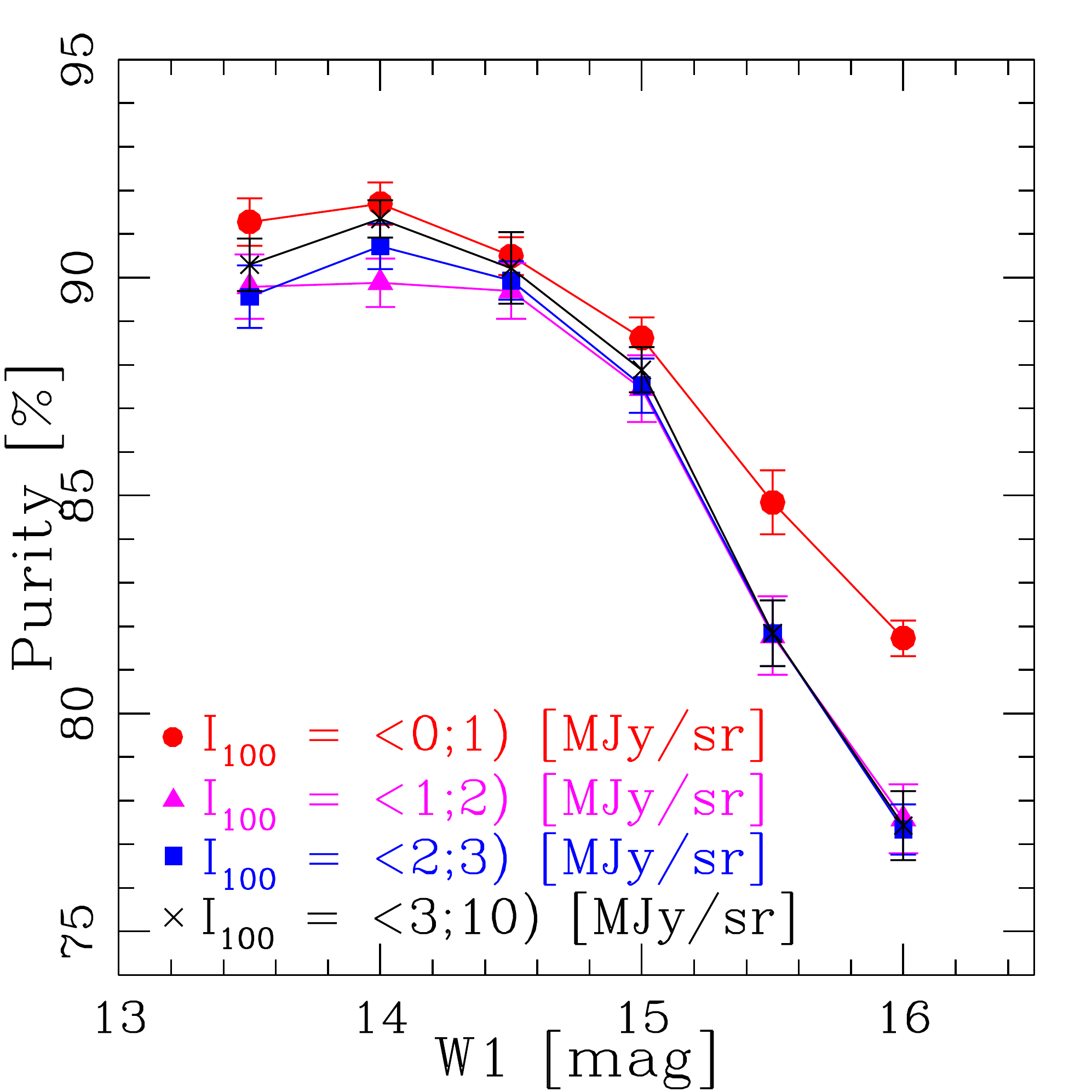}}
\end{subfigure}
\\
\begin{subfigure}{
\includegraphics[scale=0.3]{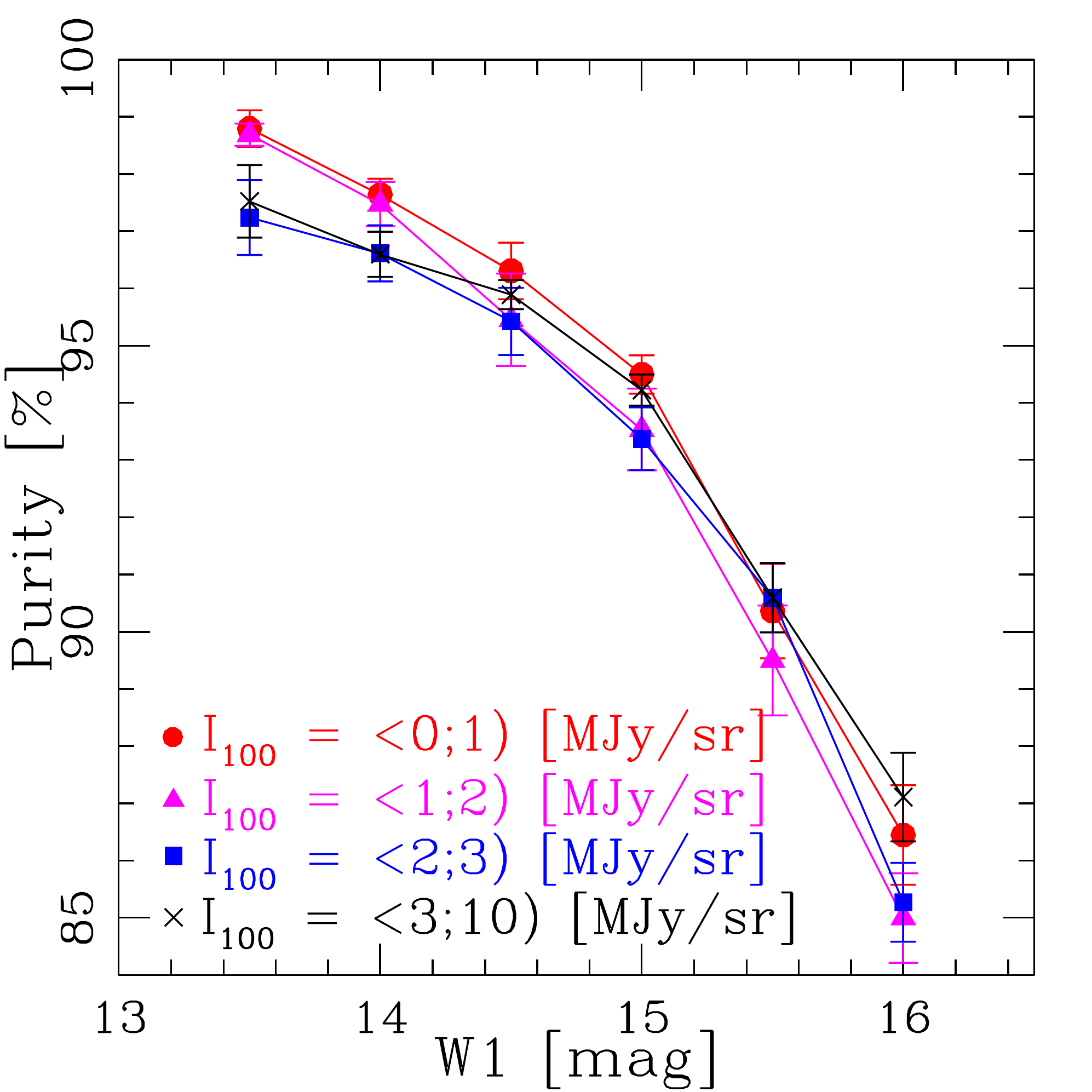}}
\end{subfigure}
\\
\begin{subfigure}{
\includegraphics[scale=0.3]{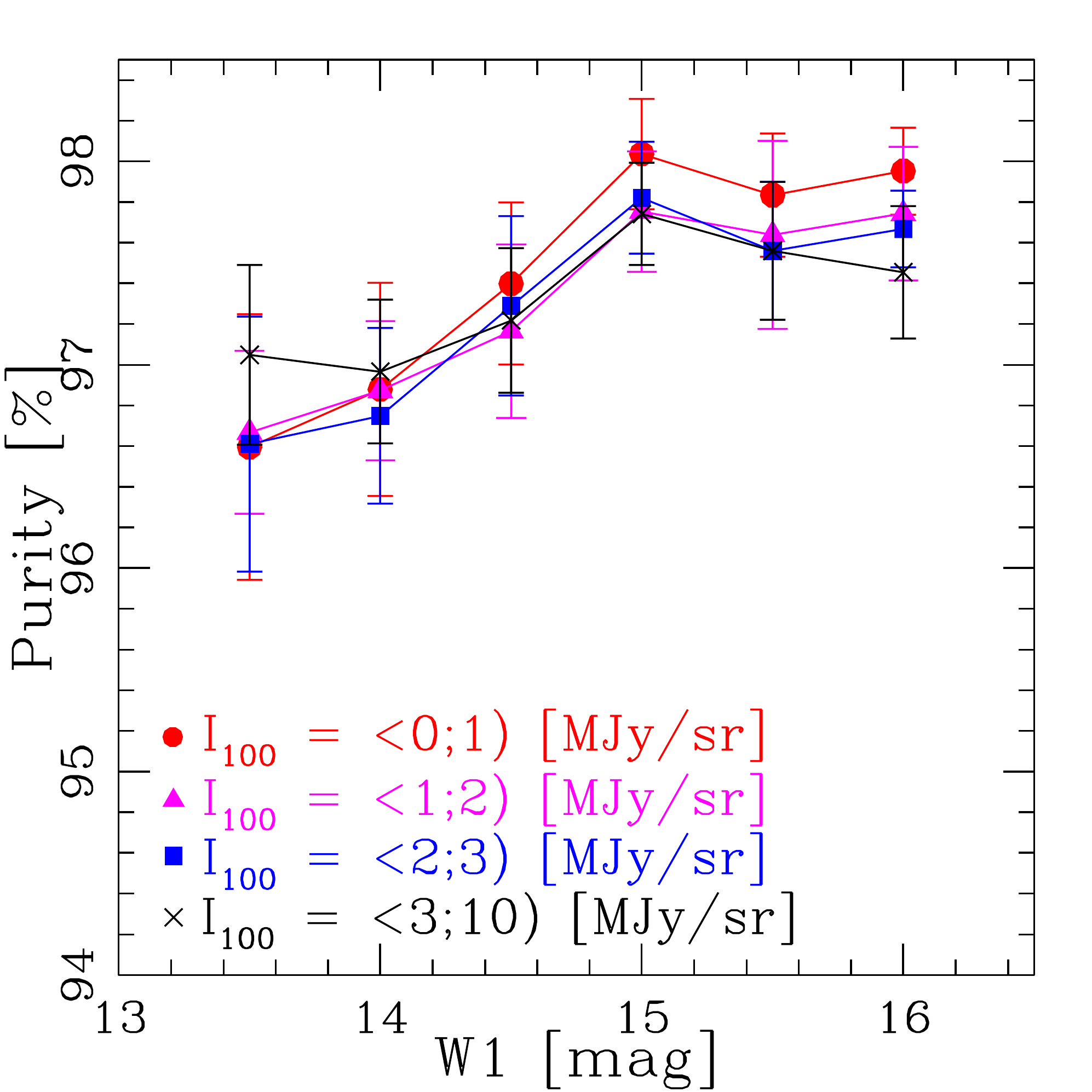}}
\end{subfigure}
\caption{Dependence of the purity on the magnitude for four $I_{100}$ bins when using three classification parameters ($W1$, $W1-W2$, and $\mtt{w1mag\_1} - \mtt{w1mag\_3}$) for galaxies (upper panel), stars (middle panel), and quasars (lower panel) from the \WS\ sample. These results are for the cross-test. }
\label{Fig: puri ext-mag dependence}
\end{figure}

Two final tests of the SVM algorithm applied to WISE\ti{}SDSS data were to check its performance against varying extinction levels and increased magnitude cut. Figs.\ \ref{Fig: compl ext-mag dependence} and \ref{Fig: puri ext-mag dependence} summarise the results, and show how completeness and purity change with varying magnitude for four $I_{100}$ bins, for the three classes of sources. Here we used smaller increments (0.5 mag) than in the other tests where it was 1 mag.

At the bright end, both completeness and purity retain very high levels of greater than $90\%$ irrespective of extinction. These numbers for stars and galaxies gradually deteriorate for fainter sources, and some dependence on the extinction starts to appear as larger magnitudes are reached. The statistics are  relatively stable  and are at very good levels for quasars (where a slight increase in purity is actually observed at the faint end). We note however that even for fainter sources, star and galaxy samples exhibit completeness of over $\sim80\%$ and purity of over $77\%$. Detailed results regarding completeness, purity, and contamination for all magnitude-extinction bins for the self-check and cross-test in the three dimensional parameter-space are presented in Tables \ref{Table: Self-check} and \ref{Table: Cross-test} in the Appendix.

We note, however, that such statistics may be partly misleading as they refer to the test sample, which is statistically consistent with the training set. The full-sky catalogue will  differ, and in particular may (and does) contain sources not represented in the training, such as  asteroids. The classifier tuned to the training set will underperform in such cases, which in particular will be reflected in low probabilities of such objects  belonging to any of the three classes used in the analysis. We discuss this further in the following section.

\section{Application of the SVM classifier to all-sky WISE}
\label{Sec: All-sky}

\begin{figure}
\centering 
\includegraphics[width=0.49\textwidth]{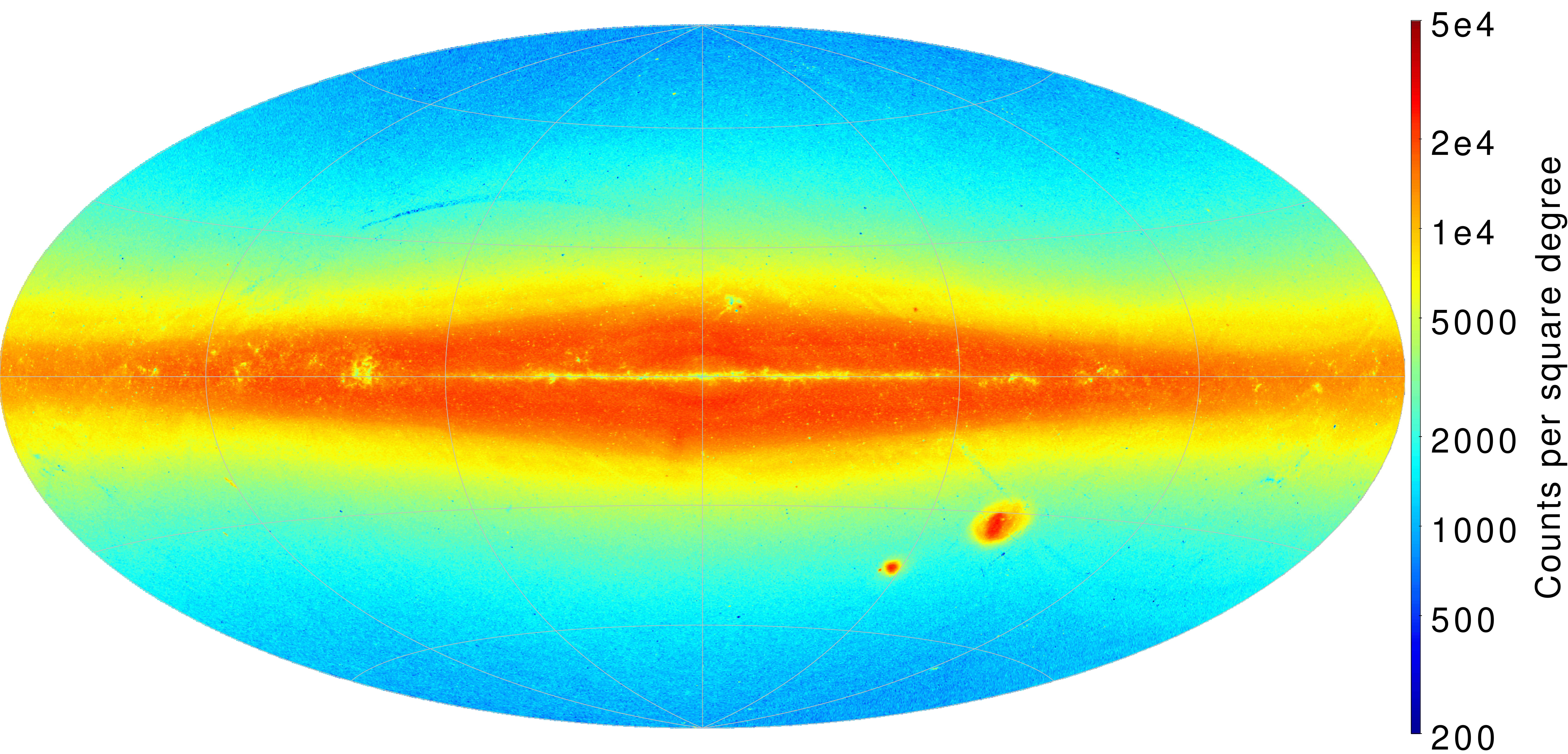}
\caption{All-sky map of 220 million star candidates identified by our classifier in AllWISE $W1<16$ mag data, in Aitoff projection in Galactic coordinates. We note the logarithmic scaling of the count colour bar.}
\label{Fig: all-sky stars}
\end{figure}

Having verified in various ways the performance of our classifier, we finally applied it to the all-sky WISE data limited to $W1<16$ mag. In this case, to tune the classifier we used the most comprehensive and general training data; we  randomly selected $10^4$ galaxies, $10^4$ stars, and $10^4$ quasars from the cross-matched \WS\ dataset with $W1<16$ mag. Thus, the trained classifier flagged 70\% / 27\% / 3\% of our WISE sources respectively as stars / galaxies / QSOs on the full sky. These numbers are consistent with the fact that stars dominate the source counts at the bright end of WISE \citep{Jarrett11,Jarrett16}; however, they should not be taken at face value. At low Galactic latitudes and in other highly crowded areas (Magellanic Clouds, Galactic extended sources such as dust clouds) the classification is highly unreliable. In addition, these numbers refer to the sources for which the probability of being of a given type was higher than that of the  other two (e.g.\ $p(\mathrm{star})>p(\mathrm{galaxy})$ \& $p(\mathrm{star})>p(\mathrm{QSO})$ for stars). However, for a considerable number of objects, especially in the galaxy and QSO classes,  the three probabilities were comparable, which means a very low level of confidence for a class assignment. Thus, to obtain galaxy and quasar candidate catalogues based on our data, the masking of problematic areas was necessary, as was additional cleanup of low-probability sources. In the case of stars this was not needed,  first, because we do expect them to be present in the highly crowded areas and second, their class assignment was the most robust: 99.4\% of the sources classified as stars had $p(\mathrm{star})>0.5$. A map of the 220 million star candidates is shown in Fig.\ \ref{Fig: all-sky stars}.  A decrease in counts at $b\sim0^\circ$ is caused by saturation and blending.

\begin{figure}
\centering 
\includegraphics[width=0.49\textwidth]{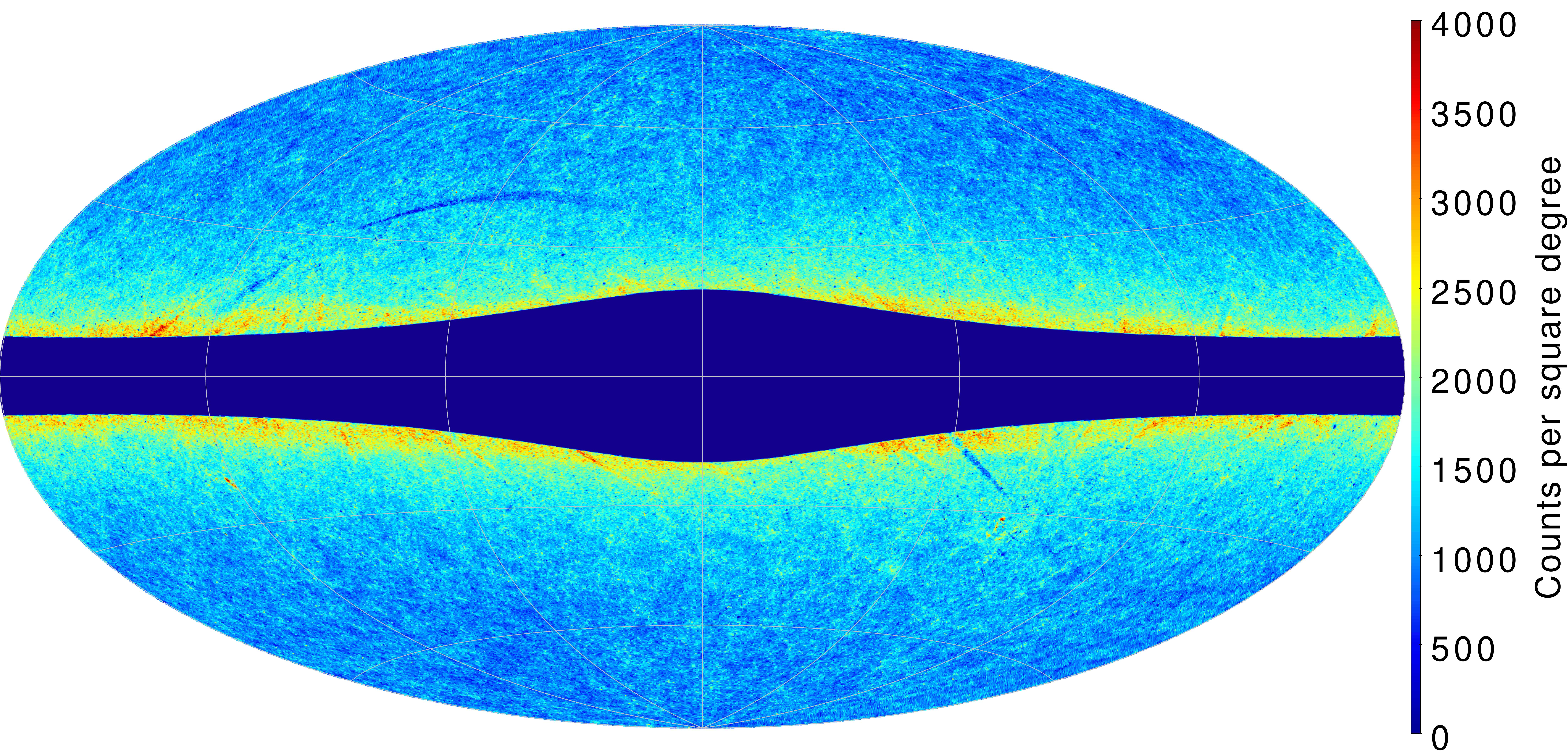}
\caption{All-sky map of 45 million galaxy candidates identified by our classifier in AllWISE $W1<16$ mag data, in Aitoff projection in Galactic coordinates. This plot shows sources over the probability threshold of $p(\mathrm{gal})>0.6$ after appropriate cleanup and masking (see text for details).}
\label{Fig: all-sky galaxies}
\end{figure}

\begin{figure}
\centering 
\includegraphics[width=0.49\textwidth]{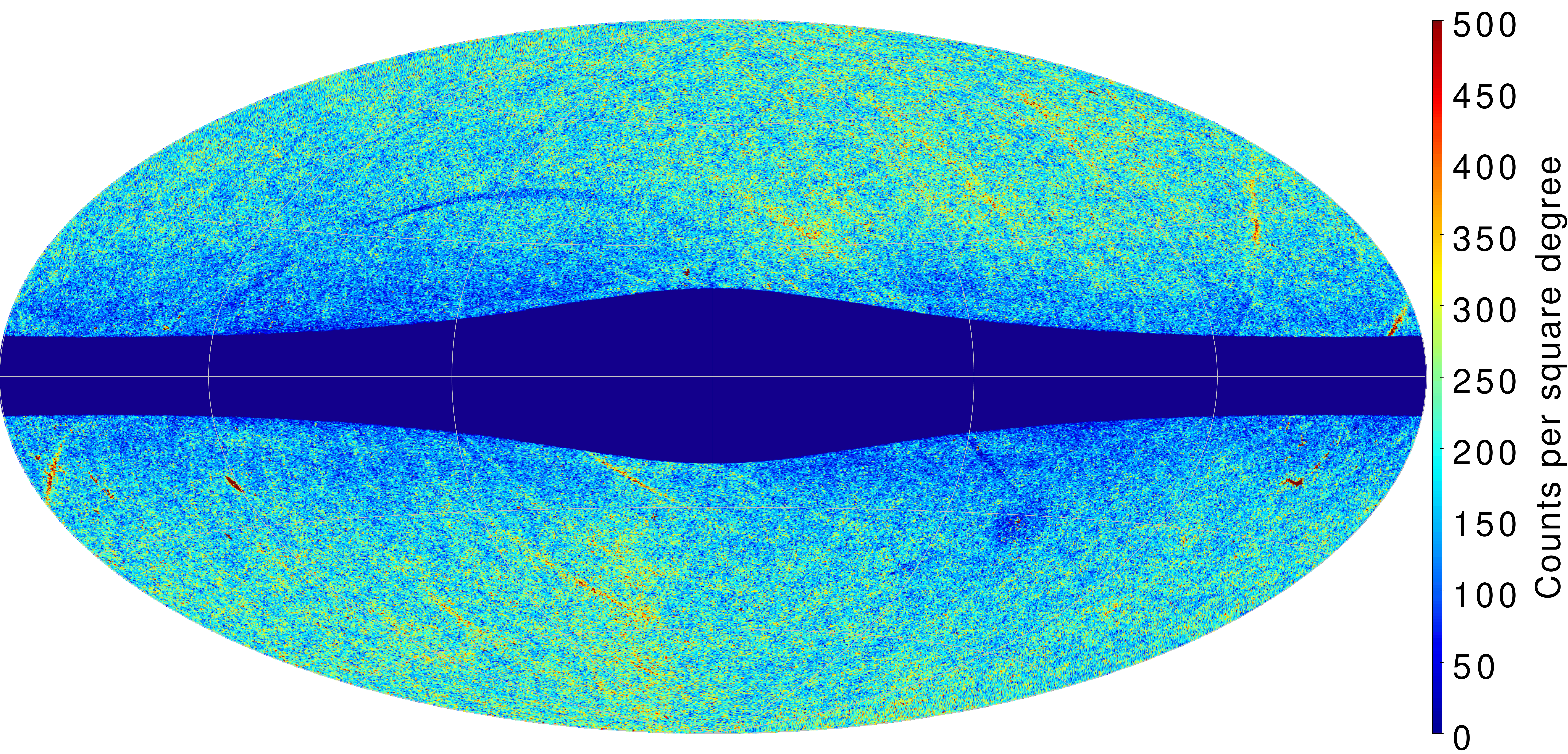}
\caption{Sources flagged by our classifier as quasar candidates in the WISE $W1<16$ mag catalogue. This sample shows 6 million objects with an SVM probability $p(\rm{QSO})>0.5$ after appropriate cleanup and masking (see text for details).}
\label{Fig: all-sky QSO}
\end{figure}

The catalogue of candidate galaxies, unlike stars, needed considerable purification. First of all, we had to cut out the most confused areas of the Galactic Plane and Bulge, using a longitude-dependent masking of the $|b|<6.5^\circ$ sources at $\ell=180^\circ$ up to $|b|<20^\circ$ near the Galactic Centre. This removed almost 30 million objects out of the $84.5\times10^6$ pre-assigned to the galaxy class all-sky. As seen in Fig.\ \ref{Fig: all-sky galaxies}, this mask could have been wider, but we leave it in this  form to emphasise classification issues at low Galactic latitudes where blends become a significant problem for star/galaxy separation in WISE.  In addition to the bright-end cut already mentioned in Sec.\ \ref{Sec: SDSS} (due to saturation and lack of bright sources in the training set), which affected a very small number of the objects, we also eliminated over 400,000 outliers at the faint end in the $W2$ band, $W2>16.1$ mag. These were located mostly near the Ecliptic Poles where WISE coverage was the highest owing to the scanning strategy. This cutout also automatically removed the sources with $W1-W2<-0.1$ mag which are most certainly stellar \citep{WISE}. In the final stage of the galaxy catalogue cleanup, we used the probabilities assigned by SVM as described in Sec.\ \ref{Sec: SVM}, and examined the sky and colour distribution of the galaxy candidates after applying different thresholds in $p(\mathrm{gal})$. More aggressive cuts in this probability lead to a more uniform distribution of the sources as a function of latitude. However, even for $p(\mathrm{gal})>0.7$ or more, differences of over 50\% in the source density remain between the Galatic Caps and $|b|\sim20^\circ$. As the $W1-W2$ colours of the objects with high galaxy probability were consistent with those of genuine galaxies, the effect of gradually increasing number counts from high to low Galactic latitudes must be related to both the stellar contamination (blending) and galaxy incompleteness going up with decreasing $|b|$. 

Fig.\ \ref{Fig: all-sky galaxies} shows an example of all-sky source distribution of galaxy candidates. Included are 45 million sources obtained after the masking, bright- and faint-magnitude cutouts, and placement of a threshold of $p(\mathrm{gal})>0.6$.  There is clear contamination at low Galactic latitudes, but the classifier seems to be working well at least for half of the sky down to $|b|=30^\circ$. The missing data in stripes on the left above the Plane and just below it on the right are due to AllWISE instrumental artefacts (saturation at the beginning of the post-cryogenic phase) as already discussed in Sec.\ \ref{Sec: WISE}.

The quasar candidates underwent similar purification to the galaxy candidates. The same cutout of the Galactic plane was first applied, which removed almost 30\% of the 9.4 million all-sky sources flagged by SVM as QSO. In addition, more aggressive bright-end cuts than in the galaxy case were necessary to avoid dangerous extrapolation from the training sample. We removed quasar candidates with $W1<10.4$ mag or $W2<10.1$ mag, as  such bright QSOs are practically non-existent in \WS. In the SVM output, they were mostly misclassified stars or blends of stars, localised chiefly at low Galactic latitudes and in the Magellanic Clouds. A further cleanup to keep only the $p(\rm{QSO})>0.5$ sources resulted in 6 million objects as pictured in Fig.\ \ref{Fig: all-sky QSO}. This number is most certainly an overestimate for the true WISE quasar population at $W1<16$ mag. In addition to the saturation-related artefacts, we note some interesting features in the map which are not seen for the galaxy candidates. First, there is a lack of sources at low Galactic latitudes, qualitatively similar to the WISE AGN distribution presented in \cite{Ferraro14}, where sources were classified based on colour cuts. Second, various WISE scanning issues are imprinted here, the most important being overdensity stripes perpendicular to the ecliptic, resulting from Moon avoidance manoeuvres (cf.\ the mask applied in \citealt{Ferraro14}). There is, however, additional spurious overdensity which seems to roughly follow the ecliptic, visible at the top right of the map and below the Bulge, to the left. This suggests some very local contamination, such as from asteroids or maybe zodiacal light, and most likely reflects the presence of a fourth type of source in addition to the three types in the training set from SDSS.

\begin{figure}
\centering 
\includegraphics[width=0.49\textwidth]{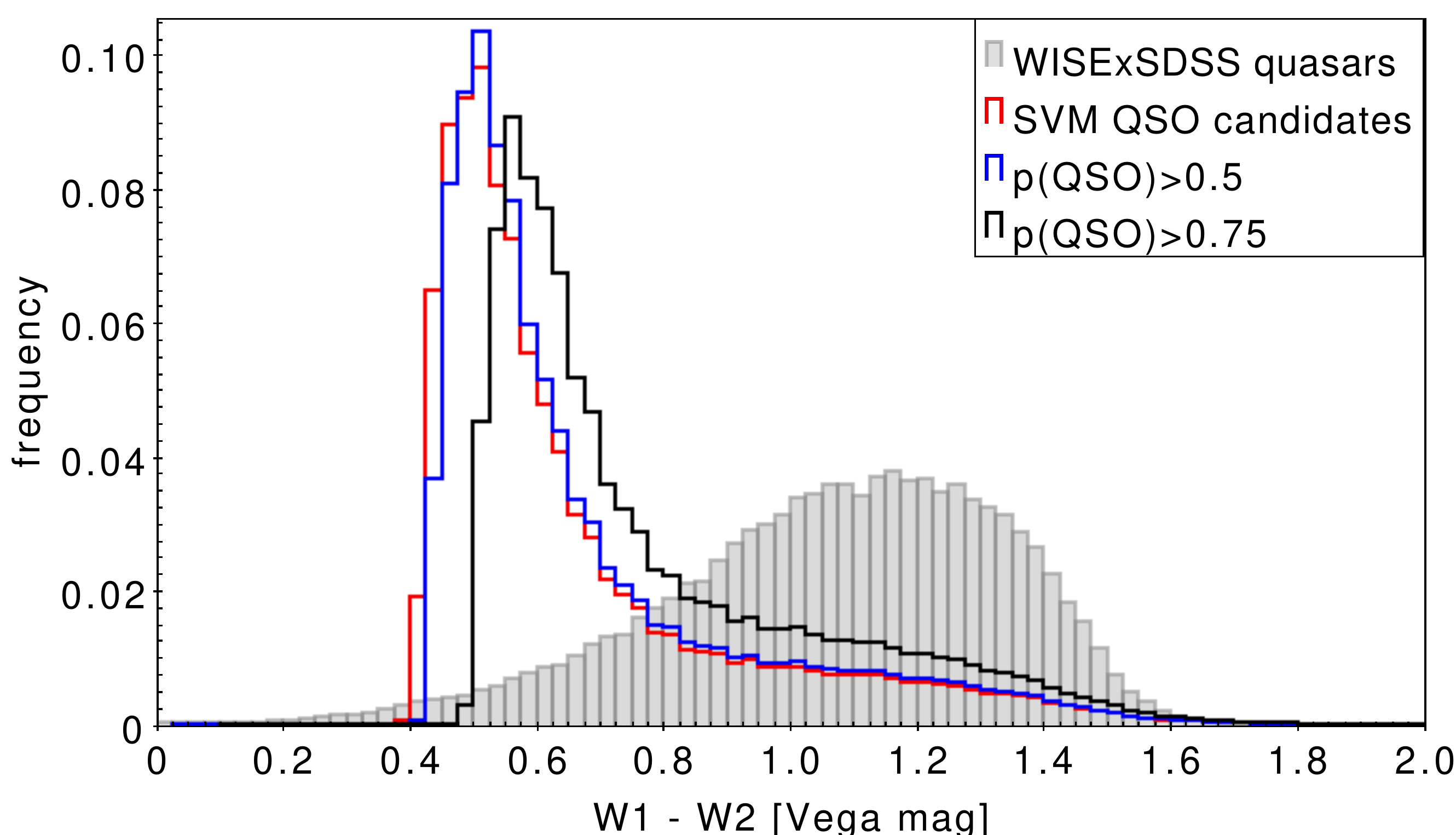} 
\caption{Distribution of the $W1-W2$ colour for \WS\ quasars (grey bars) and for SVM quasar candidates identified in all-sky WISE data, with no threshold on the SVM QSO probability (red), and for  $p(\rm{QSO})>0.5$  and $>0.75$ (blue and black, respectively).}
\label{Fig: QSO cand W1-W2}
\end{figure}

Unlike in the test phase described in Sec.\ \ref{Sec: tests}, here we do not know the `truth' to which we could compare the classifier's performance; however, some indirect a posteriori tests are  possible. The first  is the all-sky distribution, which for stars and galaxy candidates (at high latitudes) is consistent with expectations, but much less for the quasars. The second test is to verify source properties, such as colours. For identified galaxies, the $W1-W2$ colour is very  consistent with the colour found in the \WS\ training set (cf.\ Fig.\ \ref{Fig: WISExSDSS W1-W2}) when   the higher $p(\rm{gal})$ cut is applied. For quasar candidates the situation is different. Even for very high thresholds of $p(\rm{QSO})$, the peak in this colour is at $W1-W2\sim0.6$ mag rather than $\sim1$ mag as in the training set (Fig.\ \ref{Fig: QSO cand W1-W2}). In fact, the sources of expected quasar nature with $W1-W2>0.8$ mag \citep{Stern12} are only 1/5 of our QSO candidate sample. As already seen from their all-sky map, some of the contamination may come from solar system objects. As an additional verification, we checked the  location of the SVM QSO in the $W2-W3$ vs. $W1-W2$ diagram for those sources that had a $W3$ detection. The bulk of the QSO candidates are located at $3<W2-W3<4$ [mag] and $0.4<W1-W2<0.7$ [mag], which does not give a clear characterisation of their nature. Indeed, following \cite{WISE}, this is where various galaxy types overlap in this parameter space (`normal' spirals, Seyferts, strabursts, LIRGs, etc.). This also indicates that even if we had used the $W3$ parameter (which, as we emphasise, is robustly measured only for a small subset of our sources), this degeneracy between quasars and non-AGN galaxies would most likely remain.

\section{Summary and future prospects}
\label{Sec: summary}

In this paper we presented an application of a machine learning algorithm -- the support vector machines -- to classify sources in an all-sky catalogue drawn from WISE. The algorithm was trained and tested on a sample of WISE objects cross-matched with SDSS spectroscopic data, where three main types of astrophysical sources -- stars, galaxies and quasars -- had been independently identified. To optimise the performance of SVM, we first determined that a polynomial kernel of the third degree is  preferred over the traditionally used radial one. We next verified that a training sample of less than 10,000 randomly chosen sources was sufficient to obtain stable results; in addition, the  algorithm had already performed satisfactorily  for a three-dimensional parameter space ($W1$ magnitude; $W1-W2$ colour; differential aperture mag in the $W1$ channel). 

Having established the optimal set-up of the SVM method for our purposes, we performed several tests of its performance on WISE data. Here we focused on completeness and purity as a function of the limiting magnitude of the test sample, and  on their dependence on Galactic extinction. For stars and galaxies both these statistics deteriorate for increasing magnitudes, but even at the faint end they rarely fall below $\sim80\%$. On the other hand, no obvious dependence of the SVM performance on magnitude is observed for quasars. Finally, Galactic extinction does not seem to have influence on the results, although we note that the tests were limited to regions of $EBV\lesssim0.3$, outside of which there is practically no calibration data. 

We finally applied the SVM algorithm, trained on the \WS\ sample, to the full-sky WISE data flux-limited to $W1<16$ mag. About 220 million sources preselected in this way were flagged  by SVM as star candidates; the remaining objects required significant cleanup to obtain galaxy-candidate and QSO-candidate samples. This cleanup consisted in removing the brightest sources, as well as those located in the Galactic Plane and Bulge areas, for which the classification is not expected to be reliable. We also used source type probabilities provided by SVM to remove the objects of insecure classification. As a result, we obtained catalogues of 45 million galaxy candidates, as well as of 6 million QSOs. In the latter case, however, we observe significant contamination by sources consistent with dusty (non-AGN) galaxies and by probable solar system objects.

These shortcomings of our classification are related to the limitations of the training sample and  to the lack of additional classification parameters that could be reliably used for the full sample together with the three basic ones employed here.  It is possible to mitigate the former drawback  thanks to forthcoming spectroscopic data, for example   from SDSS-IV; however,  it is not clear how much it is possible to improve the latter if only WISE data are to be used for the classification on the full sky while keeping a deep and uniform sample. Measurements in WISE $W3$ (12 $\mu$m) and $W4$ (23 $\mu$m) channels would certainly help to break degeneracies that result in unreliable identification of quasars in the present approach; however, these two bands offer  much shallower and very inhomogeneous coverage  compared to $W1$ and $W2$. Some improvement in classification could also be expected if there were reliable proper motions    for a much larger sample of WISE sources than presently available, as these data would help identify at least some of the minor bodies of the solar system which most likely contaminate our current QSO sample.

In general, a natural next step in the process of classification of WISE sources is to expand the current scheme to a larger number of object classes, which will allow the creation of more robust catalogues or at least the purification of the current ones. For this to be accomplished, more classification parameters, and more comprehensive training sets will be necessary. We plan to explore this in forthcoming studies ({\color{blue}{Wypych et al., in prep.}}). Last but not least, it is possible to work on improving the training scheme itself by implementing the so-called fuzzy logic (e.g. \citealt{klirg}) into the SVM algorithm. While in the classical SVM approach all training examples are treated equally, the fuzzy logic procedure   handles the uncertainties of the classification data by weighting the training examples (e.g. \citealt{fuzzy1}; \citealt{fuzzy2}). Each training point may belong to no more then one class, but by weighting the training points Fuzzy-SVM (FSVM) can ensure that the meaningful data points will be classified correctly, while the  noisier ones will have more freedom to be misclassified in order to ensure the maximum margin benefit. This in turn   expands the classification regions in the parameter space. However, this approach heavily extends the computational time  owing to the introduction of the additional free parameter, which,  like other SVM parameters (e.g. misclassification parameter $C$ or kernel parameters), must be tuned for best performance.  While this method could help to improve the current classification, the uncertainties of the measurements of the objects considered in this work are relatively small. In view of the largely extended computational time, the FSVM was not favourable for the purpose of the current analysis, but it will be considered in our future studies of classification in  noisier WISE or other data.

\begin{acknowledgements}

We thank the referee for the helpful review. Special thanks to Mark Taylor for the TOPCAT \citep{TOPCAT} and STILTS \citep{STILTS} software\footnote{\url{http://www.star.bris.ac.uk/\~mbt/}}. Some of the results in this paper have been derived using the HEALPix package \citep{HEALPIX}\footnote{\url{http://healpix.jpl.nasa.gov/}}.

This work was supported by the Polish National Science Center under contracts \# UMO-2012/07/D/ST9/02785. MB was supported by the Netherlands Organization for Scientific Research, NWO, through grant number 614.001.451, by the  European Research Council through FP7 grant number 279396 and by the South African National Research Foundation (NRF). AP was partially supported by the Polish-Swiss Astro Project, co-financed by a grant from Switzerland, through the Swiss Contribution to the enlarged European Union. 

This publication makes use of data products from the Wide-field Infrared Survey Explorer, which is a joint project of the University of California, Los Angeles, and the Jet Propulsion Laboratory/California Institute of Technology, funded by the National Aeronautics and Space Administration.

Funding for SDSS-III has been provided by the Alfred P. Sloan Foundation, the Participating Institutions, the National Science Foundation, and the U.S. Department of Energy Office of Science. The SDSS-III web site is \url{http://www.sdss3.org/}.

SDSS-III is managed by the Astrophysical Research Consortium for the Participating Institutions of the SDSS-III Collaboration including the University of Arizona, the Brazilian Participation Group, Brookhaven National Laboratory, Carnegie Mellon University, University of Florida, the French Participation Group, the German Participation Group, Harvard University, the Instituto de Astrofisica de Canarias, the Michigan State/Notre Dame/JINA Participation Group, Johns Hopkins University, Lawrence Berkeley National Laboratory, Max Planck Institute for Astrophysics, Max Planck Institute for Extraterrestrial Physics, New Mexico State University, New York University, Ohio State University, Pennsylvania State University, University of Portsmouth, Princeton University, the Spanish Participation Group, University of Tokyo, University of Utah, Vanderbilt University, University of Virginia, University of Washington, and Yale University. 

\end{acknowledgements}

\bibliographystyle{aa}
\bibliography{biblio}

\begin{appendix}

\section{Tables with detailed results of the tests}
In this Appendix we provide tables with detailed results of the tests described in Section \ref{Sec: tests}.
Tables \ref{Table: Self-check} and \ref{Table: Cross-test} summarise the statistics of the completeness, purity, and contamination for various combinations of extinction bins and flux limits in the test sets, for the self-check and cross-test cases.

\begin{table*}
\caption{\label{Table: Self-check}
Overall classification statistics (in \%) for various combinations of extinction bins and flux limits for the self-check case (classified objects were the same as in the training sample).}
\centering
\begin{footnotesize}
\begin{tabular}{|c|c|c|c|c|c|c|}
\hline
\multicolumn{ 7}{|c|}{SELF-CHECK} \\ \hline
Magnitude limit & \multicolumn{ 6}{c|}{$W1 < 14$ mag} \\ \hline
Extinction [MJy/sr] & \multicolumn{ 3}{c|}{$\langle 0; 1)$} & \multicolumn{ 3}{c|}{$\langle 1; 2)$} \\ \hline
 & \multicolumn{1}{l|}{Completeness} & \multicolumn{1}{l|}{Purity} & \multicolumn{1}{l|}{Contamination} & \multicolumn{1}{l|}{Completeness} & \multicolumn{1}{l|}{Purity} & \multicolumn{1}{l|}{Contamination} \\ \hline
Galaxy & 95.0 & 91.8 & 8.2 & 90.1 & 89.9 & 10.1  \\ \cline{ 1- 4}\cline{ 5- 7}
Stars & 94.5 & 97.8 & 2.2  & 97.5 & 97.5 & 2.5 \\ \cline{ 1- 4}\cline{ 5- 7}
QSO & 96.9 & 97.1 & 2.9 & 97.0 & 96.9 & 3.1 \\ \hline
Extinction [MJy/sr] & \multicolumn{ 3}{c|}{$\langle 2; 3)$} & \multicolumn{ 3}{c|}{$\langle 3; 10)$} \\ \hline
 & \multicolumn{1}{l|}{Completeness} & \multicolumn{1}{l|}{Purity} & \multicolumn{1}{l|}{Contamination} & \multicolumn{1}{l|}{Completeness} & \multicolumn{1}{l|}{Purity} & \multicolumn{1}{l|}{Contamination} \\ \hline
Galaxy & 94.6 & 91.3 & 8.7  & 94.5 & 92.2 & 7.8\\ \cline{ 1- 4}\cline{ 5- 7}
Stars & 94.6 & 96.9 & 3.1  & 94.9 & 97.1 & 2.9 \\ \cline{ 1- 4}\cline{ 5- 7}
QSO & 95.8 & 97.0 & 3.0 & 97.0 & 97.3 & 2.7 \\ \hline
Magnitude limit & \multicolumn{ 6}{c|}{$W1<15$ mag} \\ \hline
Extinction [MJy/sr]& \multicolumn{ 3}{c|}{$\langle 0; 1)$} & \multicolumn{ 3}{c|}{$\langle 1; 2)$} \\ \hline
 & \multicolumn{1}{l|}{Completeness} & \multicolumn{1}{l|}{Purity} & \multicolumn{1}{l|}{Contamination} & \multicolumn{1}{l|}{Completeness} & \multicolumn{1}{l|}{Purity} & \multicolumn{1}{l|}{Contamination}\\ \hline
Galaxy & 93.3 & 89.1 & 10.9 & 92.2 & 87.8 & 12.2  \\ \cline{ 1- 4}\cline{ 5- 7}
Stars & 91.5 & 94.9 & 5.1 & 90.5 & 93.8 & 6.2 \\ \cline{ 1- 4}\cline{ 5- 7}
QSO & 97.0 & 98.2 & 1.8 & 96.6 & 98.0 & 2.0 \\ \hline
Extinction [MJy/sr]& \multicolumn{ 3}{c|}{$\langle 2; 3)$} & \multicolumn{ 3}{c|}{$\langle 3; 10)$} \\ \hline
 & \multicolumn{1}{l|}{Completeness} & \multicolumn{1}{l|}{Purity} & \multicolumn{1}{l|}{Contamination} & \multicolumn{1}{l|}{Completeness} & \multicolumn{1}{l|}{Purity} & \multicolumn{1}{l|}{Contamination} \\ \hline
Galaxy & 92.3 & 88.4 & 11.6 & 92.5 & 88.5 & 11.5 \\ \cline{ 1- 4}\cline{ 5- 7}
Stars & 91.4 & 93.9 & 6.1 & 92.6 & 94.2 & 5.8 \\ \cline{ 1- 4}\cline{ 5- 7}
QSO & 96.1 & 97.9 & 2.1 & 95.3 & 98.1 & 1.9 \\ \hline
Magnitude limit& \multicolumn{ 6}{c|}{$W1<16$ mag} \\ \hline
Extinction [MJy/sr] & \multicolumn{ 3}{c|}{$\langle 0; 1)$} & \multicolumn{ 3}{c|}{$\langle 1; 2)$} \\ \hline
 & \multicolumn{1}{l|}{Completeness} & \multicolumn{1}{l|}{Purity} & \multicolumn{1}{l|}{Contamination} & \multicolumn{1}{l|}{Completeness} & \multicolumn{1}{l|}{Purity} & \multicolumn{1}{l|}{Contamination} \\ \hline
Galaxy & 87.7 & 82.7 & 17.3 & 87.2 & 79.3 & 20.7 \\ \cline{ 1- 4}\cline{ 5- 7}
Stars & 84.4 & 88.0 & 12.0 & 81.4 & 87.1 & 12.9 \\ \cline{ 1- 4}\cline{ 5- 7}
QSO & 96.6 & 98.5 & 1.5 & 95.0 & 98.4 & 1.6 \\ \hline
Extinction [MJy/sr]& \multicolumn{ 3}{c|}{$\langle 2; 3)$} & \multicolumn{ 3}{c|}{$\langle 3; 10)$} \\ \hline
 & \multicolumn{1}{l|}{Completeness} & \multicolumn{1}{l|}{Purity} & \multicolumn{1}{l|}{Contamination} & \multicolumn{1}{l|}{Completeness} & \multicolumn{1}{l|}{Purity} & \multicolumn{1}{l|}{Contamination} \\ \hline
Galaxy & 86.6 & 78.5 & 21.5 & 87.6 & 79.0 & 21.0 \\ \cline{ 1- 4}\cline{ 5- 7}
Stars & 81.2 & 86.6 & 13.4 & 82.9 & 88.1 & 11.9 \\ \cline{ 1- 4}\cline{ 5- 7}
QSO & 94.1 & 98.1 & 1.9 & 93.3 & 98.1 & 1.9 \\ \hline
\end{tabular}
\end{footnotesize}

\end{table*}

\begin{table*}
\caption{\label{Table: Cross-test}
Overall classification statistics (in \%) for various combinations of extinction bins and flux limits for the cross-test case (classified objects were different from those in the training sample).}
\centering
\begin{footnotesize}
\begin{tabular}{|c|c|c|c|c|c|c|}
\hline
\multicolumn{ 7}{|c|}{CROSS-TEST} \\ \hline
Magnitude limit& \multicolumn{ 6}{c|}{$W1<14$ mag} \\ \hline
Extinction [MJy/sr]& \multicolumn{ 3}{c|}{$\langle 0; 1)$} & \multicolumn{ 3}{c|}{$\langle 1; 2)$} \\ \hline
 & \multicolumn{1}{l|}{Completeness} & \multicolumn{1}{l|}{Purity} & \multicolumn{1}{l|}{Contamination} & \multicolumn{1}{l|}{Completeness} & \multicolumn{1}{l|}{Purity} & \multicolumn{1}{l|}{Contamination} \\ \hline
Galaxy & 94.7 & 91.7 & 8.3 & 94.7 & 89.9 & 10.1 \\ \cline{ 1- 4}\cline{ 5- 7}
Stars & 94.3 & 97.6 & 2.4 & 93.3 & 97.5 & 2.5 \\ \cline{ 1- 4}\cline{ 5- 7}
QSO & 97.0 & 96.9 & 3.1 & 95.9 & 96.9 & 3.1 \\ \hline
Extinction [MJy/sr]& \multicolumn{ 3}{c|}{$\langle 2; 3)$} & \multicolumn{ 3}{c|}{$\langle 3; 10)$} \\ \hline
 & \multicolumn{1}{l|}{Completeness} & \multicolumn{1}{l|}{Purity} & \multicolumn{1}{l|}{Contamination} & \multicolumn{1}{l|}{Completeness} & \multicolumn{1}{l|}{Purity} & \multicolumn{1}{l|}{Contamination} \\ \hline
Galaxy & 94.2 & 90.7 & 9.3 & 93.8 & 91.3 & 8.7 \\ \cline{ 1- 4}\cline{ 5- 7}
Stars & 94.2 & 96.6 & 3.4  & 94.5 & 96.6 & 3.4 \\ \cline{ 1- 4}\cline{ 5- 7}
QSO & 95.4 & 96.8 & 3.2 & 96.5 & 97.0 & 3.0 \\ \hline
Magnitude limit& \multicolumn{ 6}{c|}{$W1<15$ mag} \\ \hline
Extinction [MJy/sr]& \multicolumn{ 3}{c|}{$\langle 0; 1)$} & \multicolumn{ 3}{c|}{$\langle 1; 2)$} \\ \hline
 & \multicolumn{1}{l|}{Completeness} & \multicolumn{1}{l|}{Purity} & \multicolumn{1}{l|}{Contamination} & \multicolumn{1}{l|}{Completeness} & \multicolumn{1}{l|}{Purity} & \multicolumn{1}{l|}{Contamination} \\ \hline
Galaxy & 92.9 & 88.6 & 11.4 & 91.8 & 87.5 & 12.5 \\ \cline{ 1- 4}\cline{ 5- 7}
Stars & 91.2 & 94.5 & 5.5 & 90.2 & 93.5 & 6.5 \\ \cline{ 1- 4}\cline{ 5- 7}
QSO & 96.6 & 98.0 & 2.0 & 96.4 & 97.8 & 2.2 \\ \hline
Extinction [MJy/sr]& \multicolumn{ 3}{c|}{$\langle 2; 3)$} & \multicolumn{3}{c|}{$\langle 3; 10)$} \\ \hline
 & \multicolumn{1}{l|}{Completeness} & \multicolumn{1}{l|}{Purity} & \multicolumn{1}{l|}{Contamination} & \multicolumn{1}{l|}{Completeness} & \multicolumn{1}{l|}{Purity} & \multicolumn{1}{l|}{Contamination} \\ \hline
Galaxy & 91.8 & 87.5 & 12.5 & 92.2 & 87.9 & 12.1 \\ \cline{ 1- 4}\cline{ 5- 7}
Stars & 90.5 & 93.4 & 6.6 & 92.0 & 94.2 & 5.8 \\ \cline{ 1- 4}\cline{ 5- 7}
QSO & 96.0 & 97.8 & 2.2 & 95.2 & 97.7 & 2.3  \\ \hline
Magnitude limit& \multicolumn{ 6}{c|}{$W1<16$ mag} \\ \hline
Extinction [MJy/sr]& \multicolumn{ 3}{c|}{$\langle 0; 1)$} & \multicolumn{ 3}{c|}{$\langle 1; 2)$} \\ \hline
 & \multicolumn{1}{l|}{Completeness} & \multicolumn{1}{l|}{Purity} & \multicolumn{1}{l|}{Contamination} & \multicolumn{1}{l|}{Completeness} & \multicolumn{1}{l|}{Purity} & \multicolumn{1}{l|}{Contamination} \\ \hline
Galaxy & 85.5 & 81.7 & 18.3 & 84.5 & 77.6 & 22.4 \\ \cline{ 1- 4}\cline{ 5- 7}
Stars & 84.0 & 86.4 & 13.6 & 80.3 & 85.0 & 15.0 \\ \cline{ 1- 4}\cline{ 5- 7}
QSO & 96.1 & 98 & 2.0 & 94.3 & 97.7 & 2.3 \\ \hline
Extinction [MJy/sr] & \multicolumn{ 3}{c|}{$\langle 2; 3)$} & \multicolumn{3}{c|}{$\langle 3; 10)$} \\ \hline
 & \multicolumn{1}{l|}{Completeness} & \multicolumn{1}{l|}{Purity} & \multicolumn{1}{l|}{Contamination} & \multicolumn{1}{l|}{Completeness} & \multicolumn{1}{l|}{Purity} & \multicolumn{1}{l|}{Contamination} \\ \hline
Galaxy & 84.9 & 77.3 & 22.7 & 86.0 & 77.4 & 22.6 \\ \cline{ 1- 4}\cline{ 5- 7}
Stars & 80.5 & 85.3 & 14.7 & 81.5 & 87.1 & 12.9 \\ \cline{ 1- 4}\cline{ 5- 7}
QSO & 93.6 & 97.7 & 2.3 & 92.9 & 97.5 & 2.5 \\ \hline
\end{tabular}
\end{footnotesize}

\end{table*}

\end{appendix}

\end{document}